\RequirePackage{rotating}
\documentclass[fleqn,usenatbib]{mnras}

\usepackage[T1]{fontenc}

\DeclareRobustCommand{\VAN}[3]{#2}
\let\VANthebibliography\thebibliography
\def\thebibliography{\DeclareRobustCommand{\VAN}[3]{##3}\VANthebibliography}

\usepackage{graphicx}	% Including figure files
\usepackage{amsmath}	% Advanced maths commands
\usepackage{amssymb}	% Extra maths symbols
\usepackage{newtxtext,newtxmath}
\usepackage{multirow}
\usepackage{lscape}
\usepackage{makecell}
\usepackage{subfigure}
\usepackage{float}
\usepackage{newtxtext,newtxmath}
\usepackage{lineno}
\setcitestyle{notesep={ }}
\makeatletter
\newcommand{\labtext}[2]{%
  \@bsphack
  \csname phantomsection\endcsname % in case hyperref is used
  \def\@currentlabel{#1}{\label{#2}}%
  \@esphack
}

\newcommand{\Msun}{{{\rm M}_\odot}}

\setcitestyle{notesep={}}

\makeatother
\title[Impact of Long-Term Flares]{Distortions in Periodicity Analysis of Blazars: The Impact of Flares}
\author[P. Pe\~nil et al.]{
P. Pe\~nil,$^{1}$\thanks{E-mail: ppenil@clemson.edu}
N. Torres$-$Alb\`a,$^{2, 3}$\thanks{E-mail: nuria@virginia.edu}
A. Rico,$^{1}$\thanks{E-mail: aricoro@clemson.edu}
M. Ajello,$^{1}$
S. Buson,$^{2}$
S. Adhikari,$^{1}$
\\
$^{1}$Department of Physics and Astronomy, Clemson University, Kinard Lab of Physics, Clemson, SC 29634-0978, USA\\
$^{2}${GECO Fellow} \\
$^{3}$Department of Astronomy, University of Virginia, P.O. Box 400325, Charlottesville, VA 22904, USA \\
$^{4}$Julius-Maximilians-Universit\"at, 97070, W\"urzburg, Germany\\
}

\date{Accepted 2025 March 21. Received YYY; in original form ZZZ}

\pubyear{2024}

\begin{document}
\label{firstpage}
\pagerange{\pageref{firstpage}--\pageref{lastpage}}
\maketitle
% Abstract of the paper
\begin{abstract}
Blazars, a unique class of active galactic nuclei, exhibit highly variable emission across the electromagnetic spectrum.  This variability frequently manifests as intense flaring events, sparking an ongoing debate in recent literature about whether these flares exhibit periodic behavior in certain sources. However, many blazars also show clear signs of stochastic, uncorrelated flares that do not follow a regular pattern.
This paper explores how the presence of one such of these stochastic flares can distort an intrinsically periodic pattern of emission in blazars. Our results demonstrate that, depending on the specific circumstances, the deviations in significance and periods can exceed 100\%. Sometimes, these deviations can be so severe that they eliminate any evidence of a periodic pattern. These findings highlight the dramatic impact that flares can have on periodicity searches. To confront this challenge, we propose an innovative approach, the Singular Spectrum Analysis method, which appears more robust against the effects of flares. As an alternative solution, we also propose the sigma clipping technique to mitigate the impact of flares. This framework offers a valuable foundation for analyzing periodicity in similar astrophysical sources that are also subject to stochastic flaring events.
\end{abstract}

\begin{keywords}
BL Lacertae objects: general - methods: data analysis - galaxies: active
\end{keywords}

\section{Introduction} \label{sec:intro}
Multiwavelength observations have revealed that supermassive black holes (SMBHs) with masses $\rm M_{\rm BH}$$>$$10^6\Msun$ are commonly situated at the centers of most galaxies \citep[e.g.,][]{cavaliere_1989}. A subset of these SMBHs actively accretes gas, converting these systems into powerful electromagnetic emitters known as active galactic nuclei \citep[e.g.,][]{wiita_lecture}. In approximately 10\% of active galactic nuclei (AGN), a pair of highly collimated, relativistic jets originates from the core of the AGN \citep[e.g.,][]{padovani_2017}. When one of these jets is aligned closely with our line of sight, the object is classified as a blazar \citep[e.g.,][]{ulrich_variability}. Blazars exhibit emission that is predominantly driven by jet radiation, displaying high variability across various timescales, from seconds to years, and spanning a wide range of wavelengths, from radio to gamma rays \citep[e.g.,][]{urry_variability, urry_multiwavelengh}.

This variability is typically associated with stochastic phenomena \citep[e.g.,][]{covino_negation, rieger_2019}. However, in some instances, studies have uncovered periodic patterns within this variability \citep[e.g.,][]{sandrinelli_2016, penil_2020, penil_2022}. Several theoretical models have been proposed to explain the underlying physical mechanisms responsible for these periodic emissions. In broad terms, these models can be categorized based on whether they suggest the presence of a single SMBH or a binary SMBH system. For single SMBH scenarios, models explain the modulation of the emission as originating from perturbances within the jet \citep[e.g.,][]{camenzind_lighthouse, mohan_magnetic_lines} or variations in the flow in the accretion disk \citep[e.g.,][]{gracia_qpo_accretion_disk, dong_accretiondisk_periodicity}. In contrast, in binary SMBH systems, such variations can be accounted for by perturbations in the accretion flow due to orbital motions \citep[e.g.,][]{tavani_binary}, or by jet precession mechanisms induced by the orbital dynamics of the SMBH pair \citep[e.g.,][]{cavaliere_binary}.

However, several challenges still remain to be addressed in the analysis of periodicity. Historically, the lack of data has been a significant issue, although it has improved as observatories have started to regularly monitor blazars \citep[e.g. the \textit{Fermi-}LAT,][]{fermi_lat}. Additionally, the selection of methods used in the search for periodicity poses its own set of challenges, such as, for instance, the robustness of analyzing irregularly-distributed time series \citep[e.g.,][]{methods_critica_PKS_0735, lomb_vdp}. Furthermore, a new type of variability has been identified in blazar emissions, characterized by an overall increase or decrease in flux over time, modulating the already identified periodic signal \citep[i.e., a trend,][]{1ES1215+303_trend, penil_mwl_pg1553}. Consequently, detrending the time series should be considered a fundamental step in data reprocessing, as trends in the signal can lead to the detection of erroneous or spurious periodicities in time series \citep[][]{detrend_welsh}. 

Another significant challenge in the search for periodicity is the presence of red noise. Red noise is a form of stochastic noise frequently encountered in blazar emission, distinguished by its greater energy content at lower frequencies \citep[][]{vaughan_2003}. Identifying potential periodic oscillations is complex, as a substantial portion of the variance is due to random fluctuations, which is one of the factors that can lead to the detection of false periodicities \citep[e.g.,][]{vaughan_criticism}. In summary, there are several factors that can introduce distortions in the periodicity search, potentially causing inconsistencies in different analyses. These inconsistencies can result in both periodic and stochastic behaviors being associated with the emissions of the same blazar \citep[e.g., PG 1553+113,][]{ackermann_pg1553, covino_negation}.

This paper investigates the impact of long-term flares, which can extend from months to years, on the search for periodicity. We introduce a methodology that involves incorporating a flare into an artificial, periodic signal. We then evaluate the alterations in the period determination and the significance level of the detected periodicity, as determined using different methods. Subsequently, utilizing $\gamma$-ray observations, we examine the existence of periodicity in a sample of blazars characterized by the presence of flares in their light curves (LCs). Finally, we suggest alternative methods, which prove effective in detecting periodicity even in the presence of stochastic flaring events.

The paper is structured as follows. Section \textsection{\ref{sec:interpretation}} describes potential interpretations of the origin of the long-term flares. Section \textsection{\ref{sec:methodology}} explains the methodology used to assess the impact of flares on the search for periodicity. In Section \textsection{\ref{sec:results}}, the results of the analysis are presented, interpreting such results. Section \textsection{\ref{sec:alterntive_methods}} outlines the methods proposed for conducting a robust periodicity analysis in the presence of flares. Section \textsection{\ref{sec:usecase}} evaluates the flares on real blazar LCs. We conclude with a summary of the primary findings and conclusions in Section \textsection{\ref{sec:summary}}.

\section{Flares in Blazars} \label{sec:interpretation}

The variability observed in blazars is closely linked to variations in the processes of particle acceleration and radiation, typically assumed to occur within a compact region in the jet. Any modification to the conditions within this region, such as changes in particle densities, external radiation fields, or disk accretion itself, can lead to intense episodes of emission. 

Blazar variability has been modeled extensively in previous works from a large variety of possible origins, usually with multiple possible explanations for each individual flare or episode. For example, rapid variability can be explained as magnetospheric gaps close to the BH event horizon \citep[e.g.,][]{Neronov2007, Levinson2011}, magnetic field reconnection \citep[or the ``jet in a jet'' scenario, e.g.,][]{Lyubarski2005, Giannios2009, Petropoulou2016}, or penetration of external objects (i.e. clouds, stars) near the base of the jet \citep[e.g.,][]{Araudo2010,Barkov2012}. These stochastic processes, which can occur on timescales as short as minutes, significantly affect the shape of LCs, often hindering the detection of genuine periodic signals with periods in the minute-to-hour range. Furthermore, the combination of smaller flux fluctuations with longer flares makes it even more difficult to distinguish between stochastic variability and true periodic behavior, complicating the analysis of blazar emission patterns (see, e.g., Figure \ref{fig:type_flare_examples}). Studies searching for short-timescale periods typically focus on specific blazars, such as S4 0954+65 and 3C 371 \citep[][]{shortscales_S4_0954+65, otero_3C_371}. Therefore, performing a comprehensive study to search for these periods across a very large number of sources would significantly enhance our understanding of blazar variability and periodicity. Such large-scale analyses could leverage extensive datasets from upcoming surveys, providing broader insights into the transient behaviors of blazars and contributing to the field of time-domain astrophysics.

Recently, however, several systematic studies have been conducted, aiming to detect long-term periodicities in large samples of \textit{Fermi-}LAT blazars. For example, \citet{penil_2020} analyzed a sample of 2000 objects with 9~yrs of \textit{Fermi} data and found 11 of them to be periodic at $>4\sigma$ of local significance. \citet{yang_carma} analyzed dozens of the most promising blazars for periodicity using Gaussian Process methods, finding significant evidence only for PG 1553+113. \citet{jorge_2022} investigated multiwavelength emissions from 15 bright $\gamma$-ray blazars monitored by the Steward Observatory, identifying 4 potential periodic candidates. Additionally, \citet{Ren2023} discovered 24 potentially periodic blazars in 12 years of \textit{Fermi}-LAT data in a sample of 35 of the most bright blazars. \citet{alba_ssa} identified 45 blazars with evidence of periodic emission from a sample of 500 objects, also using 12 years of \textit{Fermi}-LAT observations. All these systematic studies focus on longer-term periodicity, in the range of years, with the median period of the sample of \citet{penil_2020} being $2-3$~yrs. When focusing on the 24 objects with the highest local significance within the large sample, and using 12~yrs of data, 1 object was confirmed to be periodic at $\approx2\sigma$ global significance\footnote{Global significance is corrected for the number of trials, see \citep[][]{penil_2022}} \citep[][]{penil_2022}. 

Specifically, PG 1553+113 stands out not only as the best candidate to be a truly periodic blazar but also as a candidate SMBH binary. \citet{penil_mwl_pg1553} detected a rising trend in the periodic emission of this specific blazar, which modulates its oscillations. This trend is interpreted as part of a longer oscillation, the super-orbital period caused by the two BH orbiting each other. In more recent work \citet[][]{Adhikari2023}, optical LCs extending up to $\sim100$~yr recover a superorbital period of $\sim22$~yrs, matching the prediction matching the predictions of theoretical simulations \citep[][]{WS2022+}. In these, the superorbital period should be $\sim10$ times larger than the shorter-term period, $\sim2.1$~yr, in the case of an orbiting SMBH binary system. An analysis of the 1492 variable 4FGL \textit{Fermi} blazars finds similar flux trends for $\sim40$ additional blazars (Pe\~nil et al. in prep.), which is an exciting perspective for studies of merging systems and binary SMBHs, particularly in the upcoming era of gravitational wave observatories \citep[e.g. NANOGrav,][]{nanograv}, and large optical monitoring surveys \citep[e.g. Rubin Observatory,][]{rubin_2023}. 

As already mentioned, several challenges complicate the detection of periodicity, even when it is genuine. These include the presence of red noise \citep{vaughan_criticism} and gaps in LCs. In this work, we propose to consider and study the impact of another factor: stochastic flaring. As discussed above, flaring at all timescales is common in blazars \citep[e.g.,][]{flares_fermilat_nalewajko, flares_fermilat_das}. It is, therefore, expected that, even in a blazar with a genuine periodic signal, regardless of its origin, stochastic flares could appear. 

Given how all the mentioned systematic studies find periods in the $2-3$~yr range, we focus on flares that could potentially distort this signal by having a similar amplitude and thus being more easily ``mistaken'' as being of the same origin as the underlying periodic signal. We refer to these flares as ``long flares'', i.e. with a duration in the $\sim$month$-$yr scale \citep[e.g.,][]{flares_fermilat_tanaka, flares_fermilat_wang}. Indeed, a variety of theoretical works in the past have predicted the existence of such flares from a variety of physical origins.

For example, interactions between AGN jets and stars within the AGN are expected to result in the acceleration of non-thermal particles and the production of radiation up to $\gamma-$rays \citep[e.g.][]{Barkov2010,Araudo2013,delaCita2016,Banasinski2016,Vieyro2017}. In a related process, long $\gamma-$ray flares could be produced at the moment stars and/or broad line region clouds
penetrate the jet \citep[e.g.,][]{flare_star_nuria, 2019A&A...623A.101D, zacharias_long_flares, 2022icrc.confE.676Z}, or even when stars explode inside the jet as supernovae \citep[][]{VieyroSN_2019}. Long-term flares can also result from alternative scenarios associated with the intrinsic structure of the jet. For example, the propagation of shock waves along the jet can compress plasma, accelerating particles and enhancing $\gamma$-ray emission \citep[e.g.,][]{marscher_flare_shocks}. Additionally, magnetic reconnection events near the black hole can rearrange magnetic field lines, releasing energy that accelerates particles and leads to prolonged $\gamma$-ray flaring \citep[e.g.,][]{sironi_flare}.

In the following sections, we study how one such flare of a similar duration to the amplitude of the periodic signal can disrupt the detection of genuine periodicity. 

\section{Methodology} \label{sec:methodology}
To evaluate how a specific factor distorts a property, the appropriate methodology involves creating controlled scenarios where the property is known to exist and then measuring how the factor affects its detection. This approach provides valuable insights into the factor's impact and demonstrates how systematic testing can validate the reliability and robustness of the analysis methods used to identify the property. This methodology was applied in \citet{Adhikari2024} and \citet{Adhikari2023} to evaluate the impact of gaps in observed periods. 

For that purpose, we implement the following methodology to assess the impact of stochastic flares on periodic signals: we establish a test bench, which involves generating a sinusoidal signal (contaminated by random noise) and introducing a single flare.  

The sinusoidal signal is generated according to the model: 

\begin{equation} \label{eq:model_fitting}
\phi(t) = O + A\sin \bigg(\frac{2 \pi t}{T} + \theta \bigg).
\end{equation}

The parameters considered in our methodology include offset ($O$), amplitude ($A$), period ($T$), and phase ($\theta$). The values we have selected for these parameters are based on the information inferred from the sample of \citet{penil_2022}, which consists of 24 blazars displaying evidence of periodicity:
\begin{enumerate}
\item Offset: The average offset is set at $6\times 10^{-8}$ ph cm$^{-2}$ s$^{-1}$.
\item Amplitude: We use the amplitude value from the oscillations of PG 1553+113, which is $5.5\times 10^{-8}$ ph cm$^{-2}$ s$^{-1}$ (as reported in \citet{penil_2022}).
\item Phase: The default phase value is set to $0$.
\item Period: We consider two-period values, 2 and 3 years, as most of the blazars in the \citet{penil_2022} sample exhibit periodicities falling within the 2$-$3~yr range. Accordingly, we assess the influence of the number of cycles, considering 6 cycles for the 2-year period and 4 cycles for the 3-year period to match the 12~years of \textit{Fermi}-LAT data analyzed.
\end{enumerate}

The sinusoidal signals are generated with regular sampling and observing time intervals that replicate those of real LCs of blazars included in \citet{penil_2022}, specifically from August 2008 to December 2020, and utilizing 28-day binning. We contaminate the signal with red noise generated with the approach \citet{timmer_koenig_1995}. To generate this red noise, we randomly sample power-law indices in the range [0.8–1.2], consistent with the values reported in \citet{penil_2022} and \citet{bhatta_s5_0716}. Moreover, Gaussian noise, distributed as $N(0, Std\times 10^{-8})$, is added to introduce stochastic variability to the sinusoidal signal points. The standard deviation ``Std'' for this noise is chosen to ensure a detection significance of $\approx$5-5.5$\sigma$ before any flare is introduced. This chosen value serves as a baseline to assess variations in the period and significance resulting from the introduced flare.

\subsection{Type of Flares}\label{sec:types_flares}

Our simulations consider distinct amplitudes and different durations for the flaring event. This comprehensive approach allows us to evaluate the effects of varying flare intensities and duration on the detection of periodicities.

\begin{figure*}
	\centering
        \includegraphics[scale=0.22]{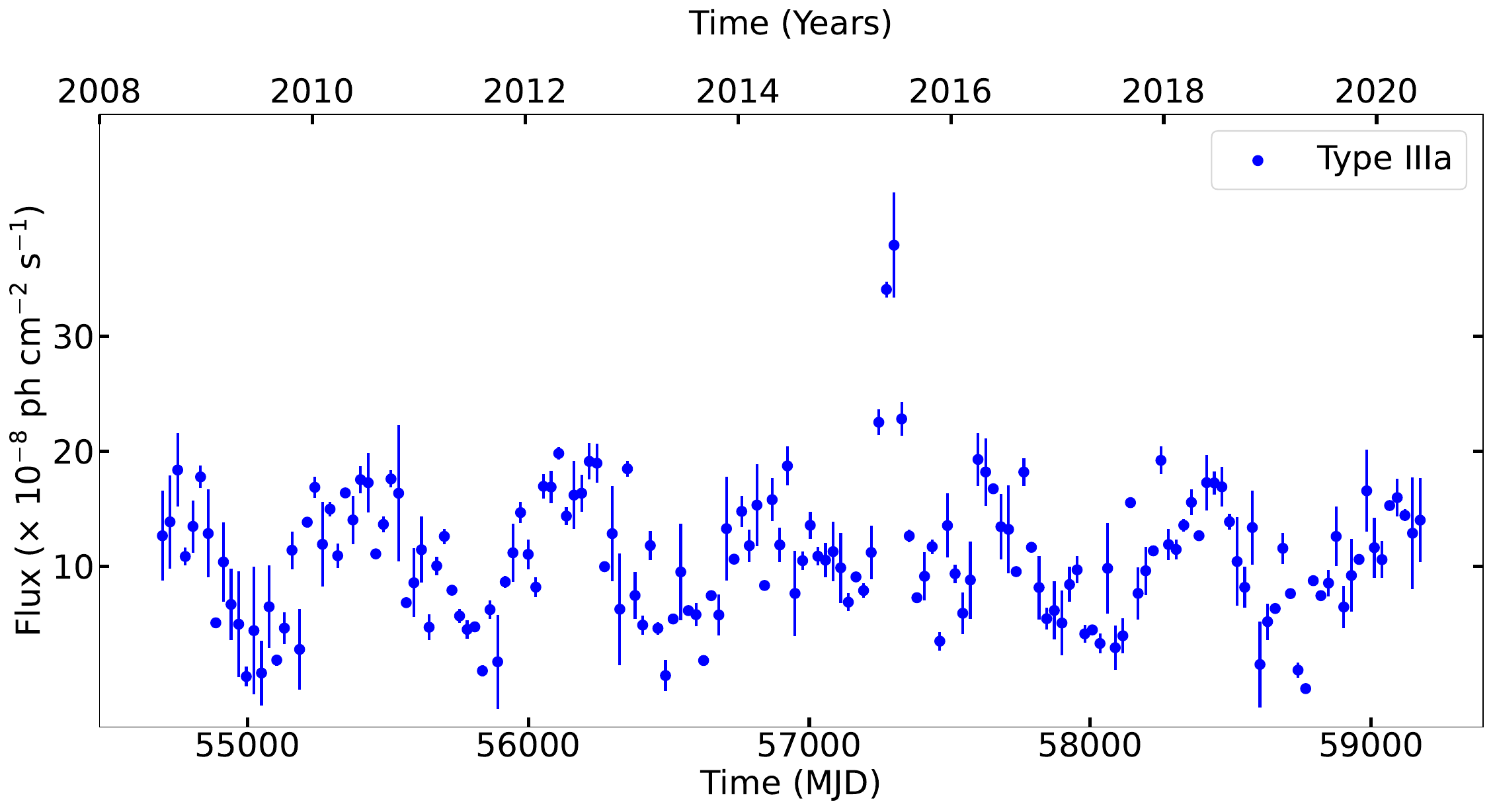} 
        \includegraphics[scale=0.22]{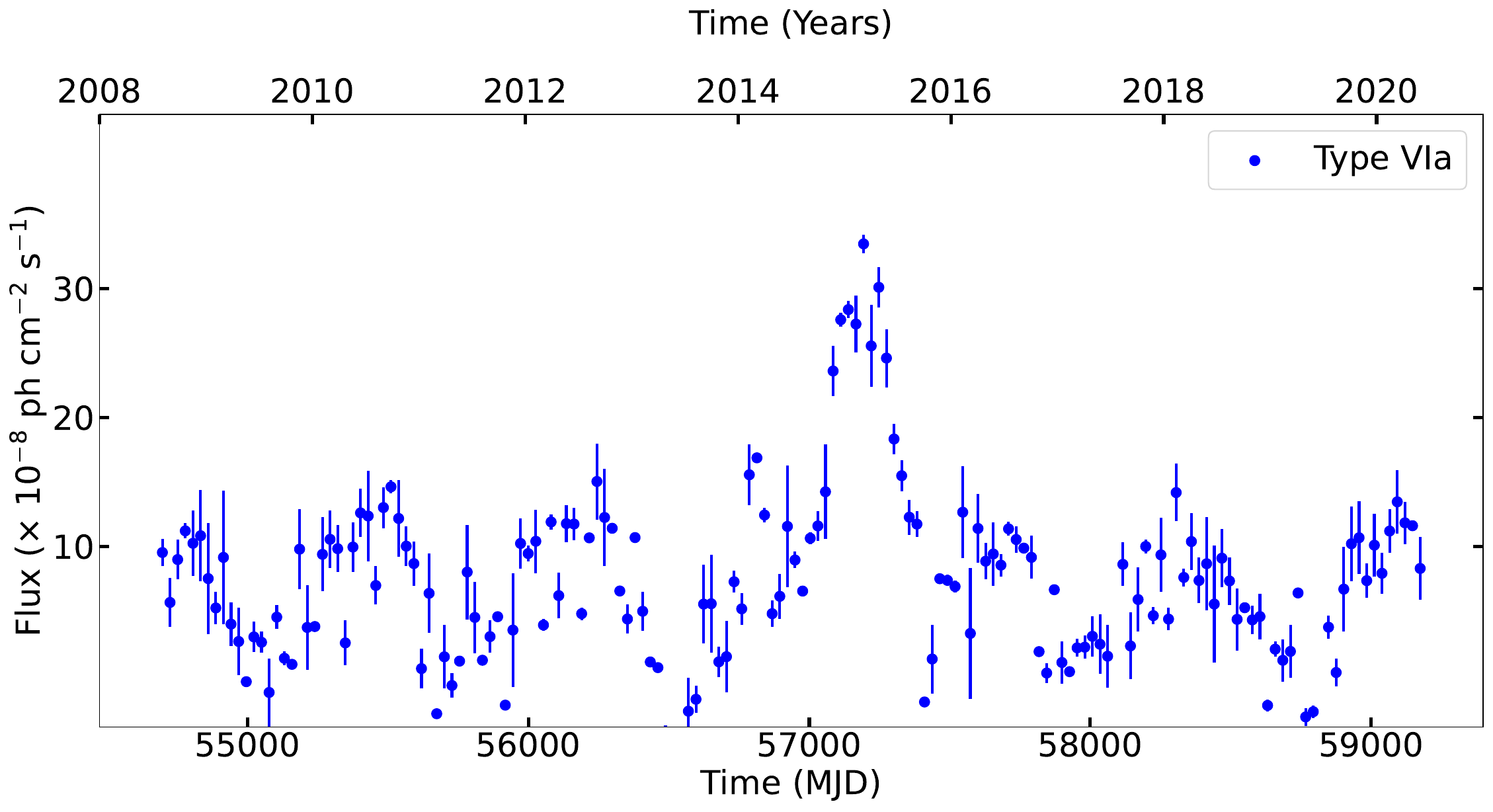}	
	\includegraphics[scale=0.22]{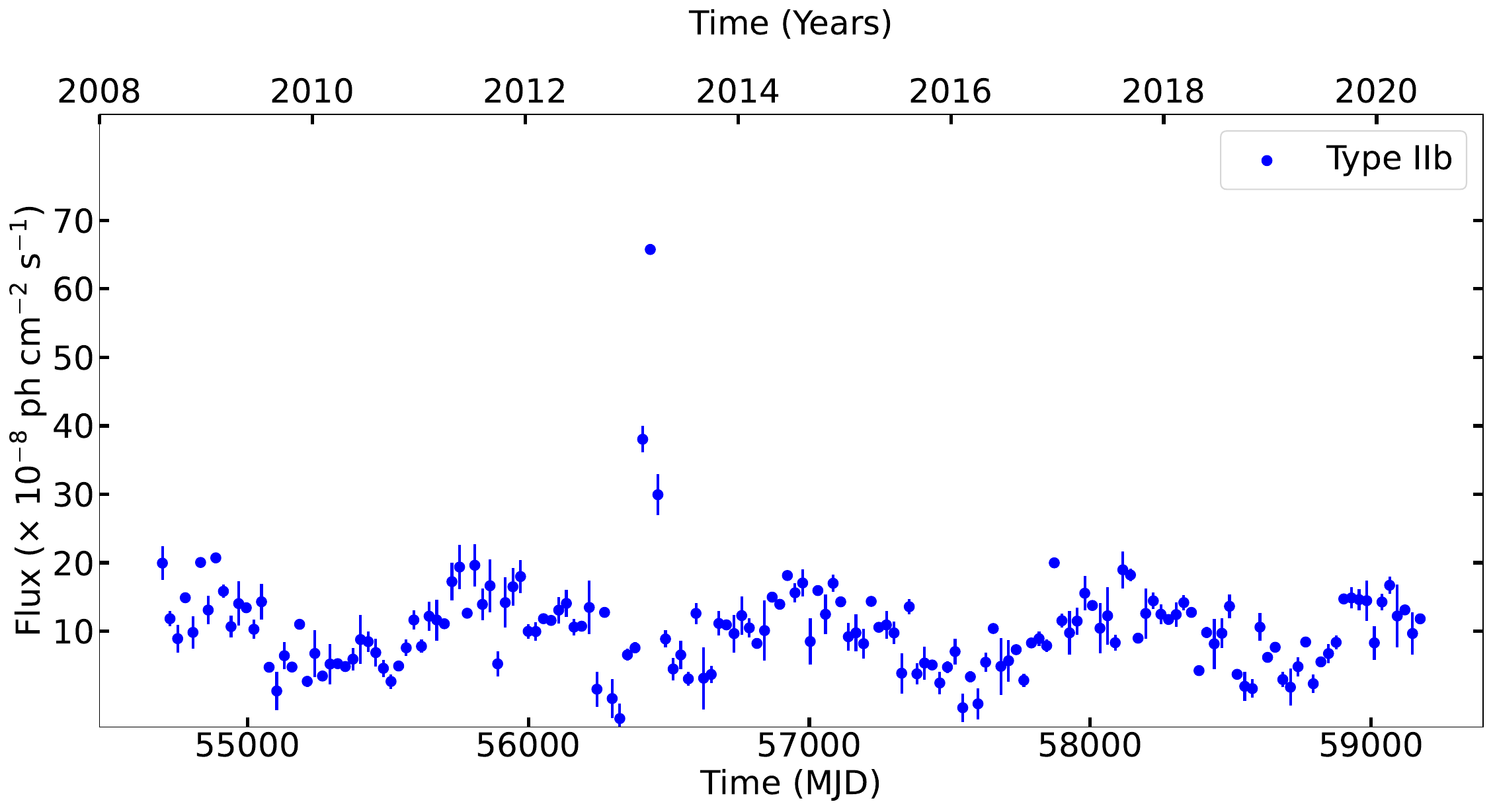}
        \includegraphics[scale=0.22]{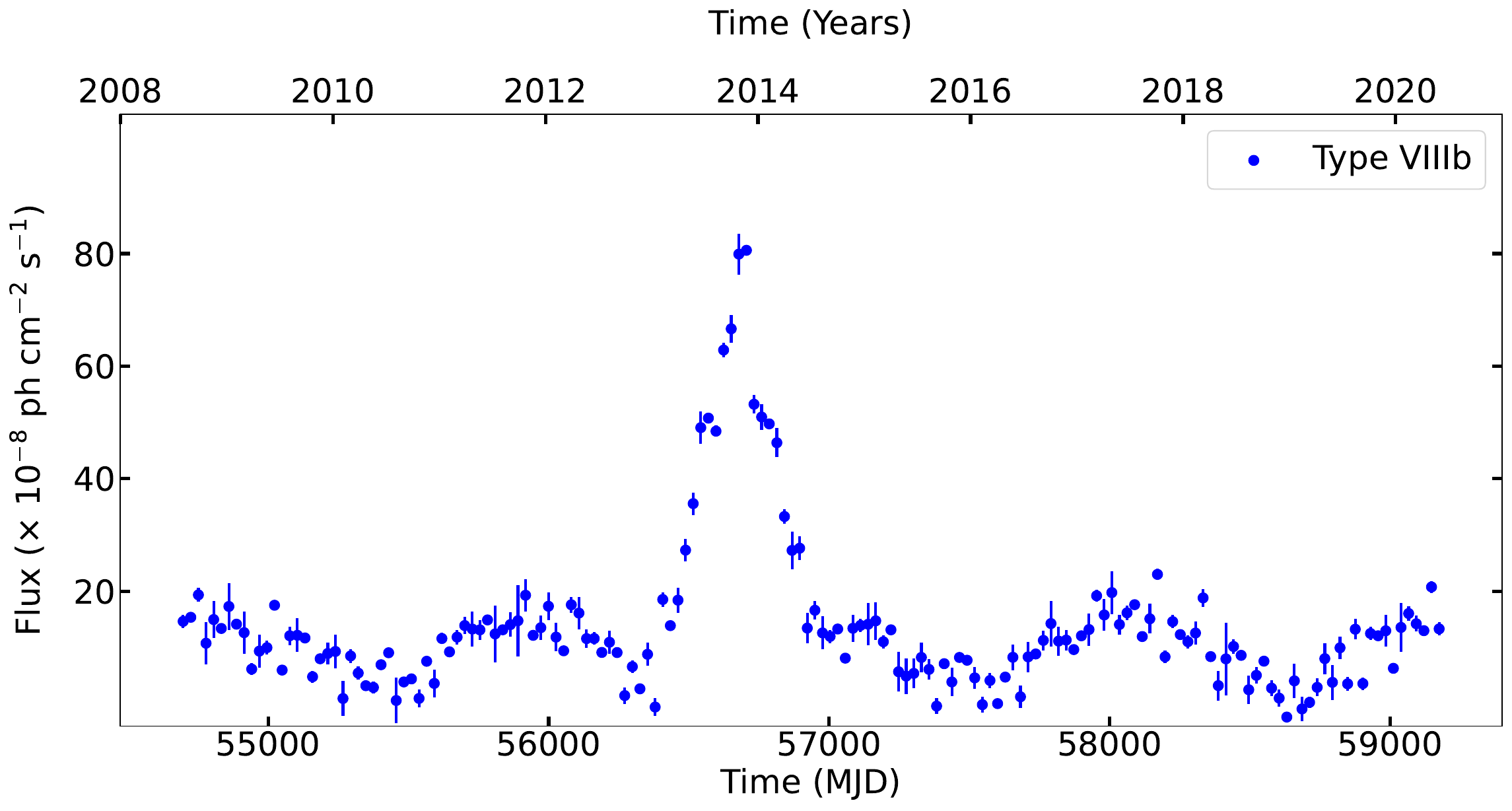}
	\caption{Examples of the type of flares. \textit{Top}: amplitude of 26.5$\times$ 10$^{-8}$ ph cm$^{-2}$ s$^{-1}$ for a signal with a period of 2 years \textit{Left}: Type IIIa flares (duration of 6 months). \textit{Right}: Type VIa flares (duration of 15 months). \textit{Bottom}: amplitude of 70.1$\times$ 10$^{-8}$ ph cm$^{-2}$ s$^{-1}$ for a signal with a period of 3 years. \textit{Left}: Type IIb flares (duration of 3 months). \textit{Right}: Type VIIIb flares (duration of 21 months) } \label{fig:type_flare_examples}
\end{figure*}

The injected flares follow a Gaussian profile, with added Poisson noise. We consider eight types of flares, depending on the duration of the flare:  
\begin{enumerate}
\item ``Type I'': 1 month, which consists of just one data point in the LC.  
\item ``Type II'': 3 months (see Figure \ref{fig:type_flare_examples})
\item ``Type III'': 6 months (see Figure \ref{fig:type_flare_examples})
\item ``Type IV'': 9 months 
\item ``Type V'': 12 months  
\item ``Type VI'': 15 months (see Figure \ref{fig:type_flare_examples}) 
\item ``Type VII'': 18 months 
\item ``Type VIII'': 21 months (see Figure \ref{fig:type_flare_examples})
\end{enumerate}

Regarding amplitude, we use the data from the \textit{Fermi} All-sky Variability Analysis (FAVA) catalog\footnote{\url{https://fermi.gsfc.nasa.gov/ssc/data/access/lat/fava_catalog/}} \citep{fava_catalog}, which is based on the integral photon flux in the 1 to 100 GeV range, using the Fourth LAT AGN Catalog\footnote{\url{https://fermi.gsfc.nasa.gov/ssc/data/access/lat/4LACDR2/}} \citep{4fgl_catalog}. We consider two amplitude levels for the flares: 
\begin{enumerate}
\item ``a'': the median-flux of flares in the FAVA catalog, which is 26.5$\times$ 10$^{-8}$ ph cm$^{-2}$ s$^{-1}$ (see Figure \ref{fig:type_flare_examples})
\item ``b'': the median of the brightest flares in the FAVA catalog, which includes flares with a flux greater than the previous median, which is 70.1$\times$ 10$^{-8}$ ph cm$^{-2}$ s$^{-1}$ (see Figure \ref{fig:type_flare_examples})
\end{enumerate}

Therefore, we perform the test for 16 different flare types, summarized in Table \ref{tab:flare_types}. All flare types are injected randomly in terms of phase angle for the majority of this work, with the exception of \textsection{\ref{sec:flare_in_phase}}  and \textsection{\ref{sec:effect_phase}}, where we study the impact of the flare phase on the periodic signal.
\begin{table*}
\centering
\caption{List of types of flares of the tests of $\S$\ref{sec:types_flares}, according to their amplitude and duration.\label{tab:flare_types}}
%\resizebox{\columnwidth}{!}
{%
\begin{tabular}{l|ccccccc}
\multicolumn{1}{c}{} & \multicolumn{2}{c}{Flare Amplitude} \\
\hline
{Duration} & 26.5$\times$ 10$^{-8}$ ph cm$^{-2}$ s$^{-1}$ &  70.1$\times$ 10$^{-8}$ ph cm$^{-2}$ s$^{-1}$ \\
\hline
\hline
1 month & Type Ia & Type Ib \\
3 months & Type IIa & Type IIb \\
6 months & Type IIIa & Type IIIb \\
9 months & Type IVa & Type IVb \\
12 months & Type Va & Type Vb \\
15 months & Type VIa & Type VIb \\
18 months & Type VIIa & Type VIIb \\
21 months & Type VIIIa & Type VIIIb \\
\end{tabular}%
}
\end{table*}

To assess the impact of flaring events on periodicity detection, we focus on a single flare. This approach allows us to isolate the specific effects of such an event on the periodic signal. Including multiple flares would introduce additional complexity, as each flare may vary in amplitude, duration, and phase, making it difficult to disentangle their combined influence from the underlying periodic signal. By concentrating on one flare, we can more precisely model how it might mimic or distort periodic behavior, particularly when its timescale closely aligns with the expected periodic signal.

Additionally, the flare is introduced randomly into the LCs to account for the stochastic nature of such events in real observations. This random placement helps us explore the potential impact of a flare occurring at any point in the LC, ensuring that the analysis captures a broad range of possible flare positions and their effects on periodicity detection.

\begin{figure*}
	\centering
        \includegraphics[scale=0.375]{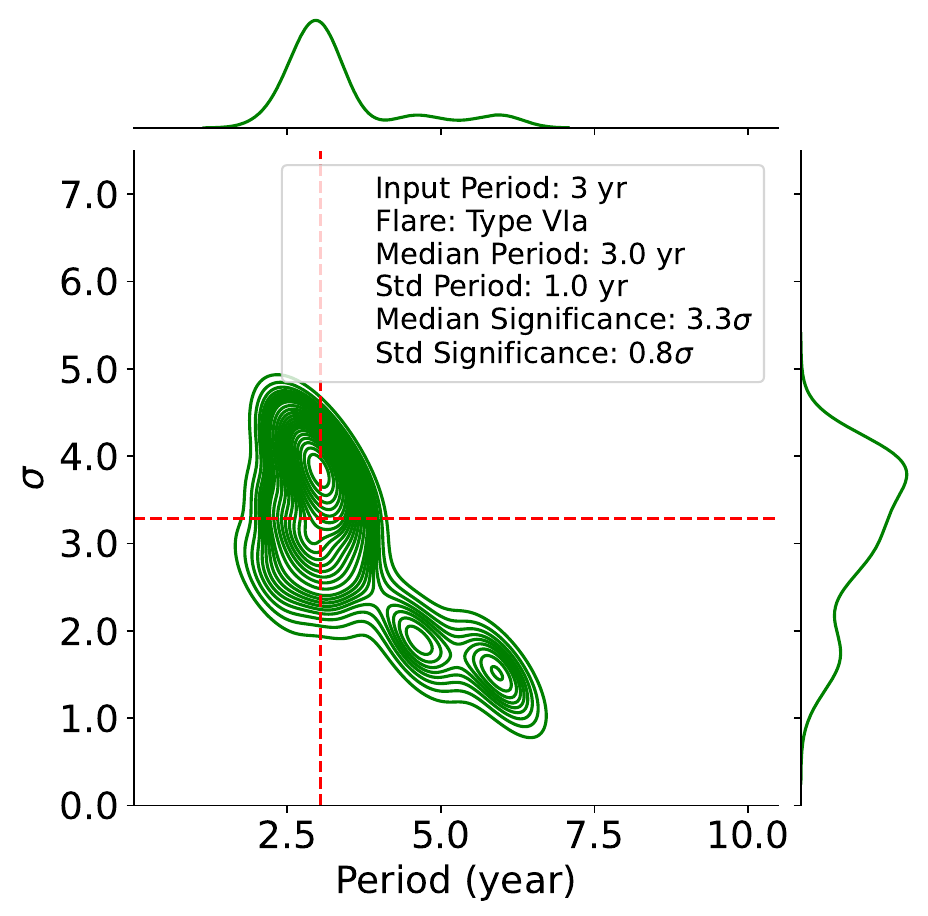}
	\includegraphics[scale=0.375]{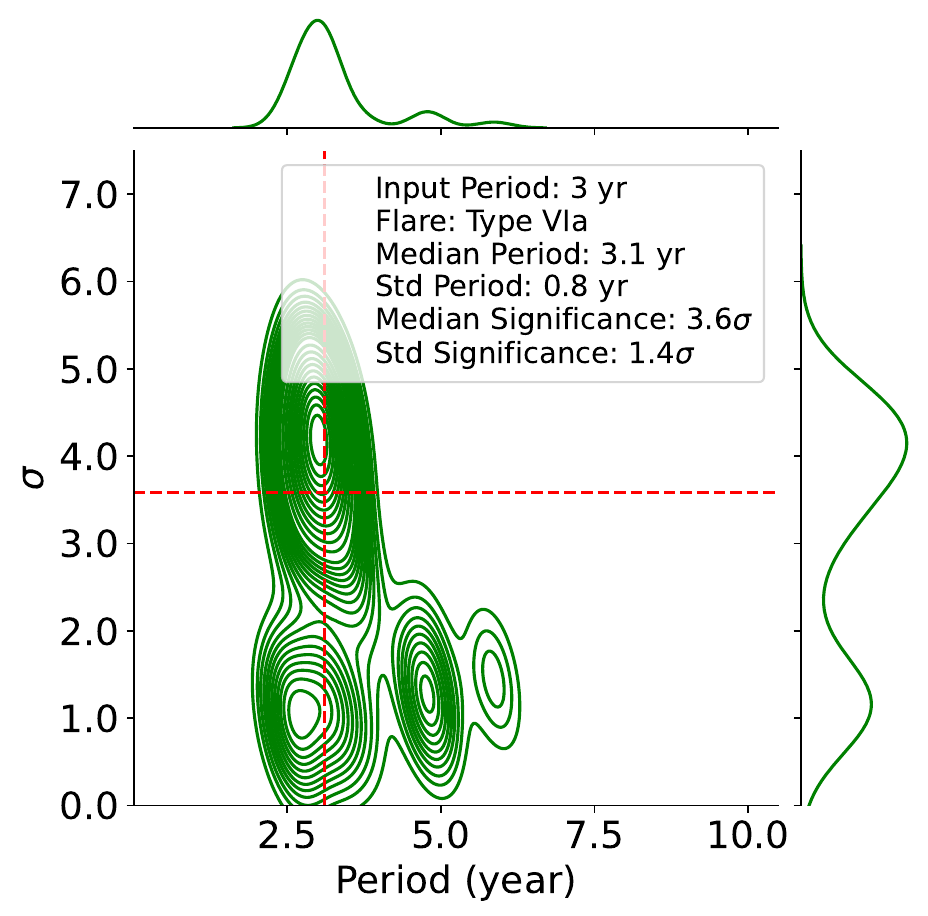}
        \includegraphics[scale=0.375]{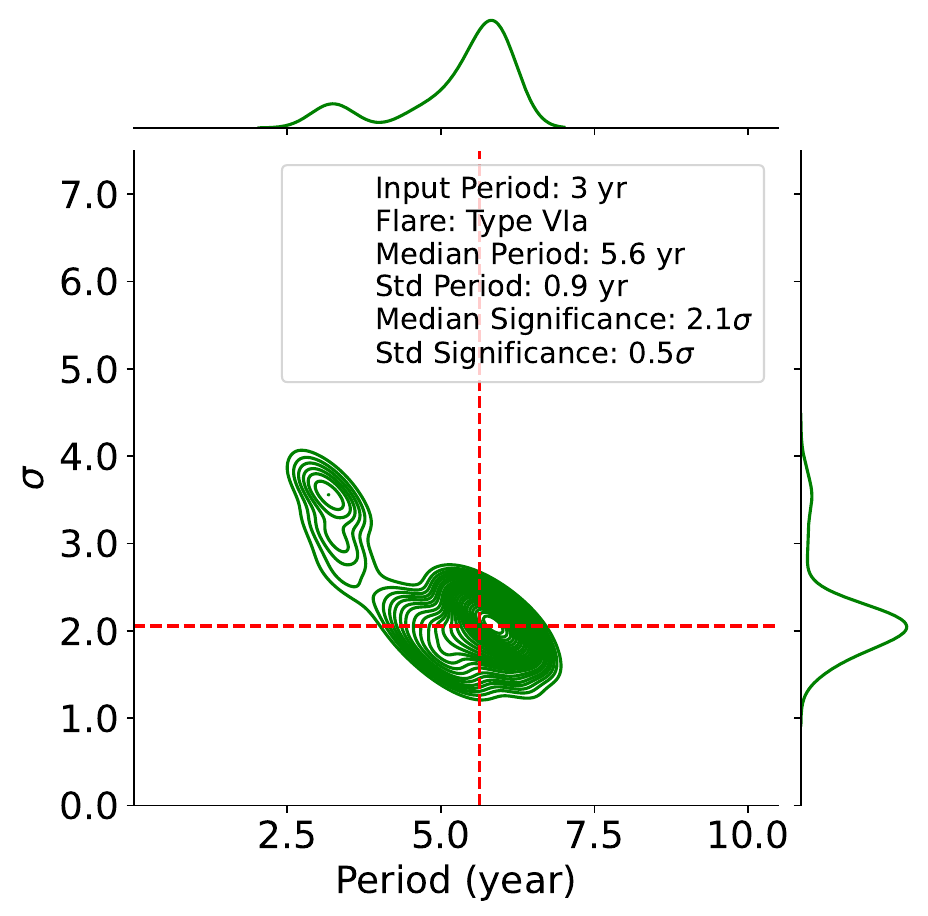}
	\includegraphics[scale=0.375]{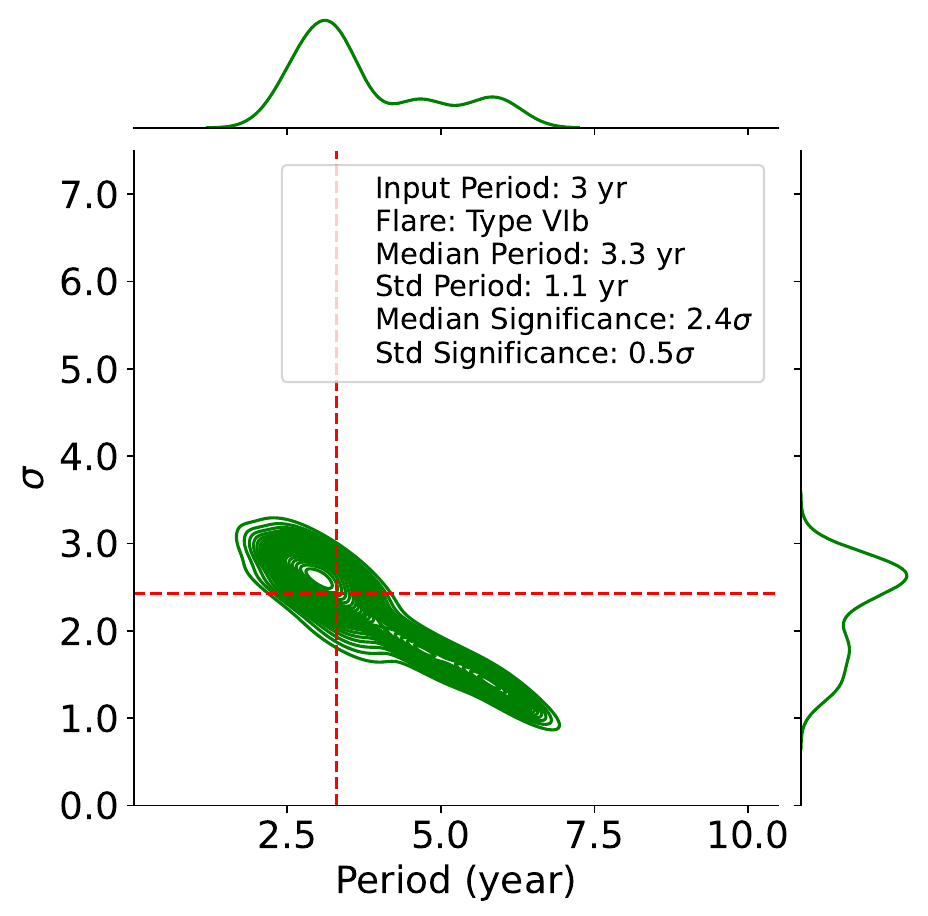}
	\includegraphics[scale=0.375]{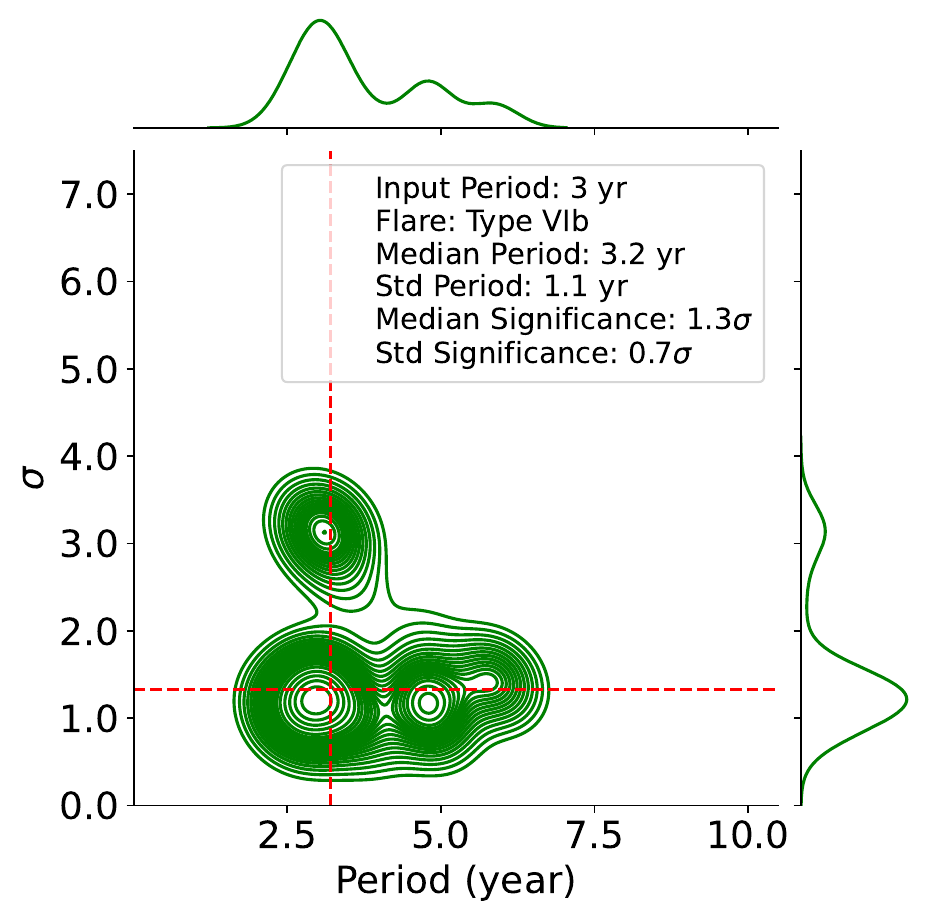}
        \includegraphics[scale=0.375]{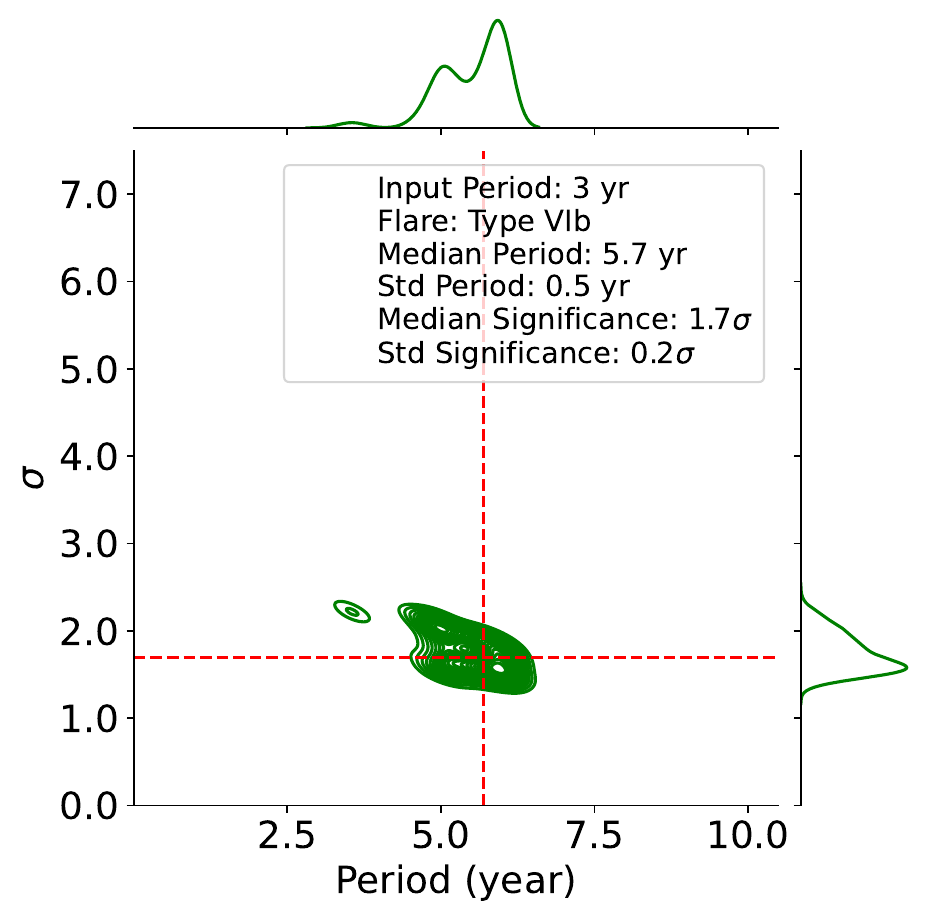}
	\caption{Example of distributions for the period and significance for a sinusoidal with a period of 3 years (see Figure \ref{fig:type_flare_examples}). \textit{Top}: Results of applying Type VIa flares to the LSP (\textit{Left}), CWT (\textit{Center}), and PDM (\textit{Right}). 
    \textit{Bottom}: Results of applying Type VIb flares to the LSP (\textit{Left}), CWT (\textit{Center}), and PDM (\textit{Right}). The ``Input Period'' refers to the signal's period used for the test. The ``Flare'' indicates the type of flare injected into the periodic signal. The ``Median Period'' represents the median of all periods resulting from the test, and the ``Std Period'' the standard deviation of the distribution of such periods. The ``Median Significance'' represents the median of the significance distribution associated with the test, and the ``Std Significance'' is the standard deviation of this significance distribution. Both LSP and CWT reveal a median period (indicated by the dotted red vertical line) consistent with the 3-year cycle of the signal. However, in PDM, the median period extends to 5.3 years, indicating the influence of the flare on this method. In terms of significance, the presence of the flare induces a notable decrease in LSP and CWT. The median significance is marked by the dotted red horizontal line.} \label{fig:flare_examples}
\end{figure*}

\section{Testing Impact Flare} \label{sec:results}
In this section, we evaluate the impact of flares using the previously described methodology, applied to a subset of the most commonly used methods for periodicity searches.

\subsection{Methods} \label{sec:methods}
The evaluation of the periodicity is conducted using 3 commonly employed methods in the literature: 

\begin{enumerate}
\item Lomb-Scargle Periodogram (LSP): This method is based on Lomb's algorithm \citep{lomb_1976} and Scargle's extension \citep{scargle_1982} for periodicity analysis.
\item Phase Dispersion Minimization (PDM): This method is employed to analyze phase variations in time series data \citep{pdm_stellingwerf}.
\item Continuous Wavelet Transform (CWT): The CWT is used to explore signal characteristics at different scales \citep{torrence_wavelet}.
\end{enumerate}

%The fake detection rate of these methods demonstrates remarkable accuracy, achieving a rate as low as 0.02\% \citep[][]{penil_2020}. This value denotes the effectiveness of the detection algorithms in minimizing false positives, ensuring a reliable performance.

To determine the significance of the results obtained from these methods, we generate 150,000 artificial signals using the approach described in \citet{timmer_koenig_1995}. These artificial signals share the same observational properties as the original sinusoidal signal, including mean, standard deviation, sampling intervals, and observing time. This significance refers to the local significance, not the global significance \citep[][]{gross_vitells_trial}. For our tests, global significance is not necessary because we are evaluating the impact of specific flares under controlled conditions rather than conducting a broad search for periodic signals. Local significance adequately addresses the goals of our analysis. 

Ultimately, this test yields a distribution of periods and their corresponding significance based on the aforementioned methods. We derive the median and standard deviation from this distribution to assess the progression of the period-significance tuple across the various types of flares previously delineated (Figure \ref{fig:flare_examples}). This approach allows us to explore whether the period of the sinusoidal signal can be accurately recovered and to understand how the significance changes under different flare conditions. By analyzing these distributions, we gain insights into the robustness of each method under the influence of flares on period detection.  

\subsection{Test Results}

\begin{figure*}
	\centering
        \includegraphics[scale=0.385]{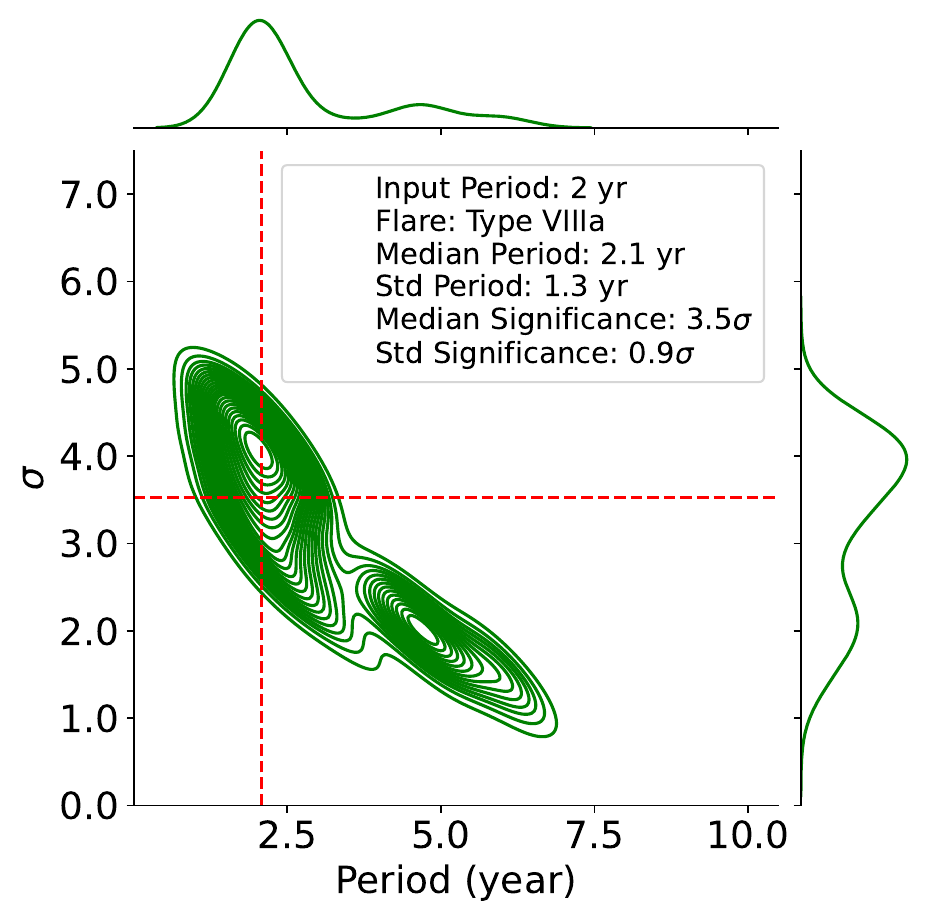}
	\includegraphics[scale=0.385]{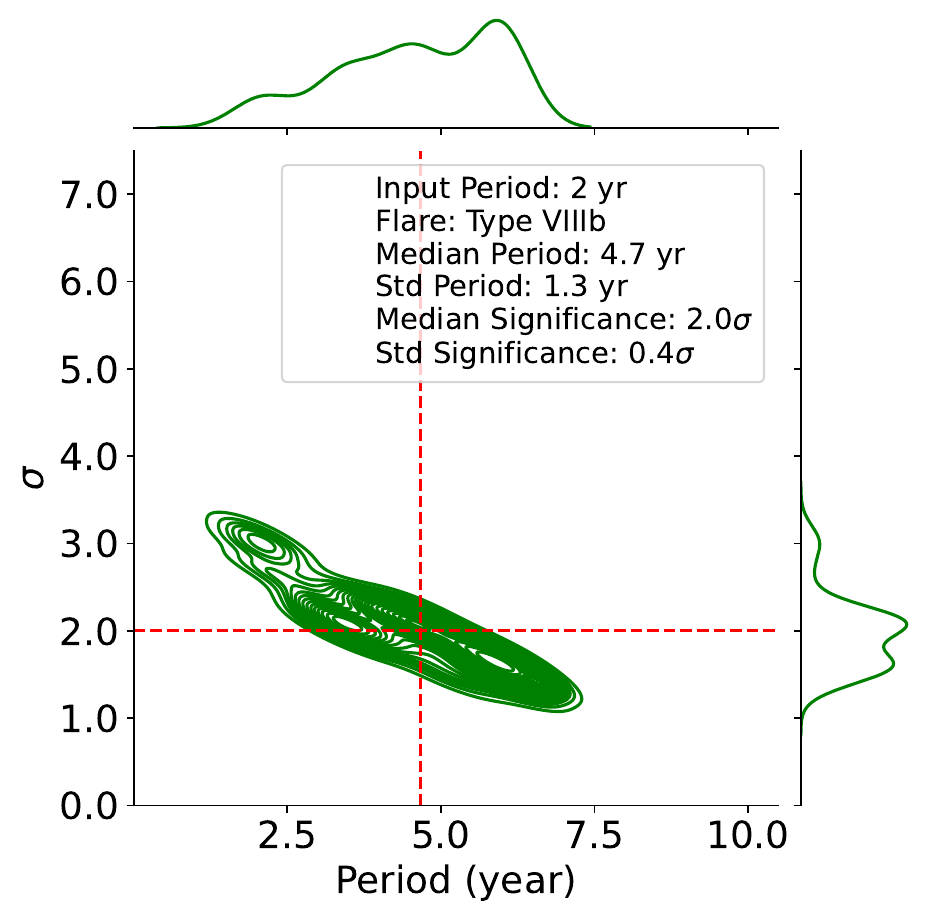}
 	\includegraphics[scale=0.385]{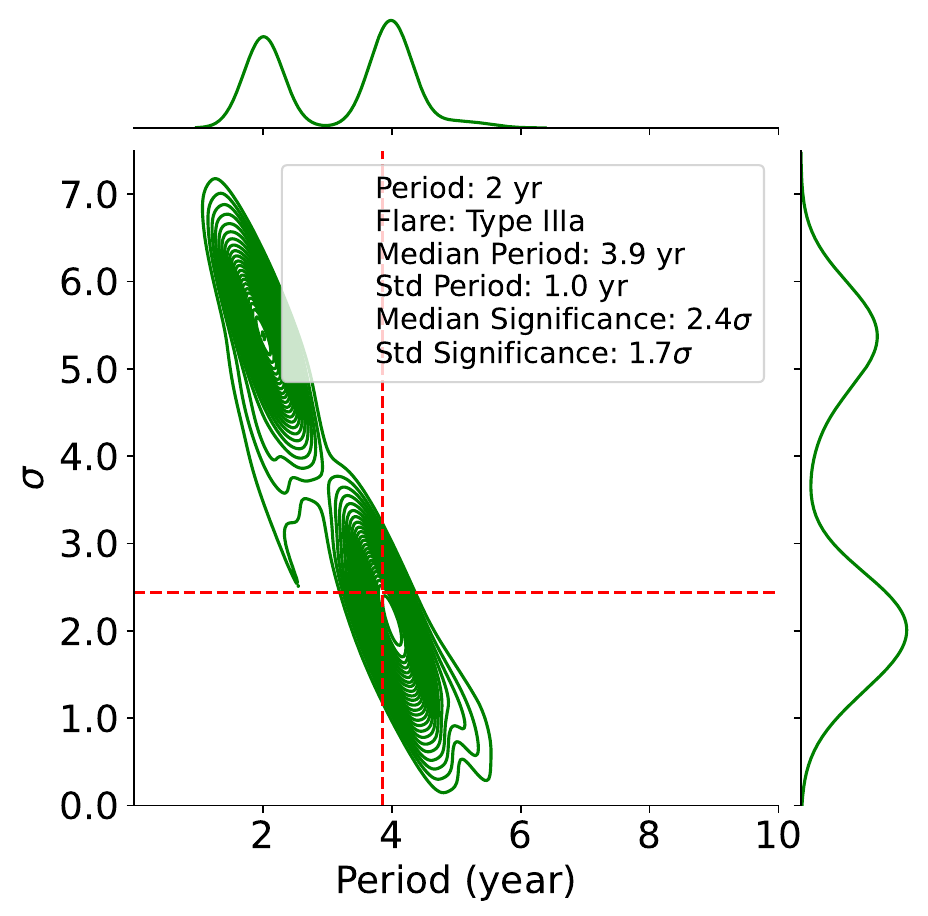}
	\includegraphics[scale=0.385]{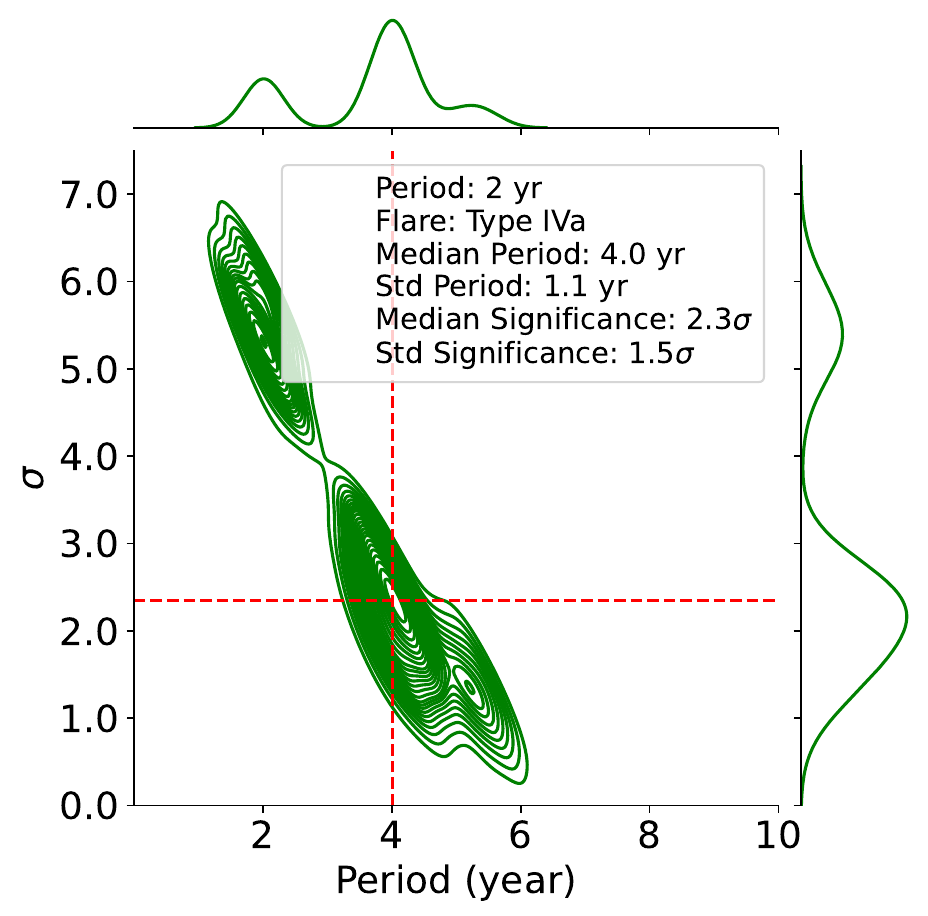}
	\caption{\textit{Top}: Example of distributions for LSP of the period and significance for a sinusoidal with a period of 2 years. \textit{Left}: VIIIa flares. \textit{Right}: Type VIIIb flares. \textit{Bottom}: Example of distributions for the period and significance for the cases Type IIIa and Type IVa flares for a sinusoidal with a period of 2 years for PDM analysis. The period distribution shows peaks place at 2 years and 4 years, where the second is associated with the harmonics of the signal. The median values for both the period and the significance of the test are indicated by the dotted red vertical and horizontal lines, respectively. The ``Input Period'' refers to the signal's period used for the test. The ``Flare'' indicates the type of flare injected into the periodic signal. The ``Median Period'' represents the median of all periods resulting from the test, and the ``Std Period'' the standard deviation of the distribution of such periods. The ``Median Significance'' represents the median of the significance distribution associated with the test, and the ``Std Significance'' is the standard deviation of this significance distribution.} \label{fig:flare_lsp_pdm}
\end{figure*}

The results of the test are shown in Table \ref{tab:experiment_general}. According to these results, the most robust method is the LSP, which accurately determines the period of the signal up to Type VII flares (duration of 18 months), with significance $\geq$3$\sigma$, in 56\%\footnote{The detection rates are calculated by dividing the number of successful detections by the total number of test cases. A detection is defined as identifying a period that matches the true signal period within a tolerance of $\pm$0.1.} of the cases (see Figure \ref{fig:flare_lsp_pdm}). The CWT also successfully identifies the signal period up to Type VII flares, although the significance in this case is $\geq$1$\sigma$. In 38\% of the cases, the periodic signal is detected with $\geq$3$\sigma$. On the other hand, the PDM method starts to deviate at Type III flares (duration of 6 months), with detections $\geq$3$\sigma$ in only 19\% of the tests. Interestingly, the PDM results exhibit two peaks in the period distribution: one corresponding to the actual signal period and the other to its harmonics (Figure \ref{fig:flare_lsp_pdm}). This phenomenon, where PDM reports harmonics of the period, was previously observed by \citet{penil_2020}. For the Type III and Type IV flares (with a duration of 9 months) shown in Figure \ref{fig:flare_lsp_pdm}, this effect explains the observed results. However, the PDM method tends to yield the least accurate results in terms of period estimation and significance. In many cases, it fails to detect periodicity in the LC, even when a true periodic signal is present. These findings underscore the substantial influence that flares exert on the reliability of the PDM method, making it less effective for identifying periodic behavior in blazars affected by flares.

The results also show the impact of the duration and amplitude of the flares. As the duration of the flares increases, the period-significance decreases, eventually leading to the inability to obtain the genuine period of the signal, as seen in Type VII and Type VIII flares (duration of 21 months) in Table \ref{tab:experiment_general}. 

The results show a significant dependence on flare amplitude. The detection rates are 88\% of cases for flares with ``a'' amplitude and 79\% of cases for those with ``b'' amplitude. These findings suggest that stronger flares have a substantial impact on the detected periodicity, even when their duration varies. 

We determine the number of additional cycles required to achieve a 5$\sigma$ detection, with a limit of 10 added cycles, as detailed in Table \ref{tab:experiment_general}. It is important to note that these findings are heavily contingent upon the type of flare. Again, the impact of both amplitude and flare duration is evident in our analysis. In many instances, inferring the required number of cycles to achieve a 5$\sigma$ detection is challenging. In 40\% of the cases, the 5$\sigma$ detection was not possible to recover. Our results indicate that the LSP method exhibits the highest capability to recover a 5$\sigma$ detection (in 79\% of the cases), whereas the PDM method performs the least effectively (recover it in 41\% of the cases).  

In conclusion, flares significantly affect the accurate identification and characterization of periodic signals. The presence of a flare can impede the detection of periodicity, eliminating the benefits of having multiple cycles in the periodic signal. Considering these results, having 2 or more stochastic flares would produce an ever larger impact on the periodicity detection capacity.

\subsection{Specific Analysis Scenario}\label{sec:flare_in_phase}
In this section, we show the results of a specific scenario for the flares. In periodic blazars, the amplitude of an oscillation can occasionally be higher than other oscillations. This could be the result of the underlying periodic emission mechanism having additional modulations or a spatial coincidence between flare and periodic emission. In any case, this phenomenon can affect the detection and analysis of a genuine periodic signal, as the higher amplitude oscillations might overshadow or obscure such signal, leading to potential inaccuracies in identifying true periodic behavior. By examining these specific scenarios, we aim to explore the impact of varying the amplitude of a flaring oscillation to measure its effect on the accurate detection of periodic signals.

\begin{figure*}
	\centering
	\includegraphics[scale=0.22]{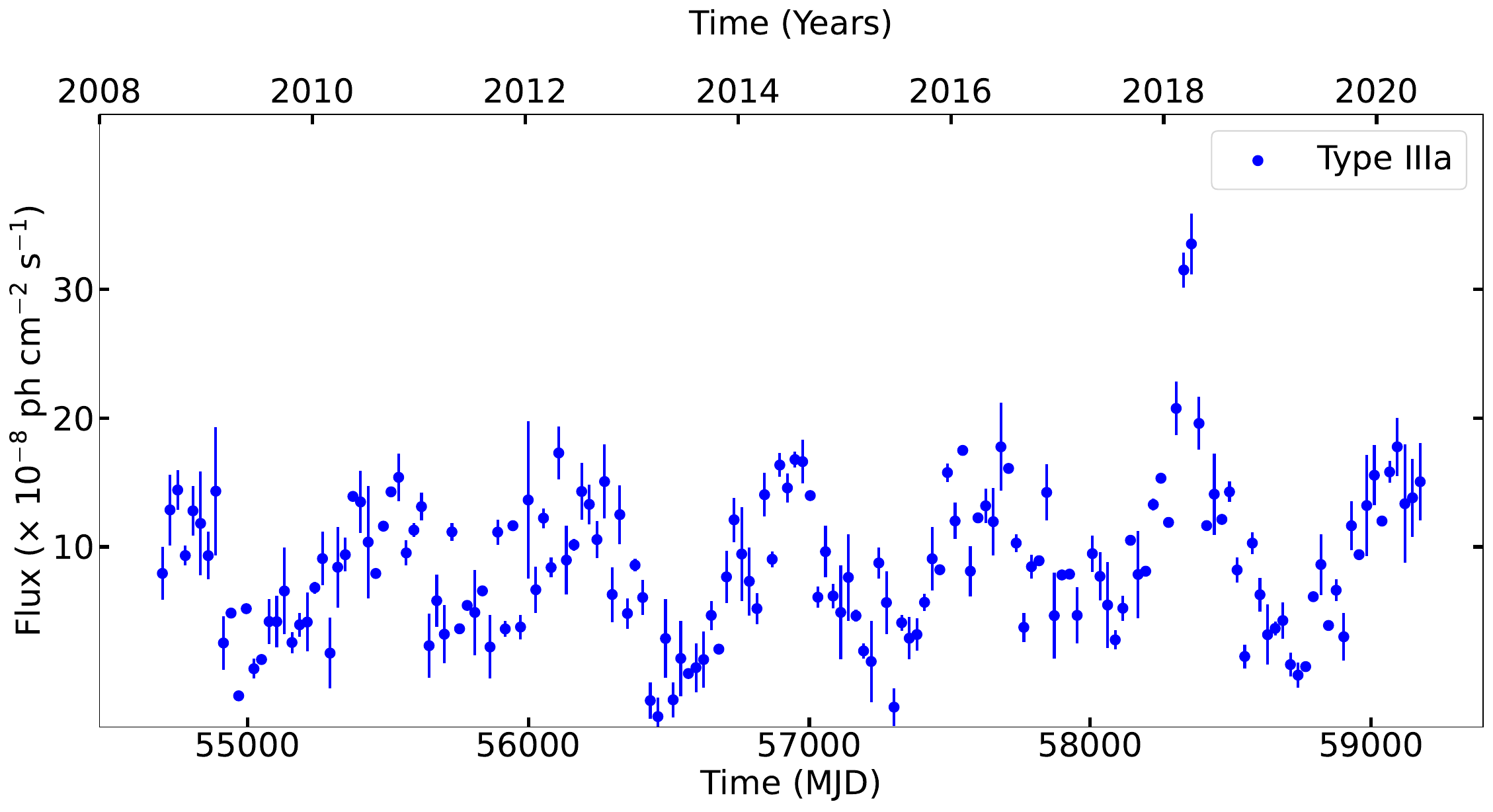}
	\includegraphics[scale=0.22]{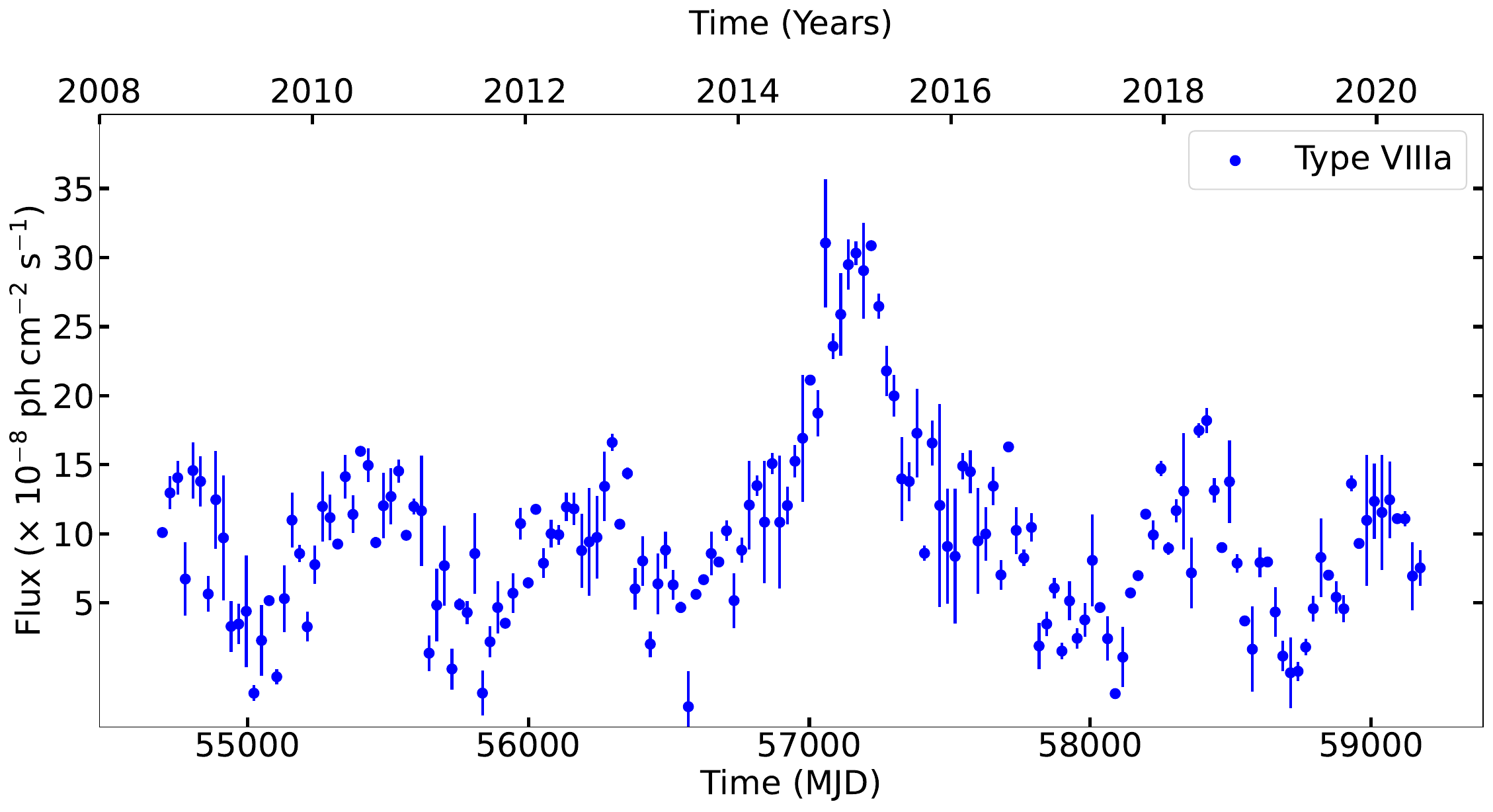}
	\includegraphics[scale=0.22]{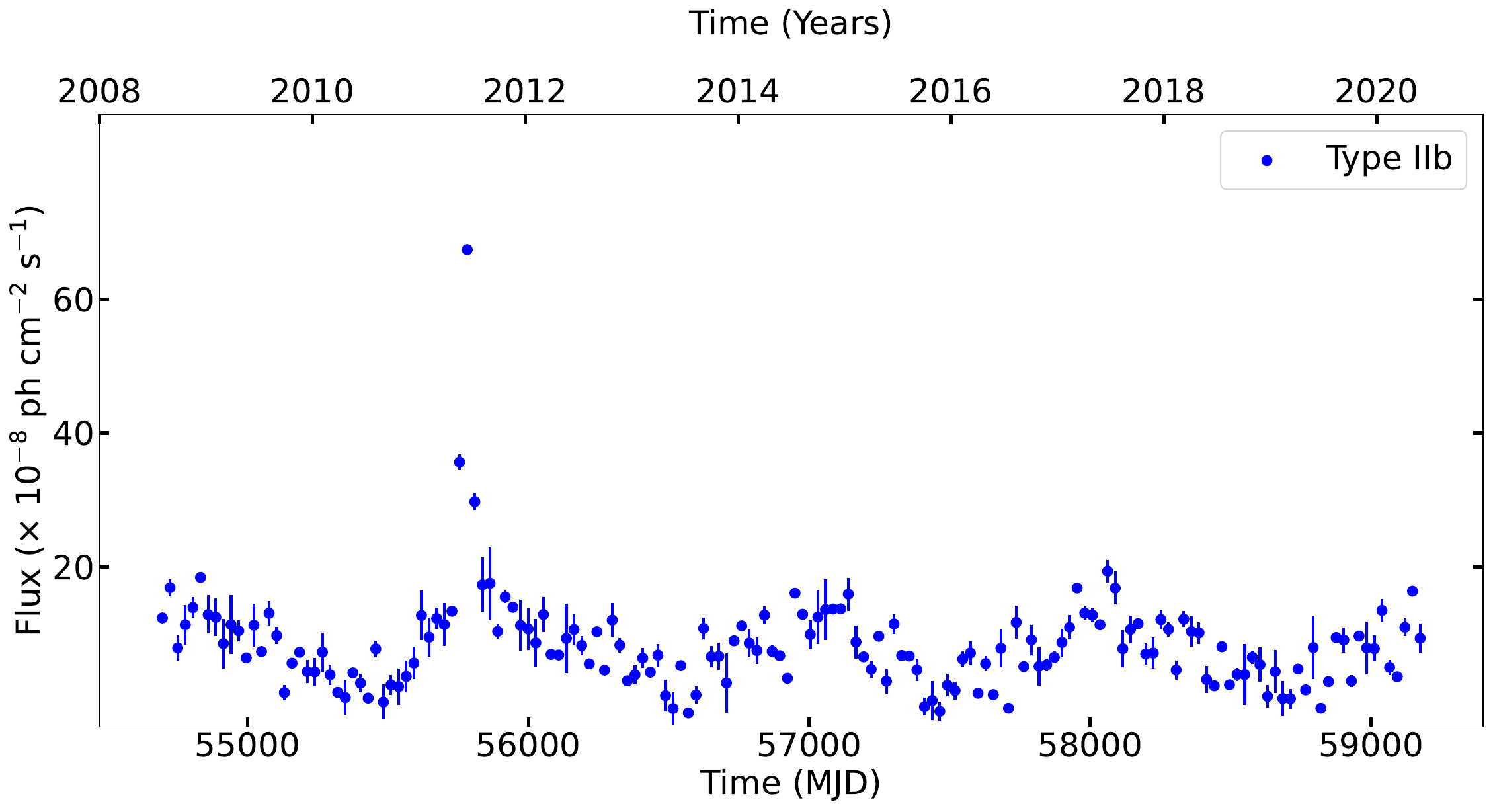}
        \includegraphics[scale=0.22]{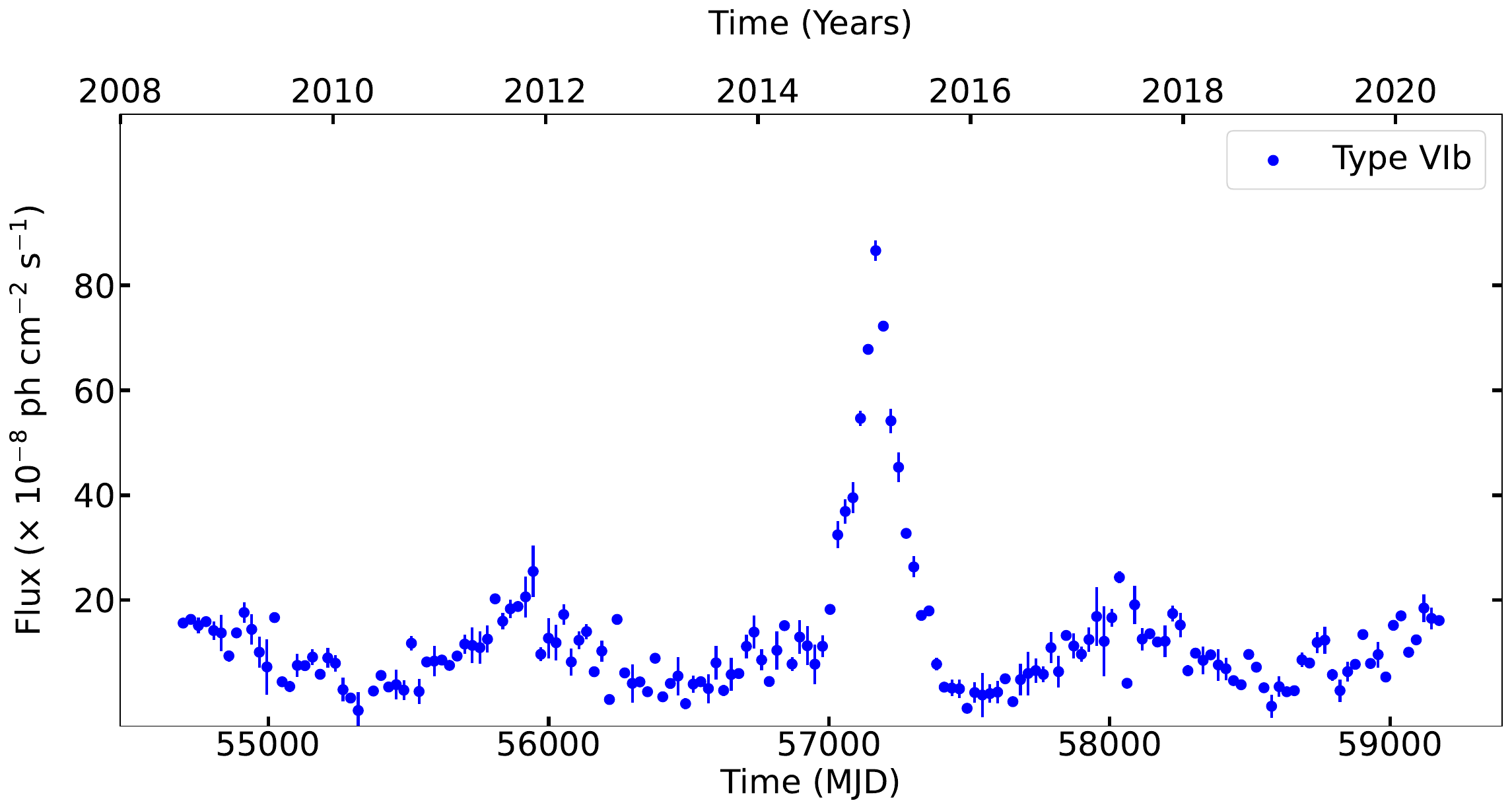}
	\caption{Examples of the test cases with different types of flares for the specific case where the flare is coincident with the oscillation. \textit{Top}: 2-year sinusoidal signals, featuring a flare amplitude of 26.5$\times$ 10$^{-8}$ ph cm$^{-2}$ s$^{-1}$. {\it Left}: Type IIIa flares (duration 6 months), {\it Right}: Type VIIIa flares (duration 21 months). \textit{Bottom}: 3-year sinusoidal signals, featuring a flare amplitude of 70.1$\times$ 10$^{-8}$ ph cm$^{-2}$ s$^{-1}$. {\it Left}: Type IIb flares (duration 3 months) {\it Right}: Type VIb flares (duration 15 months).} \label{fig:flare_examples_specific}
\end{figure*}

To understand this impact in more detail, we intentionally inject a flare that is in phase with the periodic signal at 3 distinct positions: the beginning, middle, and end of the signal (see Figure \ref{fig:flare_examples_specific}). This approach allows us to simulate different scenarios that blazars might experience, given their dynamic nature and the occurrence of high-amplitude oscillations.  

The simulation results are summarized in Table \ref{tab:experiment_coincident}. Once again, the results demonstrate the dramatic impact of a flare, even when it coincides with the oscillation of the periodic signal. The detection rate for the same period with a 5$\sigma$ significance is 14\%. In 64\% of cases, the period is not recovered. For the cases where the period is recovered, the significance is reduced by up to 75\%. There is no clear difference produced by the position of the flare in the signal, with only a slight difference observed when the flare is at the start and end of the signal in the results of CWT for flare types VII and VIII. 

Additionally, the results show that a signal with a period of 2 years is more sensitive to being affected by flares compared to a 3-year periodic signal (recovered in 32\% and 55\% of cases, respectively). This suggests that signals with shorter periods are more susceptible to the impact of flares, as the flare structure becomes more pronounced on shorter timescales. In such cases, the flare can dominate the observed variability, obscuring the underlying periodic signal. As a result, distinguishing between the flare-induced variability and true periodicity becomes more challenging. In contrast, longer-period signals are generally less affected, as their extended timescales provide more opportunities to distinguish the periodic oscillations from transient flare events. 

In terms of method robustness, the results in Table \ref{tab:experiment_coincident} reveal the following observations. The most robust method is the LSP, which provides the same period in 49\% of the tests, with a maximum reduction in significance of 50\%. The CWT method reports the same period in 44\% of the cases but with a maximum reduction in significance of 75\%. The PDM method presents the worst results, reporting the same period in 30\% of the cases, with a maximum reduction in significance of 50\%. These findings underscore the differential performance of the methods in terms of robustness under various flare conditions, with the LSP generally being the most reliable.

Finally, we assess whether the presence of a flare contributes to an increase in significance compared to the pure periodic signal. To evaluate this, we examine the scenarios outlined in Table \ref{tab:experiment_coincident}, selecting those where the inferred period matches that of the original LC and the significance remains below the 5$\sigma$ threshold. We focus our analysis on the LSP, as it yields the most reliable results. Our findings indicate no enhancement in significance due to the flare, suggesting that its presence does not amplify the significance compared to periods without flare activity. This outcome underscores that flares, while influential in other contexts, do not inherently boost the periodic signal's significance in this framework.

In conclusion, the presence of flares significantly complicates the detection and analysis of periodic signals, particularly for signals with shorter periods. The LSP method demonstrates superior robustness compared to CWT and PDM, making it the preferred choice for analyzing periodic signals in the presence of flares. Therefore, LSP should be the first option if any flare is presented in the LC for analysis. 

\subsection{Effect of the Flare Phase}\label{sec:effect_phase}
In the previous section, we evaluated the effect of the presence of a flare in phase with the periodic signal. In this section, we examine how the flare phase influences the period and significance of periodic signal detection. Intuitively, both the detected period and significance should vary substantially depending on whether the flare is injected in phase or out of phase with the periodic signal. To systematically analyze this effect, we inject a flare at different phase angles relative to the peak of the periodic signal, covering a full phase range of [0–2$\pi$]. This ensures that we account for all possible alignments between the flare and the periodic signal over a complete oscillation cycle.

\begin{figure*}
	\centering
	\includegraphics[scale=0.151]{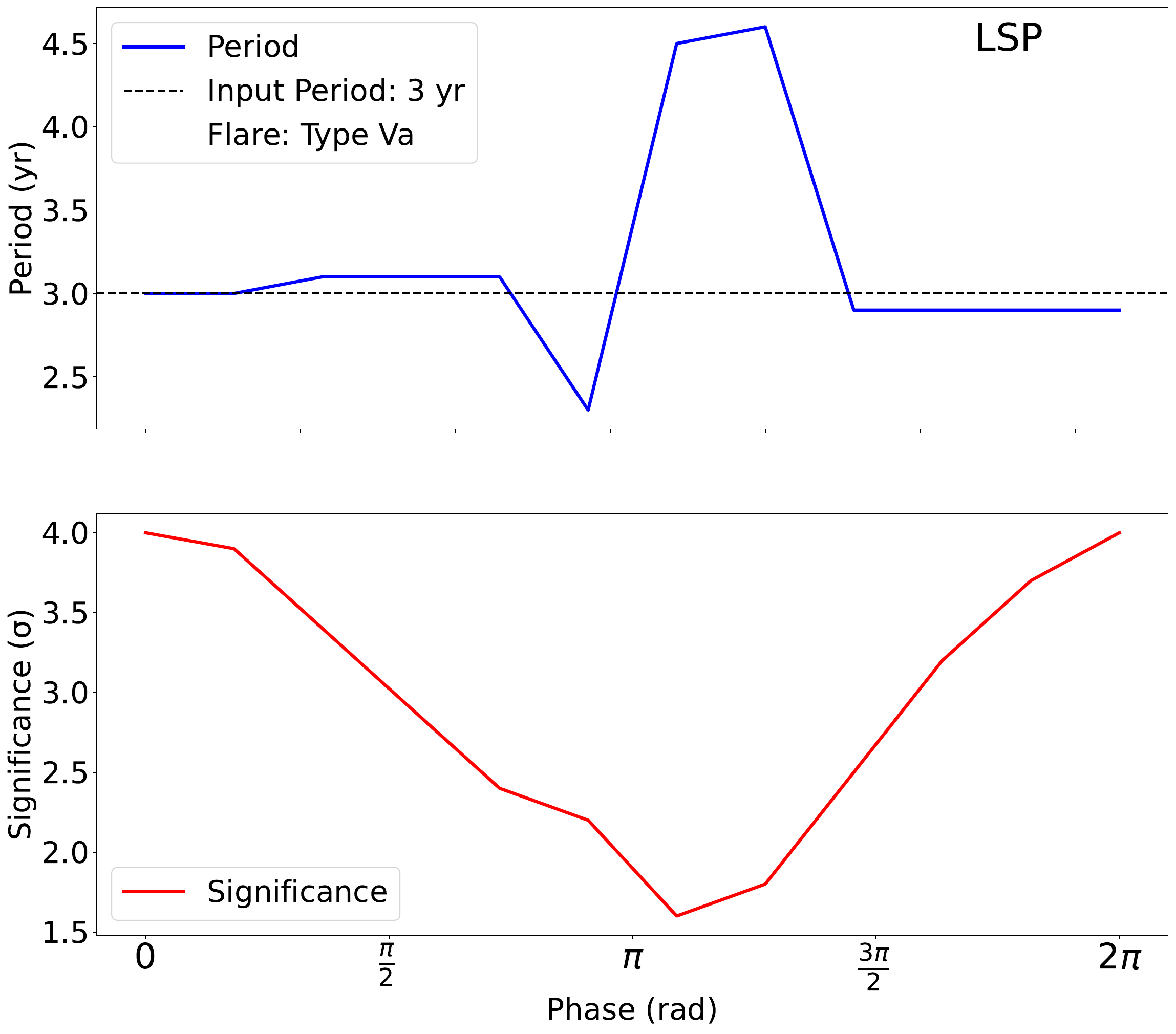}
        \includegraphics[scale=0.151]{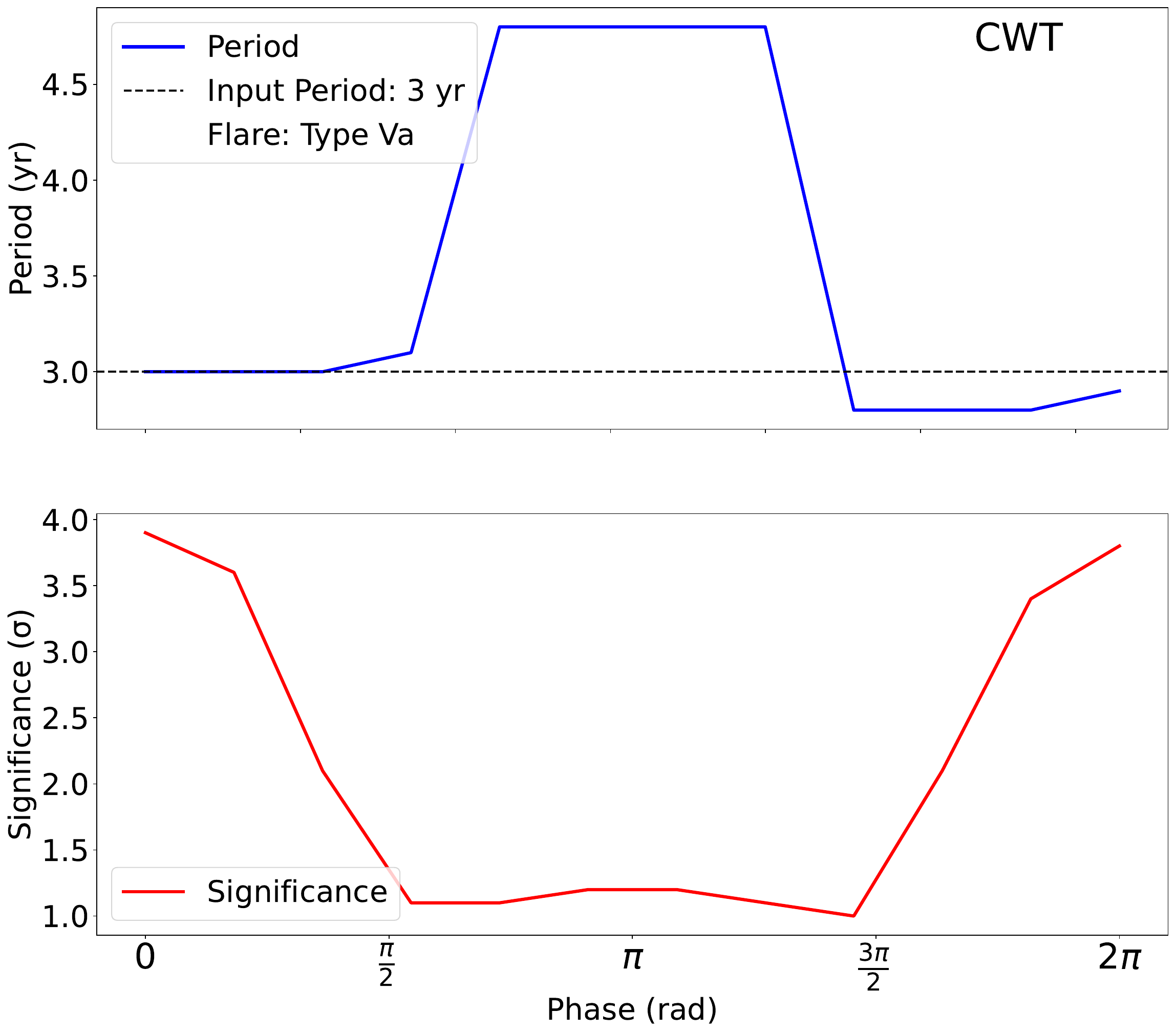}
        \includegraphics[scale=0.151]{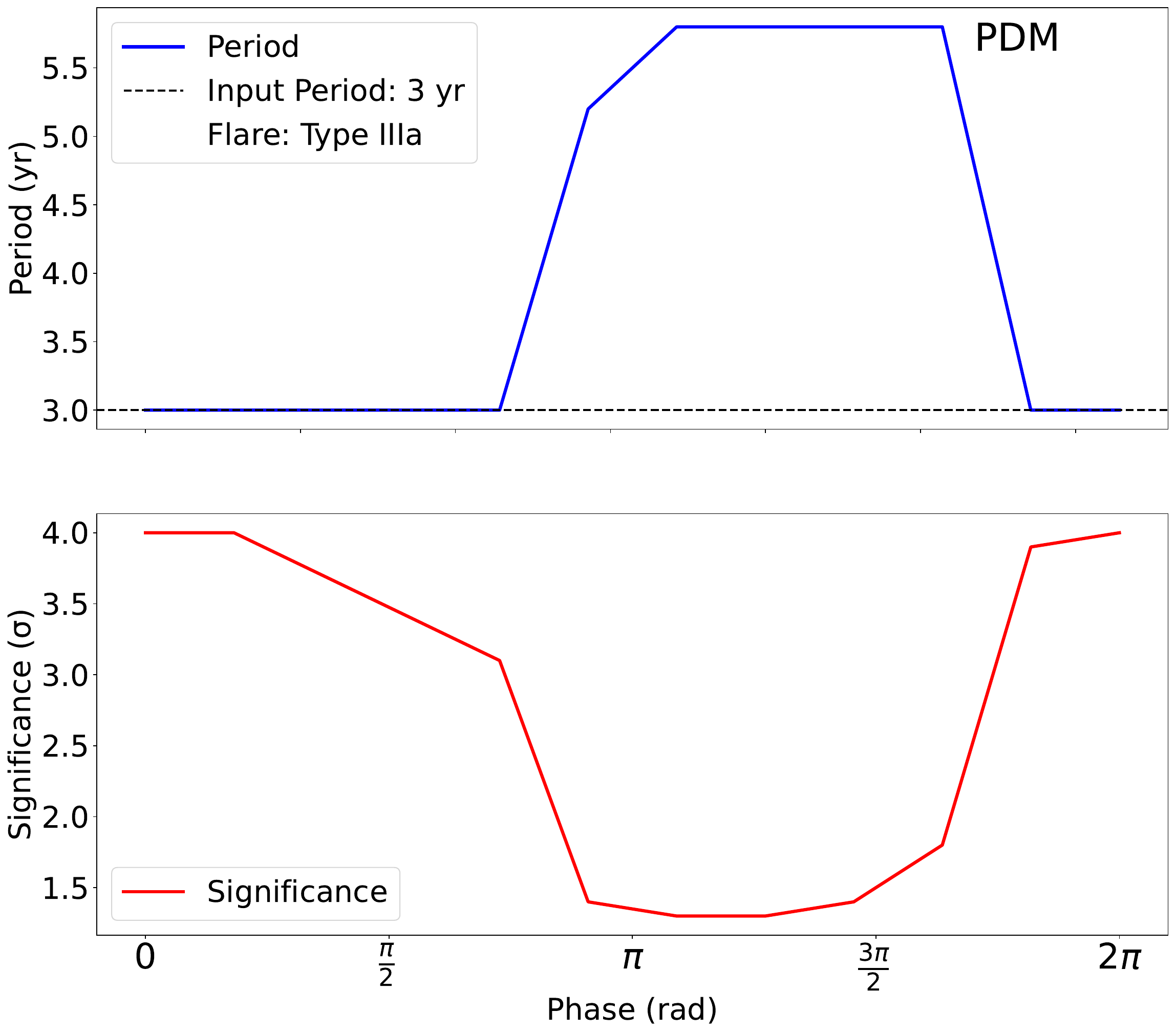}
	\caption{Example of the effect of the phase of the injected flare on the derived period and significance: \textit{Left}: LSP test for a 3-year signal period with a Type Va flare. \textit{Center}: CWT test for a 3-year signal period with a Type Va flare. \textit{Right}: LSP test for a 3-year signal period with a Type IIIa flare.} \label{fig:effect_phase}
\end{figure*}

We apply this test to the LSP, CWT, and PDM methods, selecting a specific flare type that exhibits significant variations in both period and significance. This approach allows us to assess how each method responds to different flare phases and whether certain phase alignments lead to systematic biases in period estimation or significance enhancement. The results of this analysis are presented in Figure \ref{fig:effect_phase}. For LSP, the presence of a flare significantly affects the inferred period and reduces significance only when the flare is in clear antiphase (around $\pi$). In contrast, for CWT and PDM, the impact becomes significant at approximately $\frac{\pi}{2}$
phase alignment. These findings highlight the crucial role of the flare phase in periodic signal detection, with LSP being the most robust to phase-induced distortions, making it the most reliable method in this context.

\subsection{Discussion} \label{sec:discussion}

The results of our simulations indicate that the addition of stochastic flares into a periodic LC significantly affects the capability of the conventional methods to recover the input periodic signal. The reasons behind the different impacts of flares in the different methods have to do with how such methods infer the frequencies presented in the real data. In this section, we discuss the different methods used, how efficiently they recovered the input periodicity, and why they were affected in different ways. 

\subsubsection{Lomb-Scargle Periodogram}

The LSP is a technique used for performing frequency domain analysis. It provides an estimate of the power spectral density of a signal, which describes how the power or amplitude of a signal is distributed across different frequencies. One of the key strengths of the LSP is its ability to handle unevenly sampled data effectively, addressing a limitation of the traditional Fourier transformation (FT). The LSP is particularly well-suited for detecting periodic signals within time series data that follow a sinusoidal shape. Its fundamental premise is built on the assumption that each frequency within the analysis range can be most effectively represented by a sinusoidal model \cite{lomb_vdp}:

\begin{equation} \label{ldp_fit}
    y(t;f) = A_f \sin(2\pi f(t-\phi_f))
\end{equation}

The parameters $A_f$ and $\phi_f$ are determined through a least-squares fitting process applied to the real data. Subsequently, the frequency spectrum is generated by minimizing the $\chi^2(f)$ value for each frequency while considering these fitted parameters. This approach, as described in \cite{lomb_vdp}, enables us to take into account both the amplitude (represented by $A_f$) and the phase (represented by $\phi_f$) of a sinusoidal signal at the specific frequency of interest.

The presence of a flare can introduce various challenges that impact the LSP's ability to detect periodic signals accurately. Flares can disrupt the LSP's sinusoidal modeling process because the algorithm needs to account for these exceptional flux outliers. This influence can lead the algorithm to be overly influenced by the high amplitude of the flares. Specifically, the fitting of parameters like $A_f$ and $\phi_f$ can be affected, causing distortion in the frequency that minimizes the $\chi^2(f)$ value and, consequently, introducing spurious frequencies into the periodogram. Additionally, LSP is not well-suited for non-stationary signals (one whose statistical properties, such as mean, variance, and autocorrelation, change over time), presenting detection limitations in this context. The flare variance can dominate, increasing the overall variance and thus reducing the power of the genuine periodic components relative to the noise. These effects of the flares can explain the results of flare types VII and VIII of Table \ref{tab:experiment_general} and Table \ref{tab:experiment_coincident}.

Furthermore, flares have the potential to intensify the effects of red noise. This heightened red noise contribution becomes particularly pronounced in the low-frequency portion of the spectrum. Flares are generally broadband in nature, meaning they do not correspond to a single periodic signal but rather affect a wide range of frequencies. Flares can also cause spectral leakage, where energy from the flare's primary frequency spreads to adjacent frequencies. This leakage can mask the periodicity. These scenarios can be applied to other methods, such as CWT. 

\subsubsection{Continuous Wavelet Transform}
The CWT is defined through the convolution of the data with a mother wavelet, typically the Morlet wavelet, as described in \citet{torrence_wavelet}. This choice of wavelet is suitable for mitigating issues associated with the analysis of finite data sets, such as spectral leakage, where signal energy at a specific frequency spreads into adjacent frequencies due to the discrete nature of Fourier analysis.

The convolution process in CWT involves the discrete Fourier transformation \citep[][]{torrence_wavelet}. Similar to the LSP, CWT relies on the FT. Consequently, it decomposes a signal into a combination of simple sine and cosine waves to extract information about frequency, amplitude, and phase. Just as observed with LSP, the presence of a flare can disrupt this decomposition of sine and cosine components, leading to the frequency spectrum changes that we have discussed in the previous section.

Moreover, flares, especially when they occur at or near the edges of the time series dataset, can introduce edge effects in the wavelet transform. These effects have the potential to distort the analysis, particularly in terms of the temporal localization of different spectral components, particularly visible in the wavelet power spectrum as spurious peaks near the boundaries of the signal (see Table \ref{tab:experiment_coincident}). Additionally, such edge effects can also arise due to the finite length of the analyzed time series \citep[][]{torrence_wavelet}. 

The CWT is specifically designed to handle non-stationary signals \citep[][]{torrence_wavelet}, which are signals where statistical properties like mean, variance, and autocorrelation change over time. However, flares represent extreme instances of non-stationarity. Thus, flares can lead to sudden alterations in the power spectrum, making it challenging to interpret the results and distinguish genuine periodic components, as evident in the results of the longer-duration flares types VI-VIII.  

Finally, LSP and CWT differ in that LSP searches for periodicity across the entire dataset, whereas wavelets analyze how periodicity evolves over time. Due to the noise included in the data, the detected signal frequency may slightly deviate from the true periodicity in wavelet analysis, leading to a broader distribution and, consequently, a lower significance. This fact, combined with the presence of the flare, also can result in observing CWT's lower significance in our tests.

\subsubsection{Phase Dispersion Minimization}
The PDM algorithm involves the segmentation of data into bins, where the amplitude variance is computed within each bin. These individual variances are then combined and compared to the overall variance of the entire dataset. Periods are inferred when the ratio of variance within a bin to the total variance is minimized. When this ratio is not minimized, it remains close to approximately 1.

Flares, being outliers in terms of brightness and duration, can disproportionately affect dispersion statistical measures. PDM assumes a consistent variance across phases, but flares violate this assumption. Flares cause significant deviations in the amplitude of the signal. These deviations introduce outliers that affect the variance calculation within each bin. The variance within each bin becomes less representative of the periodic signal's true variance, leading to incorrect period identification. For example, in the case of PKS 0208$-$512 (see $\S$\ref{sec:usecase}), the variance of the LC is initially measured at 3.5$\times$10$^{-14}$ $(ph~cm^{-2} s^{-1})^{2}$. However, after removing the flare, the variance reduces to 1.4$\times$10$^{-14}$ $(ph~cm^{-2} s^{-1})^{2}$. This significant change in variance illustrates how flares can impact the period detection process. This fact explains the significant distortion of the flares on the PDM detection capacity shown in \ref{tab:experiment_general} and Table \ref{tab:experiment_coincident}, resulting in PDM having the lowest accuracy among the methods tested.

\section{Proposed Methods for a Robust Analysis Against Flares} \label{sec:alterntive_methods}
The methods employed in this study have limitations in periodicity analysis, particularly when dealing with flare-affected data, as demonstrated in the previous sections. 

In time-series analysis, a critical preprocessing step involves filtering spurious factors that can distort the analysis. Detrending the time series is a recommended approach to remove long-term trends that may obscure periodicity analysis or lead to false detections \citep[][]{mcquillan_trend_fake_detection, hippke_spline_2019}. Another common method is the application of filters designed to reduce specific types of variability, such as the Savitzky–Golay filter \citep[][]{savitzky_golay_filter}, which suppresses low-frequency variability while preserving the signal's overall trend. Alternatively, Fourier-based approaches can be utilized, followed by the application of low-pass or high-pass filters to eliminate undesired variability \citep[e.g.,][]{wu_filters, albentosa_highpass}. These diverse techniques offer a toolkit for addressing contamination in time-series data. In this study, we employ two such preprocessing approaches to mitigate the impact of flares: Singular Spectrum Analysis (SSA) and Sigma Clipping.

\subsection{SSA analysis}\label{sec:ssa_results}
\begin{figure*}
	\centering
        \includegraphics[scale=0.21]{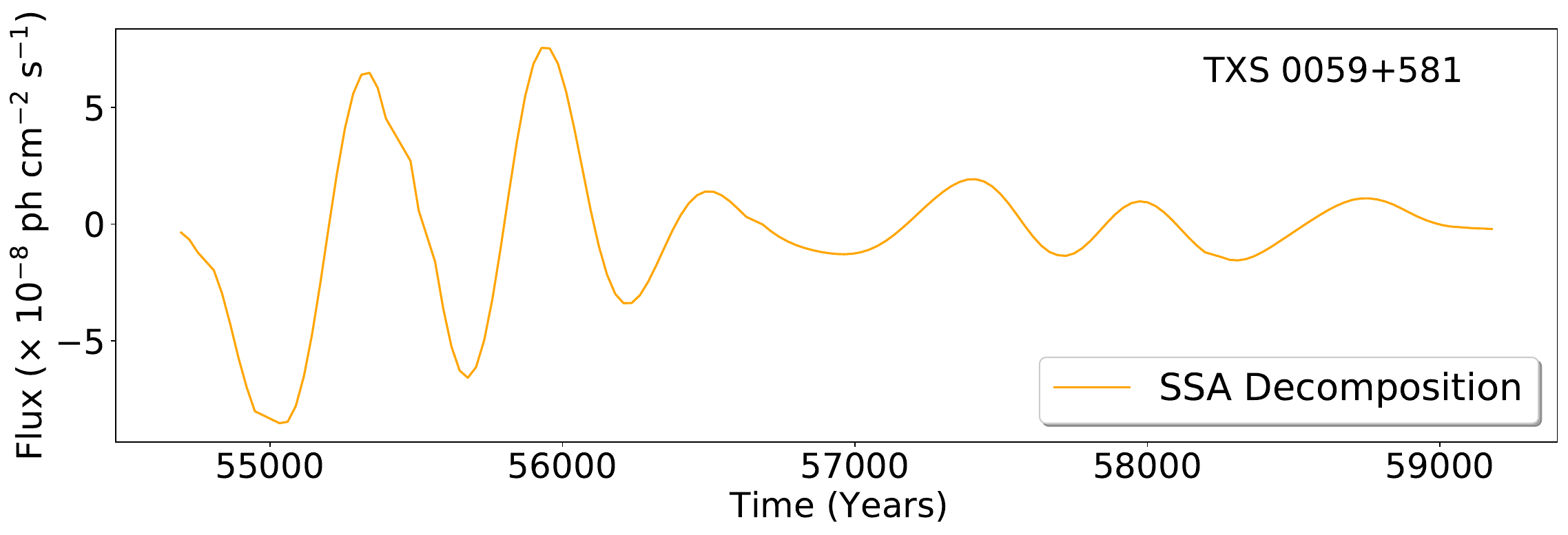}
        \includegraphics[scale=0.21]{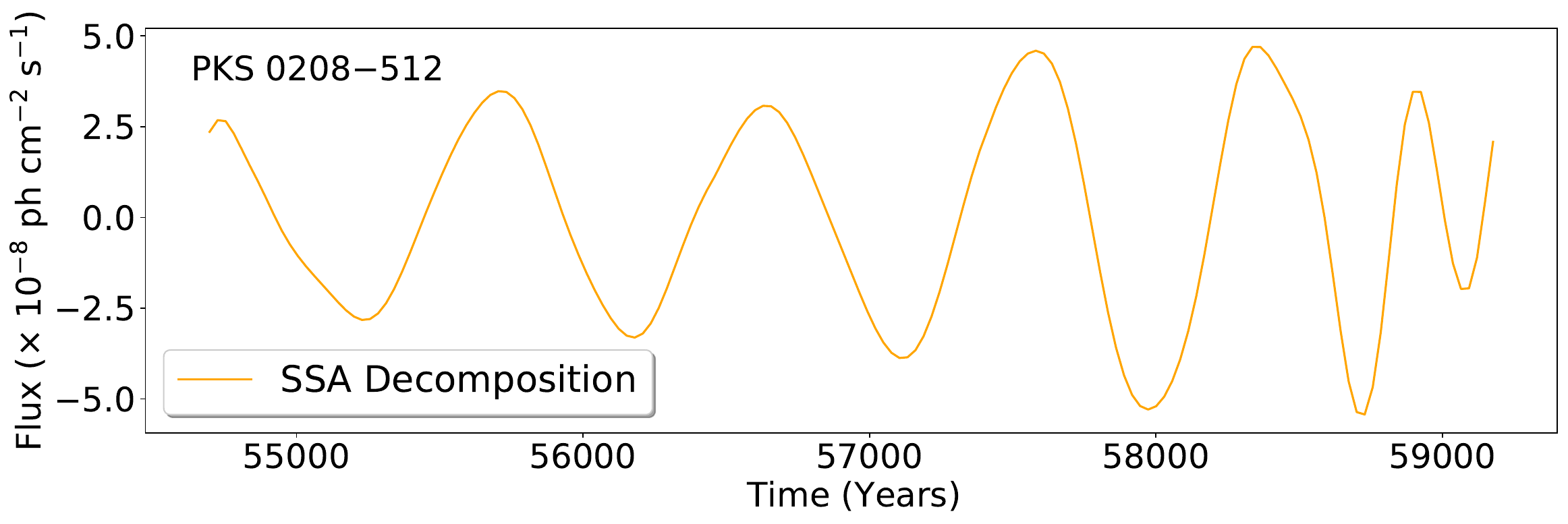}
	\includegraphics[scale=0.21]{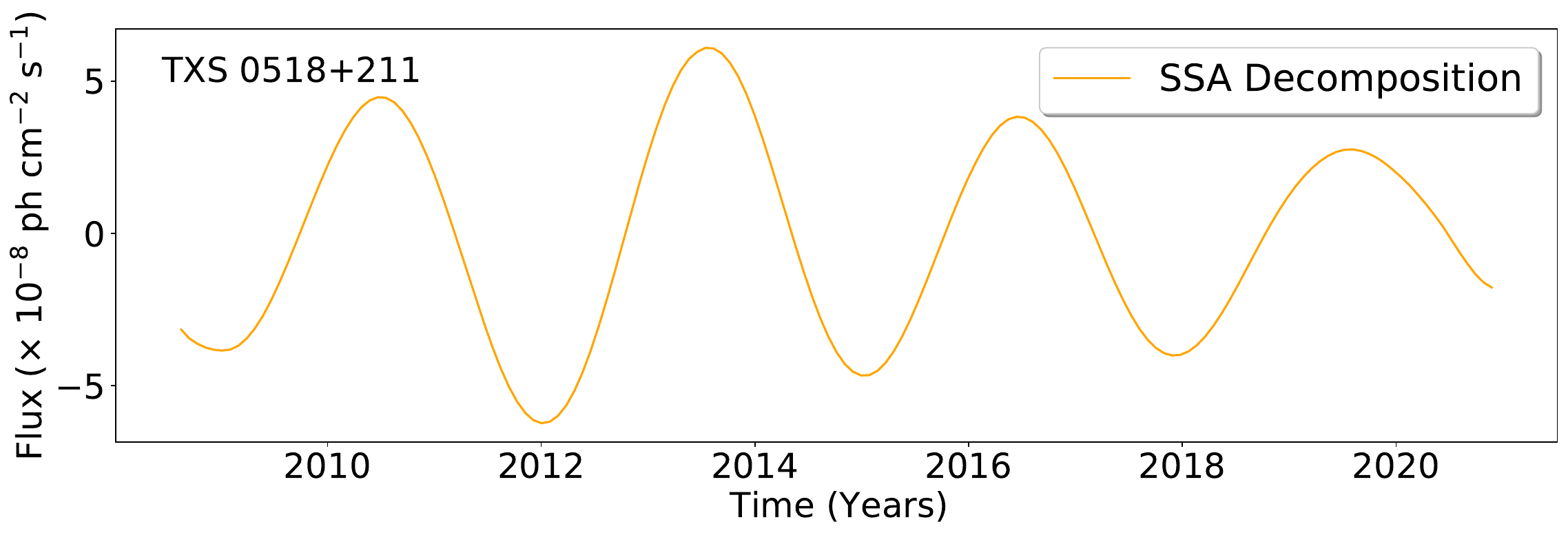}
        \includegraphics[scale=0.21]{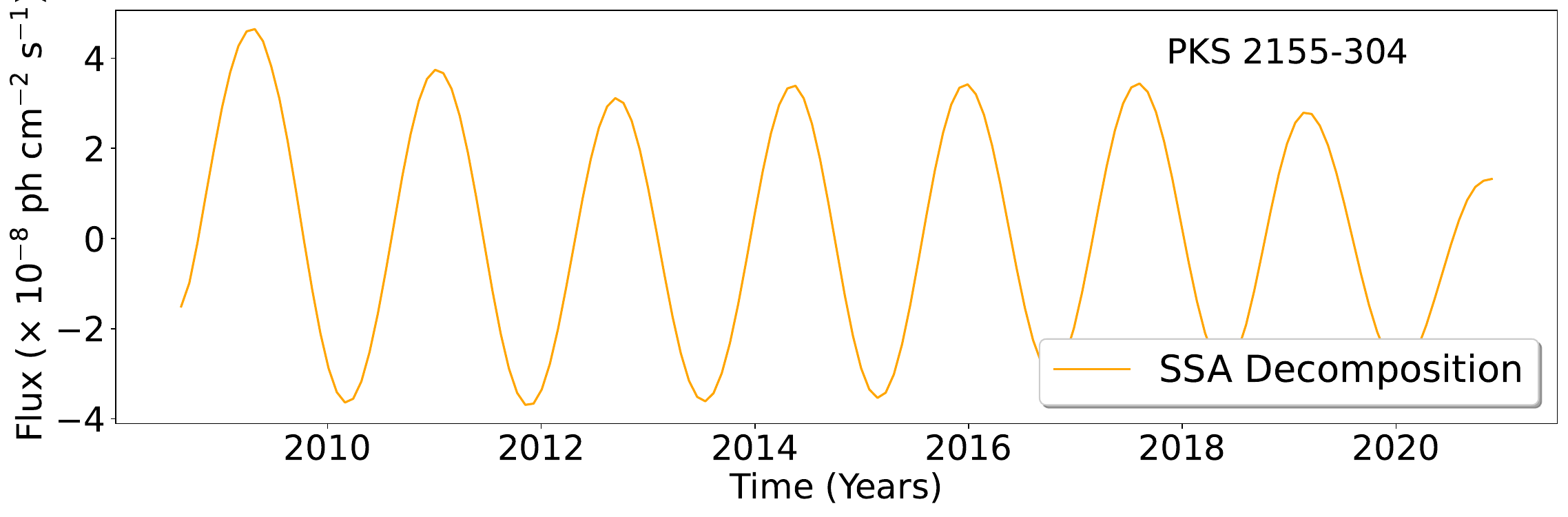} 
	\caption{SSA decomposition showing the underlying oscillatory structure. \textit{Top}: TXS 0059+581 and PKS 0208$-$512 (see Figure \ref{fig:fermi_blazars}). \textit{Bottom}: TXS 0518+211 and PKS 2155$-$304 (see Figure \ref{fig:fermi_blazars}). The flux axis shows negative values because it represents only the oscillatory component, excluding the overall emission behavior of the source. As a result, the oscillatory component is centered around zero, reflecting deviations from the mean rather than the total flux.} 
    \vspace{-0.3cm}
 \label{fig:ssa_blazars}
\end{figure*}

\subsection{New Method for Periodicity Analysis} \label{sec:ssa}
To face the challenges of the long-term flares, we propose the Singular Spectrum Analysis \citep[SSA][]{ssa_greco, SSA_algorithm}. SSA was originally proposed for analyzing astronomical time series by \citet{golyandina_ssa}. \citet{alba_ssa} performed the first systematic search for periodicity using SSA in blazars. SSA operates by decomposing a time series into its constituent sub-components, allowing for the reconstruction of the underlying time series while isolating and excluding the random (noise) component. This unique feature of SSA enables us to extract the oscillatory patterns present in the original LCs, effectively removing the interference of stochastic phenomena like noise and flares (see Figure \ref{fig:ssa_blazars}). For a more detailed technical discussion, we refer readers to \citet{golyandina_ssa} and \citet{alba_ssa}, which provide comprehensive insights into the SSA method. 

Subsequently, we apply SSA to the periodic artificial LCs contaminated by flares, focusing on extracting and analyzing the oscillatory component isolated by SSA. Next, we apply LSP to this oscillatory component to determine the period and its associated significance, following the methodology outlined by \citet{alba_ssa}. This two-step approach improves our ability to detect periodic signals in blazars by effectively separating the underlying oscillations from noise and flare-induced distortions.

%Finally, the fake detection rate of SSA is $0.13 \%$ for a significance $\geq$3$\sigma$ \citep[][]{alba_ssa}. This low rate underscores the robustness of SSA in distinguishing true signals from noise. 

\subsubsection{Testing Against Flares}
We performed the same test for SSA as implemented for the LSP, CWT, and PDM. This involved simulating periodic signals distorted by noise and introducing a flare with the characteristics described in $\S$\ref{sec:types_flares}. The results are presented in Table \ref{tab:experiment_ssa} and Table \ref{tab:experiment_coincident_SSA}.

\begin{table*}
\centering
\caption{
List of parameters for the BPL model of the blazars of our sample.  The parameter ``$A$'' is the normalization ($rms^{2}/yr^{-1}$), $\nu_{b}$ the bending frequency, and $\alpha$ the spectral index. The Poisson noise is represented by ``$C$'' ($rms^{2}/yr^{-1}$). The table includes the period (and significance) obtained in \citet{penil_2022} (denoted as \textit{Period*}), the period resulting from using SSA, and the periods obtained for LSP, CWT, PDM and SSA applying the sigma clipping technique denoted by $\#$. The periods are in years. \label{tab:psd_parameters}}
\resizebox{\textwidth}{!}
{%
\begin{tabular}{l|cccc|c||c||cccc}
\hline
\hline
Association Name & $C$ & $A$ & $\alpha$ & $\nu_{b}$ & Period* & SSA & LSP$\#$ & CWT$\#$ & PDM$\#$ & SSA$\#$ \\
\hline
TXS 0059+581 & $8.48\mathrm{x}10^{-3}$ & 6.53$\mathrm{x}10^{-2}$$\pm$6.90$\mathrm{x}10^{-3}$ & 3.84$\pm$0.28 & 0.77$\pm$0.09 & 4.0$\pm$0.6 (0.8$\sigma$) & 1.8$\pm$0.2 (1.2$\sigma$) & 4.0$\pm$0.6 (1.1$\sigma$) & 1.0$\pm$0.1 (1.0$\sigma$) & 3.7$\pm$0.5 (1.3$\sigma$) & 1.8$\pm$0.2 (1.1$\sigma$) \\
\hline
PKS 0208$-$512 & $6.14\mathrm{x}10^{-4}$ & 6.03$\mathrm{x}10^{-2}$$\pm$1.34$\mathrm{x}10^{-3}$ & 2.66$\pm$0.17 & 0.86$\pm$0.05 & 3.8$\pm$0.5 (0.1$\sigma$) & 2.6$\pm$0.2 (2.2$\sigma$) & 2.5$\pm$0.2 (2.0$\sigma$) & 2.8$\pm$0.4 (2.5$\sigma$) & 2.6$\pm$0.3 (2.7$\sigma$) & 2.6$\pm$0.3 (2.6$\sigma$) \\
\hline
TXS 0518+211 & $2.35\mathrm{x}10^{-3}$ & 1.14$\mathrm{x}10^{-1}$$\pm$2.93$\mathrm{x}10^{-2}$ & 2.76$\pm$0.18 & 0.82$\pm$0.09 & 3.1$\pm$0.4 (2.6$\sigma$) & 3.1$\pm$0.4 (3.2$\sigma$) & 3.1$\pm$0.4 (2.7$\sigma$) & 3.2$\pm$0.4 (3.1$\sigma$) & 3.2$\pm$0.3 (2.9$\sigma$) & 3.0$\pm$0.3 (3.8$\sigma$) \\
\hline
PKS 2155$-$304 & $1.62\mathrm{x}10^{-3}$ & 4.75$\mathrm{x}10^{-2}$$\pm$1.03$\mathrm{x}10^{-2}$ & 0.88$\pm$0.09 & 1.19$\pm$0.22 & 1.7$\pm$0.1 (3.3$\sigma$) & 1.7$\pm$0.1 (4.1$\sigma$) & 1.7$\pm$0.1 (3.2$\sigma$) & 1.7$\pm$0.1 (3.1$\sigma$) & 1.7$\pm$0.2 (2.9$\sigma$) & 1.7$\pm$0.1 (4.1$\sigma$)\\
\hline
\hline
\end{tabular}%
}
\end{table*}

Regarding the general case, Table \ref{tab:experiment_ssa} demonstrates that SSA improves the results compared to the other analyzed methods. Specifically, SSA reports 5$\sigma$ detections in 31\% of cases, compared to 12\% considering all the results for the methods LSP, CWT and PDM (Table \ref{tab:experiment_general}). The identification of the periodic at a signal significance $\geq$3$\sigma$ occurs in 81\% of the cases, while it only occurs in 56\% of cases for LSP, which is the best scenario among the other methods. The improvement with SSA is more notable in the number of cycles required to recover a 5$\sigma$ detection. Only in one case was it not possible to recover the detection with SSA, whereas with LSP, it was not possible in 3 cases (see Table \ref{tab:experiment_general}). In the best case, SSA requires 2 cycles, in the worst case, 9 cycles are required. Regarding LSP, it requires 3 cycles for the best case, and the worst case is 10 cycles. As mentioned in $\S$ \ref{sec:results}, flares have a greater impact on shorter periods due to the increased distortion they produce. These results also highlight the robustness of SSA, which consistently delivers more reliable outcomes, even in cases where flare-induced distortions are more pronounced. 

In the specific test scenario where a flare is introduced coinciding in phase with an oscillation at the start, middle, and end of the periodic signal, Table \ref{tab:experiment_coincident_SSA} shows that SSA is, again, more robust. For the 2-year period signal, SSA accurately reports the real period in 85\% of cases and achieves 5$\sigma$ detections in 31\% of cases. For the 3-year period signal, SSA reports the real period in 55\% of cases and achieves 5$\sigma$ detections in 31\% of cases. In comparison, the best-performing method among the others for the 2-year period signal is LSP, which detects the original period in 69\% of cases and achieves 5$\sigma$ detections in 14\% of cases. For the 3-year period signal, LSP detects the original period in 34\% of cases and achieves 5$\sigma$ detections in 12\% of cases. The reduction in significance with SSA is up to 50\%, which is similar to LSP. 

\subsection{Sigma Clipping Technique} \label{sec:sigmaclipping}
As an alternative method, we evaluate the sigma clipping technique \citep[e.g.,][]{pietka_clipping, fan_sigma_clipping}. The sigma clipping technique identifies and removes outliers in datasets. It works by iteratively calculating the median and standard deviation (sigma) of a dataset and then excluding data points that deviate by more than a specified number of standard deviations from the median. This process is repeated until no further outliers are identified, resulting in a dataset where outliers are effectively excluded. In our test, the role of outliers is played by the data points of the flares. In our study, we use the traditional clipping threshold of 3$\sigma$ and iterating until convergence\footnote{We use the function ``sigma\_clip'' of Astropy} \citep[e.g.][]{pietka_clipping}.

As a consequence, these outliers are removed from the LCs, resulting in the introduction of gaps. As shown in \citet{penil_2020}, these gaps can distort the accurate determination of the period and the associated significance. The impact of the gaps depends on their percentage in the LC, becoming significant in terms of both period non-detection and reduced significance when the gap percentage exceeds 50\%. In our analysis of the gap distribution in the simulated cases, the largest median gap percentage is 11\% (in the Type VIIIb flare case). This percentage is low enough to have no significant impact on the distortion of the period or the significance of our tests. 

Additionally, the CWT is not recommended for analyzing unevenly distributed LCs. The CWT assumes uniformly sampled data, and when applied to unevenly spaced observations, it can introduce artifacts or distortions in the period detection. However, for the sake of consistency in this study, we apply the CWT to unevenly distributed LCs, assuming that potential distortions are negligible due to the limited, small rate of gaps produced by the sigma clipping. For real analysis scenarios, we recommend using the weighted wavelet Z-transform \citep[][]{foster_wwz}, which is better suited for handling unevenly sampled data. 

\subsubsection{Testing Against Flares}
We conduct the same tests as those applied to LSP, CWT, and PDM, but applying first the sigma clipping, with the results presented in Table \ref{tab:experiment_general_sigma_cipping}. All three methods produced similar outcomes in terms of 5$\sigma$ detections, achieving approximately 50\% detection rates across the cases. This represents a 30\% improvement compared to SSA alone (results of Table \ref{tab:experiment_ssa}) and a 75\% improvement compared to cases where the LC was analyzed by LSP, CWT, and PDM but not applying the sigma clipping technique (results of Table \ref{tab:experiment_general}). The identification of the periodic signal at a significance $\geq$3$\sigma$ occurs in 100\% of the cases, while only applying SSA was 81\%. Regarding the number of extra cycles required to recover the 5$\sigma$ detection, the maximum is 3, denoting the significant improvement regarding SSA. 

We also evaluate the scenario where the flare coincides with the oscillation of the periodic signal at the beginning, middle, and end of the LC. The results, shown in Table \ref{tab:experiment_coincident_sigma_clipping}, indicate that for the 2-year period signal, the true period is detected in 87\% of cases (comparable to SSA) and achieves 5$\sigma$ detections in 68\% of cases, a 50\% improvement over SSA. For the 3-year period signal, the true period is reported in 81\% of cases (a 30\% improvement over SSA) and achieves 5$\sigma$ detections in 76\% of cases, representing a 60\% improvement compared to SSA.

These results underscore the significant effectiveness of applying sigma clipping to mitigate the impact of flares on periodicity detection. Sigma clipping effectively removes outlier data points caused by flares, which distort the analysis and obscure the underlying periodic signals. This improvement is particularly evident in the performance of the PDM method. As noted in $\S$\ref{sec:results}, PDM was previously the least effective method due to its sensitivity to flare-induced distortions. However, with the application of sigma clipping, PDM now delivers the best results, outperforming both LSP and CWT in detecting the true periodic signal. This demonstrates that sigma clipping not only enhances the accuracy of detection methods but also transforms the performance of otherwise less robust techniques, leading to more reliable identification of the true period and a marked improvement in the significance of the results.

\begin{figure*}
	\centering
	\includegraphics[scale=0.23]{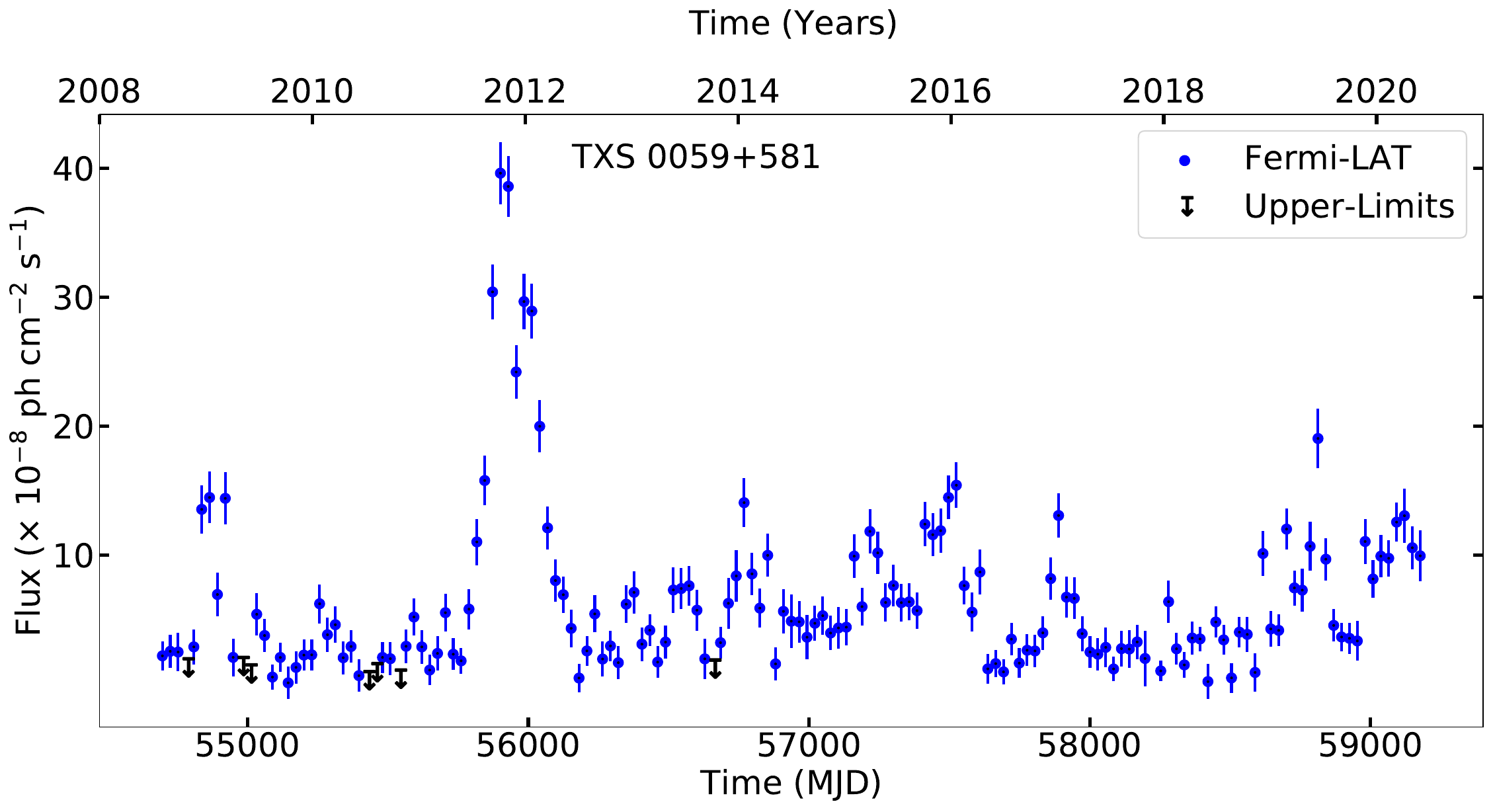}
	\includegraphics[scale=0.23]{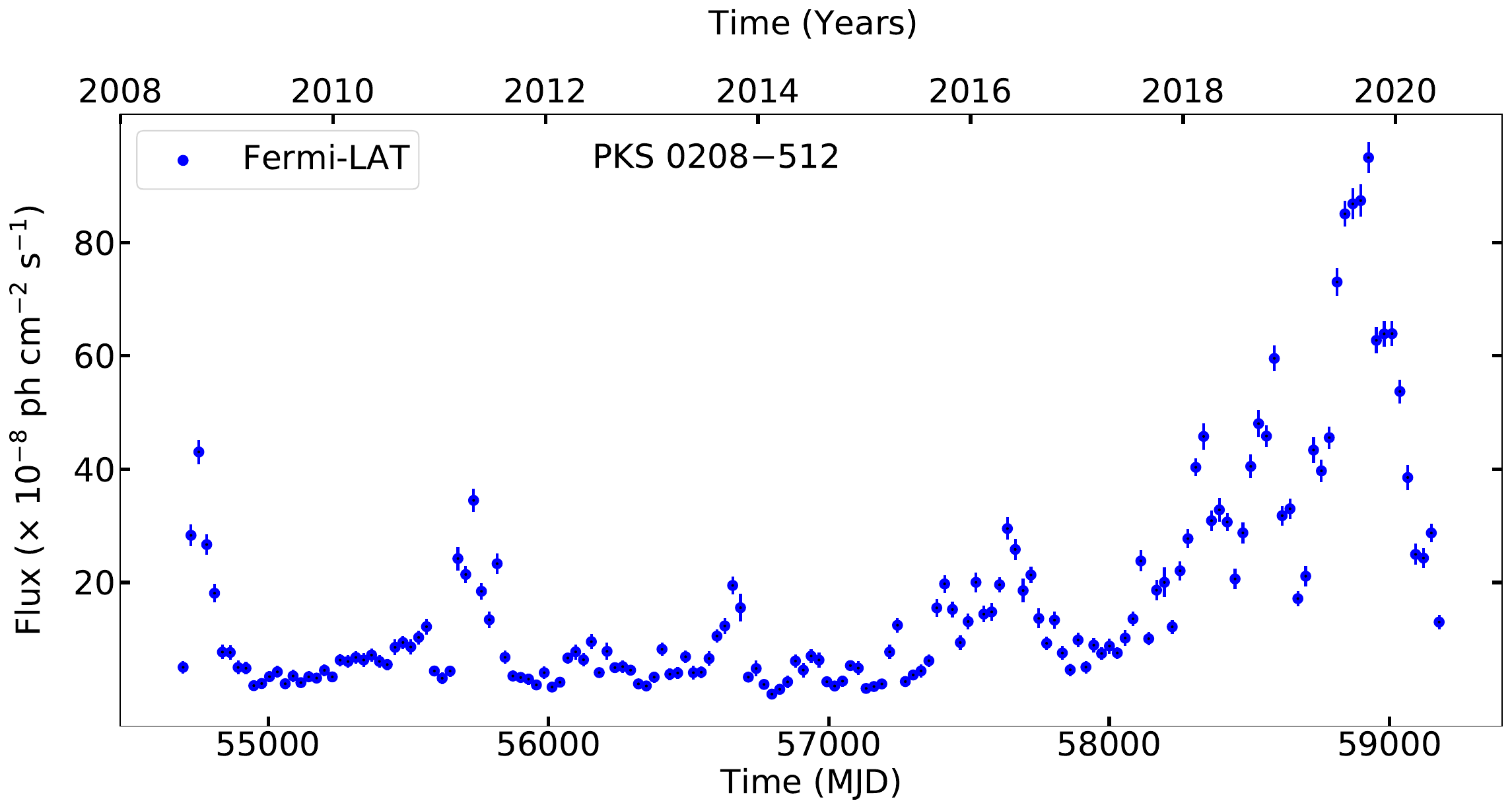}
        \includegraphics[scale=0.23]{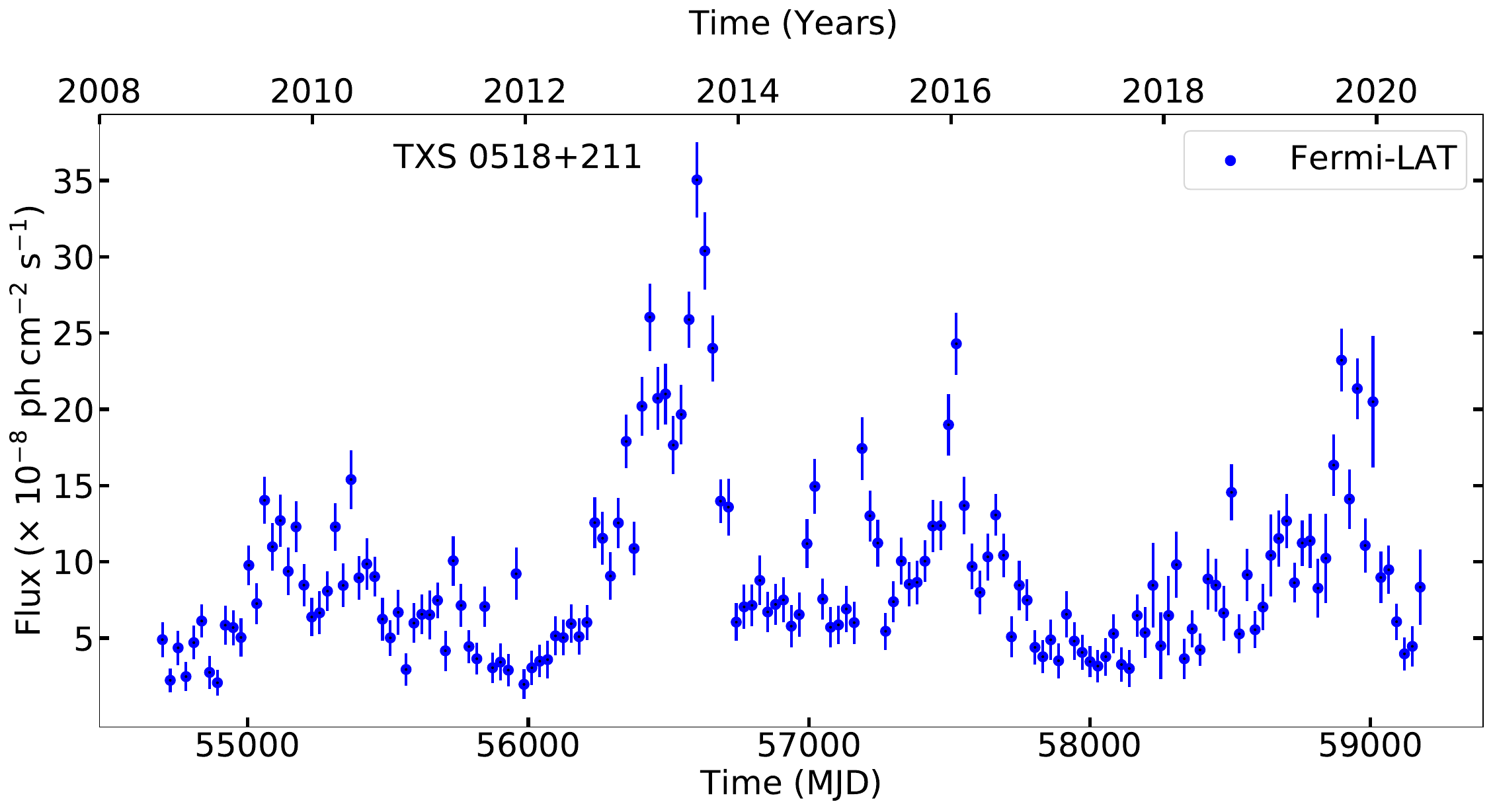}
        \includegraphics[scale=0.23]{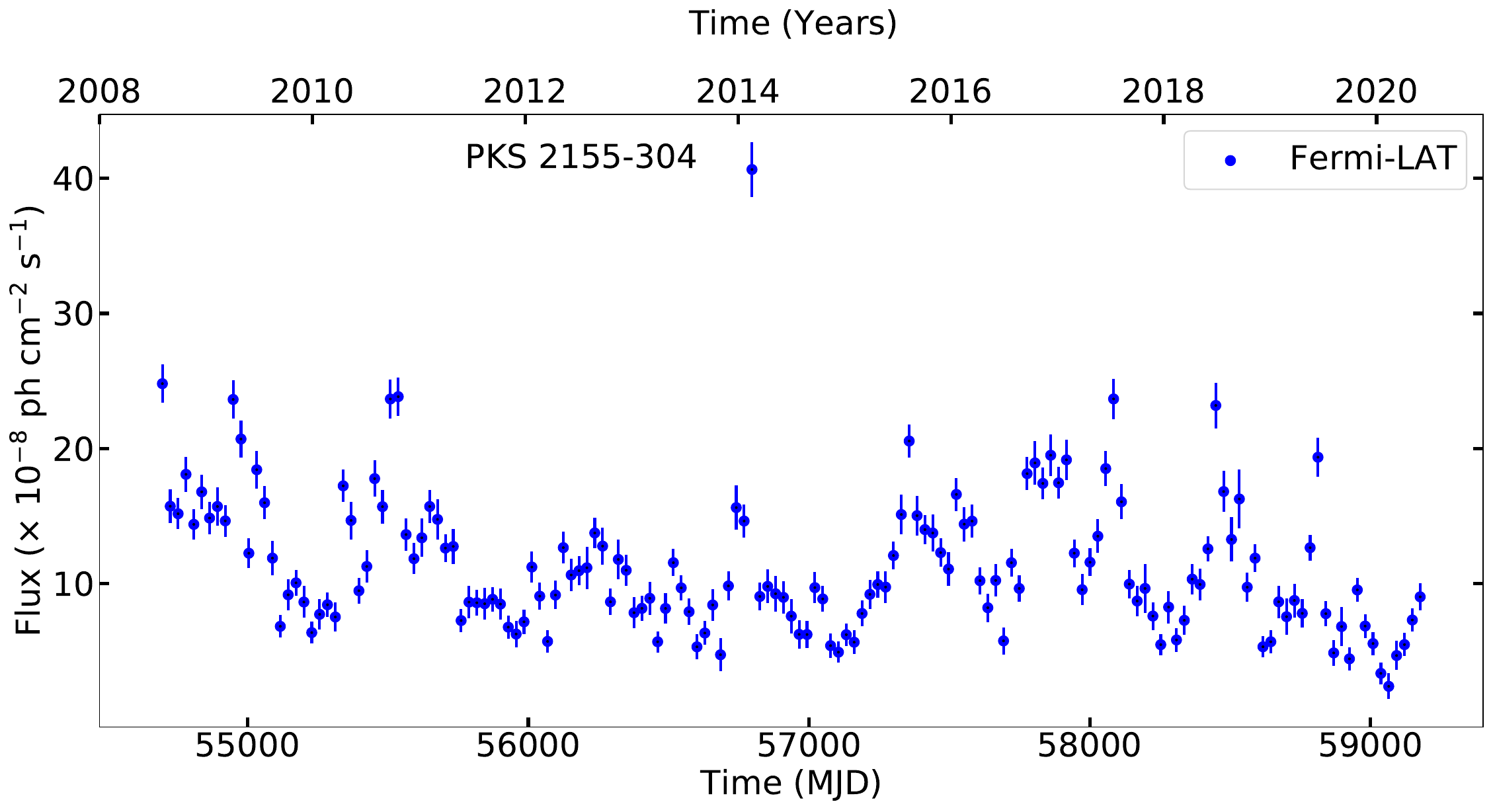}
	\caption{{\it Top}: \textit{Fermi}-LAT LCs of TXS 0059+581 and PKS 0208$-$512. {\it Bottom}: \textit{Fermi}-LAT LCs of TXS 0518+211 and PKS 2155$-$304.} \label{fig:fermi_blazars}
\end{figure*}

\section{Real Use Cases} \label{sec:usecase}
We apply the previous results to real data. Specifically, we select four blazars that present flares in their \textit{Fermi}-LAT emissions, with a periodicity that was previously reported in their $\gamma$-ray emissions in \citet{penil_2022}. The blazars are:
\begin{enumerate}
    \item TXS 0059+581: FSRQ; {\it z}=0.644; period of 4.0 yr (0.8$\sigma$) 
    \item PKS 0208$-$512: FSRQ; {\it z}=1.003; period of 3.8 yr (0.1$\sigma$) 
    \item TXS 0518+211: BLL; {\it z}=0.108; period of 3.1 yr (2.6$\sigma$)
    \item PKS 2155$-$304: BLL; {\it z}=0.116; period of 1.7 yr (3.1$\sigma$)
\end{enumerate}

\begin{figure*}
	\centering
	\includegraphics[scale=0.22]{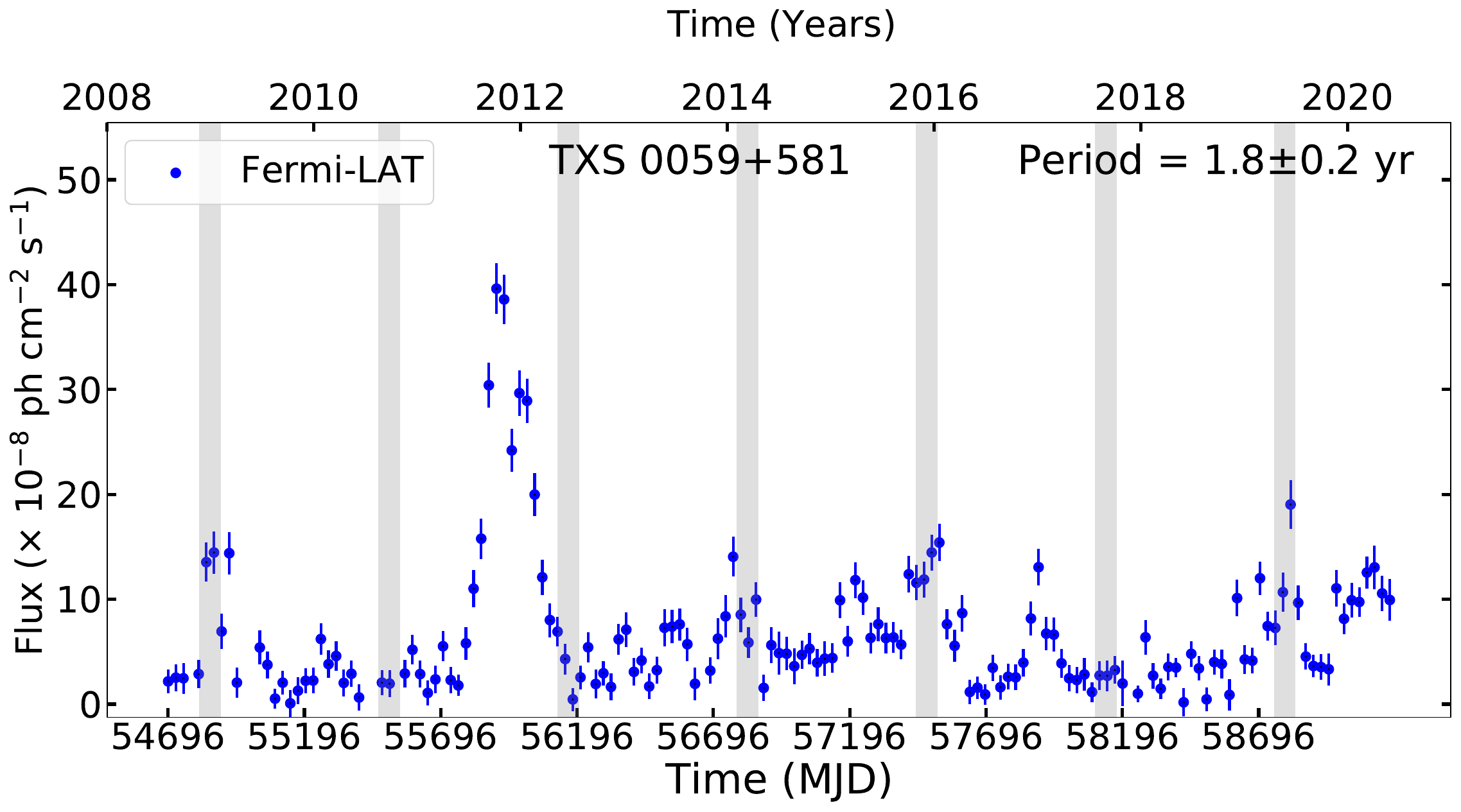}
        \includegraphics[scale=0.22]{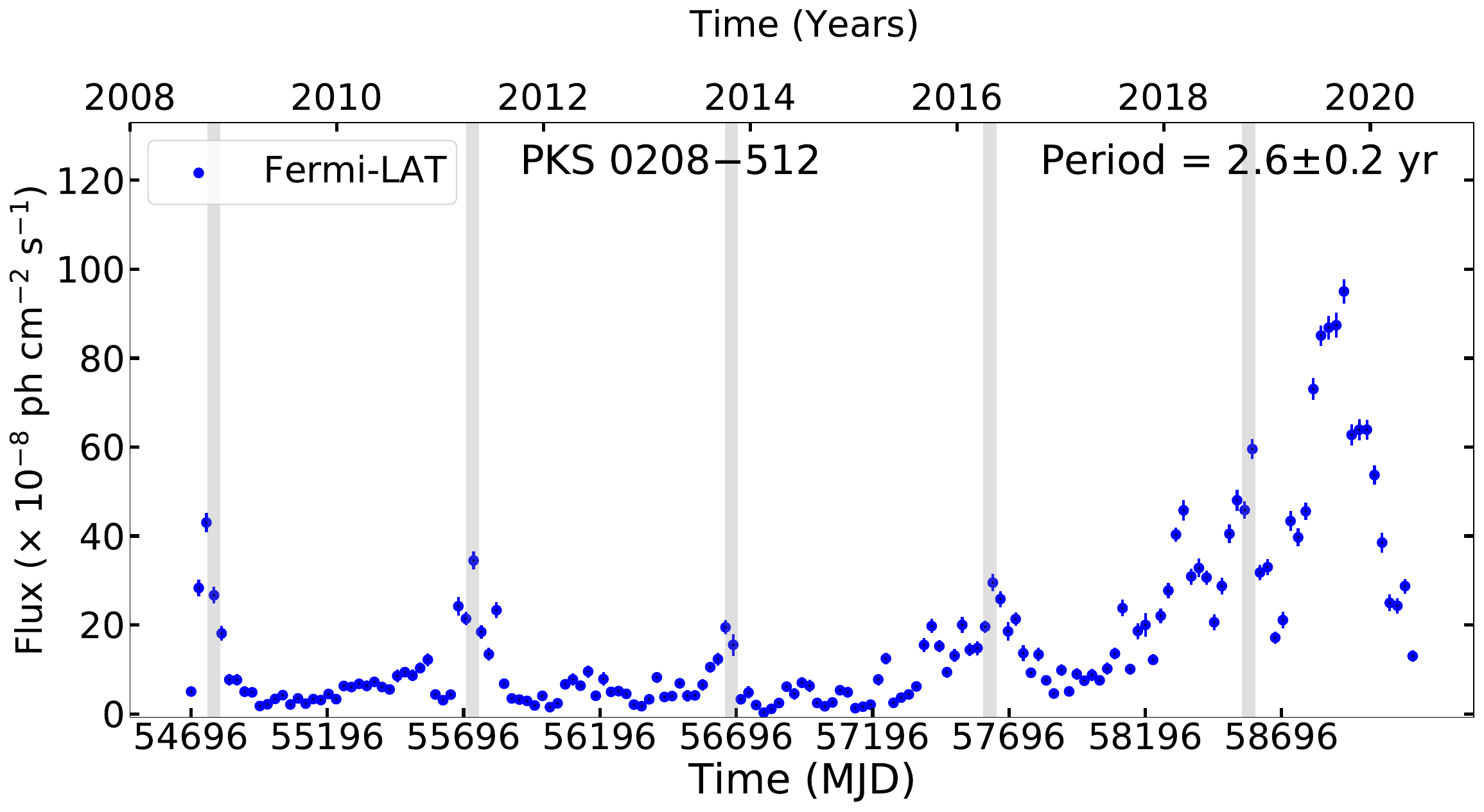}
        \includegraphics[scale=0.22]{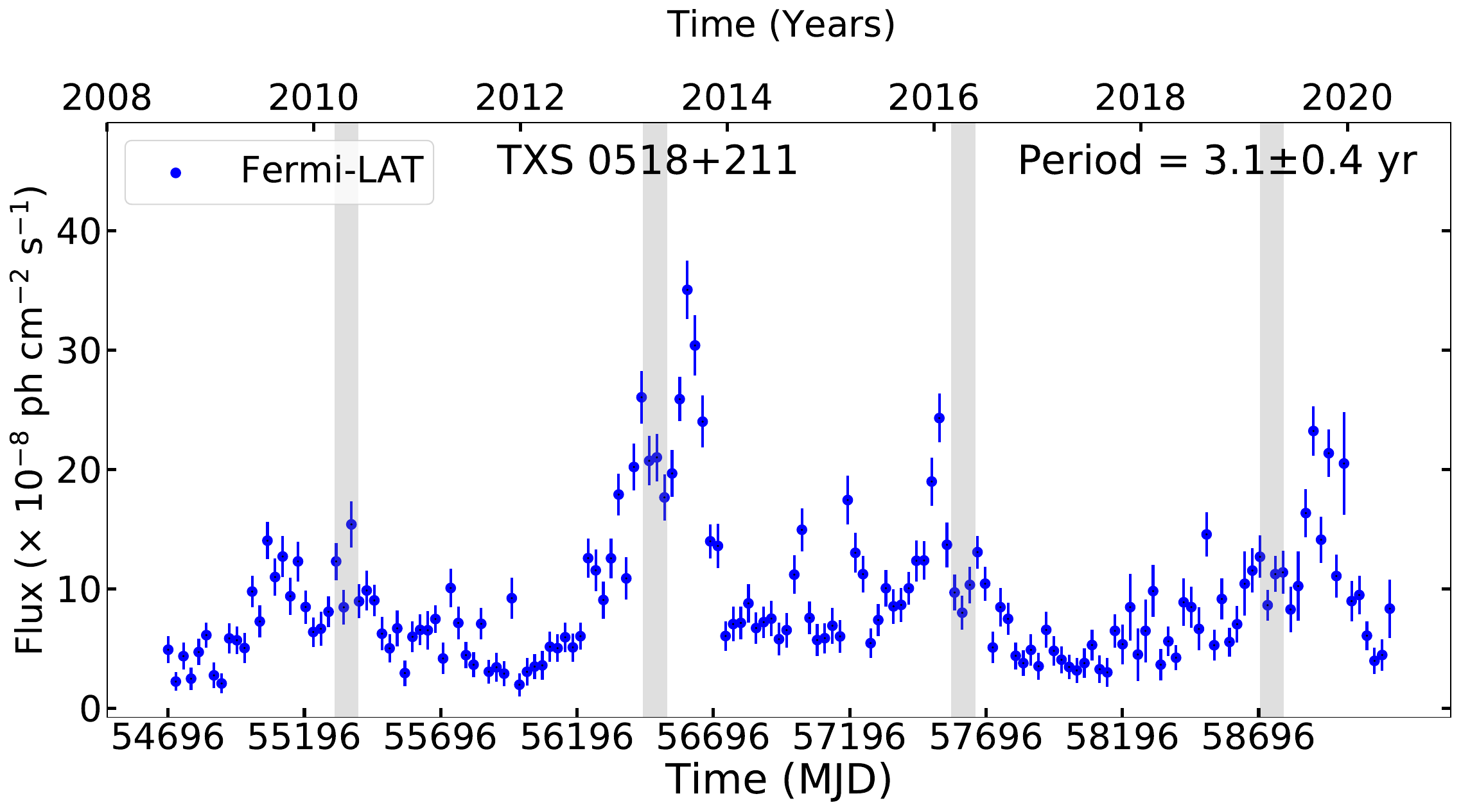}
        \includegraphics[scale=0.22]{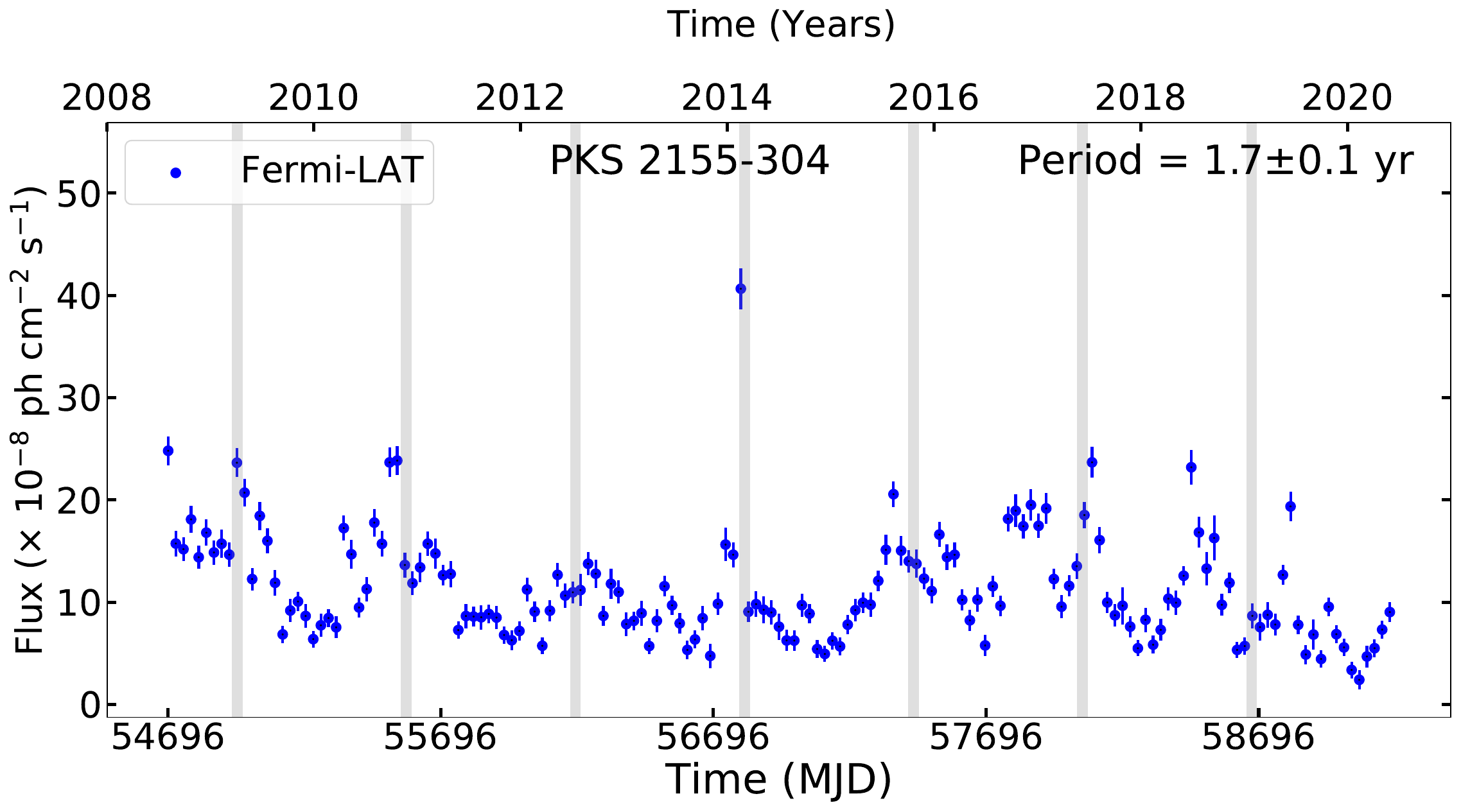}
 	\caption{Light curves of TXS 0059+581, PKS 0208$-$512, TXS 0518+211 and PKS 2155$-$304. The gray vertical bars approximate high-flux periods suggested by the period inferred by SSA. The width of the gray bars indicates the uncertainty in the periodic signal.} 
 \label{fig:pks0208}
\end{figure*}

The LCs shown in Figure \ref{fig:pks0208} were analyzed in \citet{penil_2022} and in \citet{alba_ssa}. These objects present different flaring events that can be associated with the categories presented in Table \ref{tab:flare_types}. In the case of TXS 0059+581, a notable flare with a flux 250\% higher than the other high-emission states occurs approximately in the middle of the LC, which can be considered a Type Vb flare. The situation with PKS 0208$-$512 presents a different scenario. A flare occurs at the end of the LC, which is a Type VIIIb flare. In the case of TXS 0518+211, a mid-light curve flare exhibits a flux $\approx$100\% higher than the high-state average of the other oscillations, being categorized as Type IIIa flare. Finally, PKS 2155$-$304 has a short-duration flare in the middle of the signal, with a flux of 150\% higher than the flux associated with the oscillations. It is categorized as Type Ib flare.   

We employ SSA to investigate the periodic behavior in the $\gamma$-ray emissions of the blazars in our sample. Following the two-step procedure outlined in $\S$\ref{sec:ssa}, we first extract the oscillatory component of the LC using SSA, as shown in Figure \ref{fig:ssa_blazars}. Next, to search for periodicity, we apply the LSP, following the approach described by \citet{alba_ssa}. To ensure the reliability of this method, we estimate the significance by simulating 1,000,000 artificial LCs. The artificial LCs are generated following the technique introduced in \citet{emma_lc}, maintaining the same power spectral density (PSD) and probability density function as observed in real blazar LCs.  

We adopt a similar approach to the one presented in \citet{penil_2022} for fitting the PSD. Specifically, we utilize the bending-power law \citep{chakraborty_bending_power_law} with Poisson Noise (``$C$''), with the parameters normalization (``$N$''), bending frequency (``$\nu_{b}$'') and spectral index (``$\alpha$''):

\begin{equation} \label{eqn:bending} 
  P(\nu) = A \left( 1 + \left\{ \frac{\nu}{\nu_{b}} \right\}^{\alpha} \right)^{-1} + C, 
\end{equation}

The specific values of such parameters for the blazars are presented in Table \ref{tab:psd_parameters}. The results of applying SSA are also presented in Table \ref{tab:psd_parameters}. The results of our analysis reveal distinct periods for the blazars in our sample:
\begin{enumerate}
\item For TXS 0059+581, we identify a period of 1.8$\pm$0.2 years with a significance level of 1.2$\sigma$ (see Figure \ref{fig:pks0208}). 
\item PKS 0208$-$512 exhibits a period of 2.5$\pm$0.2 years, accompanied by a significance of 2.2$\sigma$ (see Figure \ref{fig:pks0208}). 
\item In the case of TXS 0518+211, we find a period of 3.1$\pm$0.4 years, with a significance level of 3.2$\sigma$ (see Figure \ref{fig:pks0208}). 
\item In the case of PKS 2155$-$304, we find a period of 1.7$\pm$0.1 years, with a significance level of 4.1$\sigma$ (see Figure \ref{fig:pks0208}). 
\end{enumerate}

Comparing these results with those presented in \citet{penil_2022}, we can infer the following conclusions:
\begin{enumerate}
\item For TXS 0059+581, our analysis yields a different period from \citet{penil_2022}, indicating that the previously suggested periodic behavior is not genuine.
\item PKS 0208$-$512 shows significant differences in our findings compared to \citet{penil_2022}. While our period of 2.6 years aligns with the result in \citet{penil_2020}, which used the first 9 years of \textit{Fermi}-LAT observations, thus missing the last, bright flare (see Figure \ref{fig:fermi_blazars}). As predicted by the simulations in previous sections, the presence of the flare introduces uncertainty. Therefore, this blazar could be a case of genuine periodicity being hidden by a flaring event. 
\item In the case of TXS 0518+211, our analysis confirms the same period as \citet{penil_2020, penil_2022}. We achieve an increase of 16\% of the level of significance, reinforcing the evidence of the inferred periodicity for this blazar.
\item Similar to the previous blazar, PKS 2155$-$304 analysis obtains the same period with a significant increase in the significance, 27\%, regarding \citet{penil_2022}. 
\end{enumerate}

In relation to the results presented by \citet{alba_ssa}, our findings reveal consistent values in terms of period and significance for TXS 0059+581, which is not included among the blazars with significance $\geq 2.0\sigma$. For PKS 2155$-$304, \citet{alba_ssa} obtained a period of $1.65 \pm 0.1$ with a significance of $4.5\sigma$, aligning well with our results.

For PKS 0208$-$512 and TXS 0518+211, \citet{alba_ssa} derived periods of $2.47 \pm 0.26$ and $2.83 \pm 0.38$, respectively, which are compatible with those reported by us. However, we observe discrepancies in the associated significance levels, which \citet{alba_ssa} determined to be $4.5\sigma$ for PKS 0208$-$512 and $4.8\sigma$ for TXS 0518+211. These variations in significance could be attributed to differences in the parameters estimated in the PSD modeling. This fact directly impacts the properties of the simulated LCs used to estimate significance, potentially leading to variations in the derived significance values \citep{chakraborty_bending_power_law}. Additionally, these discrepancies may arise from differences in the statistical frameworks employed to calculate significance. For instance, \citet{alba_ssa} adopted the approach outlined in the Appendix of \citet{oneill_2022}. This method differs from the approach used in this study, which applies a distinct procedure for evaluating the significance of periodic signals.

\subsection{Sigma Clipping Analysis}
Alternatively, the sigma-clipping technique is applied as a preprocessing step to analyze the flares of four blazars in the context of their flares using LSP, CWT, and PDM. The results, presented in Table \ref{tab:psd_parameters}, are consistent with those obtained using SSA in terms of both period and associated significance. These findings confirm significant evidence of periodicity for TXS 0518+211 and PKS 2155$-$304. For PKS 0208$-$512, the inferred period of 2.6 years adds further uncertainty regarding the presence of periodic behavior in this blazar, suggesting that its periodicity remains inconclusive. Regarding the percentage of gaps generated by the sigma clipping process for these objects, the highest percentage of gaps is observed in TXS 0059+581 and PKS 0208$-$512, both reaching 8.7\%, while the lowest is recorded for PKS 2155$-$304 at 0.6\%. These gap rates are within acceptable thresholds, as they are unlikely to introduce significant distortions into the periodicity analysis \citep[][]{penil_2020}.

\subsection{Sigma Clipping Combined with SSA}

Finally, we apply sigma clipping in combination with SSA, resulting in enhanced detection performance. The updated results, summarized in Table \ref{tab:psd_parameters}, remain consistent with those discussed in $\S$\ref{sec:ssa_results} but show an improvement in the significance levels for two sources. For PKS 0208$-$512, the significance level increases from $2.2\sigma$ to $2.6\sigma$, reflecting the impact of refined data processing. Similarly, TXS 0518+211 demonstrates an enhancement from $3.2\sigma$ to $3.8\sigma$. These improvements highlight the utility of sigma clipping as a preprocessing strategy in periodicity analysis, particularly when combined with SSA.

\section{Summary} \label{sec:summary}
In this study, we have conducted a comprehensive analysis of the impact of long-term flares on periodicity analyses. Specifically, we have assessed how long-term flares with varying properties and positions within an LC affect three of the most commonly employed methods in the literature for periodicity detection, LSP, CWT, and PDM. Our objective was to understand how flares can alter the determination of the period and significance levels in periodicity analysis. 

Our analysis demonstrates that incorporating preprocessing steps enhances the detection of periodic signals. Specifically, applying SSA significantly improves the performance of the LSP, CWT, and PDM by effectively mitigating the effects of flares. These results underscore SSA's robustness and its suitability for signal analysis under challenging conditions.

Additionally, sigma clipping as a preprocessing step enhances the performance of LSP, CWT, and PDM by reducing flare-induced distortions. This method proves effective for analyzing periodic signals in the presence of flares, yielding results comparable to SSA while confirming periodicity. However, sigma clipping introduces gaps into the LC, which can potentially distort the analysis if a substantial portion of the dataset is affected. Careful implementation is essential to minimize these gaps, as extensive data loss can hinder the detection of periodicity or lead to false detections, thereby compromising the reliability of the results.

To apply and validate our findings, we examined the LCs of four blazars, namely TXS 0059+581, PKS 0208$-$512, TXS 0518+211, and PKS 2155$-$304. The results obtained from these blazars align closely with the outcomes observed in our test signals. This reaffirms the influence of long-term flares on periodic analysis methods and the potential distortions they introduce.

\subsection {Software}
\begin{enumerate}
        \item Astropy \citep{astropy_2013, astropy_2018}
	\item PyAstronomy \citep{PyAstronomy}
	\item PyCWT, \url{https://pypi.org/project/pycwt/} 
	\item SciPy \citep {SciPy}
	\item Simulating light curves \citep{connolly_code}
        \item Singular Spectrum Analysis \url{https://www.kaggle.com/code/jdarcy/introducing-ssa-for-time-series-decomposition}
\end{enumerate}

\section{Acknowledgements}

P.P and M.A acknowledge funding under NASA contract 80NSSC20K1562. S.B. acknowledges financial support by the European Research Council for the ERC Starting grant MessMapp under contract no. 949555. 

\section{Data Availability}

All the data used in this work are publicly available or available on request to the responsible for the corresponding observatory/facility.

\bibliographystyle{mnras}
\bibliography{literature} % if your bibtex file is called example.bib

\clearpage

\appendix

\renewcommand{\thetable}{A\arabic{table}}
\section*{Appendix}\label{sec:appendix}
\subsection{Tables}
This section includes the tables with the results of the different tests performed in this study. 
\begin{table*}
\centering
\caption{Results of the tests when a single flare is injected at random phase angles within a periodic signal. The evaluation of periodicity detection for signals with periods of 2 and 3 years, considering different types of flares shown in Table \ref{tab:flare_types}. In bold, we highlight the tests where at least two methods report a compatible period with a significance of $\geq$5$\sigma$. A compatible period is defined as achieving a period that matches the signal period within a tolerance of $\pm$0.1. All periods are expressed in years. We also estimate the number of cycles required to achieve a period detection significance of 5$\sigma$ (only performed for the periodic signal with a period of 2 years). The symbol \textit{--} denotes that a 5$\sigma$ detection was achieved in the initial analysis, while \textit{X} values indicate that it was not possible to obtain the estimation within a limit of 10 cycles.
\label{tab:experiment_general}}
%\resizebox{\columnwidth}{!}
{%
\begin{tabular}{ccccccccc}
\hline
\hline
Period [yr] & Flux of Flare & LSP & Cycles & CWT & Cycles & PDM & Cycles \\
\hline
		\multirow{4}{*}{2 yr} & 
                            \makecell{Type Ia \\ Type Ib} &
                            \makecell{\textbf{2.00$\pm$0.02 (5.1$\pm$0.4$\sigma$)} \\ 2.00$\pm$0.04 (3.0$\pm$0.6$\sigma$)} & 
                            \makecell{-- \\ 3 } &
                            \makecell{\textbf{2.00$\pm$0.02 (5.0$\pm$0.3$\sigma$)} \\ 2.00$\pm$0.03 (2.2$\pm$0.4$\sigma$)} & 
                            \makecell{-- \\ 7 } &
                            \makecell{\textbf{2.00$\pm$0.03 (5.2$\pm$0.4$\sigma$)} \\ 2.00$\pm$0.05 (3.6$\pm$0.9$\sigma$)} &
                            \makecell{-- \\ 2 } \\ \cline{2-8} 
                            & \makecell{Type IIa \\ Type IIb} &
                            \makecell{\textbf{2.00$\pm$0.02 (5.2$\pm$0.4$\sigma$)} \\ 2.0$\pm$0.4 (3.0$\pm$0.8$\sigma$)} & 
                            \makecell{-- \\ 5 } &
                            \makecell{\textbf{2.00$\pm$0.03 (5.0$\pm$0.3$\sigma$)} \\ 2.0$\pm$0.5 (2.2$\pm$0.6$\sigma$)} & 
                            \makecell{-- \\ 10 } &
                            \makecell{\textbf{2.00$\pm$0.05 (5.2$\pm$0.6$\sigma$)} \\ 2.0$\pm$0.9 (2.1$\pm$0.7$\sigma$)} &
                            \makecell{-- \\ 4 } \\ 
                            \cline{2-8}
                            & \makecell{Type IIIa \\ Type IIIb} &
                            \makecell{2.0$\pm$0.1 (4.0$\pm$0.9$\sigma$) \\ 2.0$\pm$0.7 (2.4$\pm$0.7$\sigma$)} &
                            \makecell{3 \\ 9 } &
                            \makecell{2.0$\pm$0.2 (3.2$\pm$0.9$\sigma$) \\ 2.0$\pm$0.9 (1.6$\pm$0.8$\sigma$)} & 
                            \makecell{4 \\ X } &
                            \makecell{3.9$\pm$1.0 (2.4$\pm$1.0$\sigma$) \\ 4.2$\pm$0.7 (1.5$\pm$0.4$\sigma$)} &
                            \makecell{2 \\ X } \\ 
                            \cline{2-8}
                            & \makecell{Type IVa \\ Type IVb} &
                            \makecell{2.0$\pm$0.4 (4.1$\pm$0.9$\sigma$) \\ 2.1$\pm$0.9 (2.6$\pm$0.7$\sigma$)} & 
                            \makecell{3 \\ 9 } &
                            \makecell{2.0$\pm$0.6 (3.4$\pm$1.0$\sigma$) \\ 2.1$\pm$0.9 (1.1$\pm$0.8$\sigma$)} & 
                            \makecell{4 \\ X } &
                            \makecell{4.0$\pm$1.1 (2.3$\pm$1.5$\sigma$) \\ 4.4$\pm$0.6 (1.7$\pm$0.3$\sigma$)} &
                            \makecell{8 \\ X } \\ 
                            \cline{2-8}
                            & \makecell{Type Va \\ Type Vb} &
                            \makecell{2.0$\pm$0.7 (3.8$\pm$0.9$\sigma$) \\ 2.1$\pm$1.1 (2.6$\pm$0.6$\sigma$)} & 
                            \makecell{3 \\ 10 } &
                            \makecell{2.0$\pm$0.8 (3.0$\pm$1.3$\sigma$) \\ 2.2$\pm$1.2 (1.0$\pm$0.7$\sigma$)} &
                            \makecell{5 \\ X } &
                            \makecell{4.2$\pm$0.8 (2.2$\pm$0.8$\sigma$) \\ 5.0$\pm$0.6 (1.8$\pm$0.3$\sigma$)} &
                            \makecell{9 \\ X } \\ 
                            \cline{2-8}
                            & \makecell{Type VIa \\ Type VIb} &
                            \makecell{2.0$\pm$1.0 (3.6$\pm$1.1$\sigma$) \\ 2.1$\pm$1.3 (2.7$\pm$0.7$\sigma$)} &
                            \makecell{4 \\ X } &
                            \makecell{2.0$\pm$0.9 (2.8$\pm$1.3$\sigma$) \\ 2.7$\pm$1.3 (1.0$\pm$0.5$\sigma$)} & 
                            \makecell{6 \\ X } &
                            \makecell{4.2$\pm$0.7 (2.1$\pm$0.7$\sigma$) \\ 5.2$\pm$0.6 (1.8$\pm$0.3$\sigma$)} &
                            \makecell{10 \\ X } \\ 
                            \cline{2-8}
                            &\makecell{Type VIIa \\ Type VIIb} &
                            \makecell{2.1$\pm$1.0 (3.5$\pm$1.0$\sigma$) \\ 4.1$\pm$1.4 (1.9$\pm$0.6$\sigma$)} & 
                            \makecell{5 \\ X } &
                            \makecell{2.1$\pm$1.1 (1.3$\pm$1.1$\sigma$) \\ 4.8$\pm$1.1 (1.1$\pm$0.3$\sigma$)} &
                            \makecell{7 \\ X } &
                            \makecell{4.4$\pm$0.6 (2.0$\pm$0.4$\sigma$) \\ 5.3$\pm$0.6 (1.8$\pm$0.3$\sigma$)} &
                            \makecell{X \\ X } \\ 
                            \cline{2-8}
                            & \makecell{Type VIIIa \\ Type VIIIb} &
                            \makecell{2.1$\pm$1.3 (3.5$\pm$0.9$\sigma$) \\ 4.7$\pm$1.3 (2.0$\pm$0.4$\sigma$)} & 
                            \makecell{5 \\ X } &
                            \makecell{2.2$\pm$1.3 (1.2$\pm$1.0$\sigma$) \\ 4.8$\pm$0.6 (1.2$\pm$0.3$\sigma$)} & 
                            \makecell{8 \\ X } &
                            \makecell{4.4$\pm$0.6 (1.9$\pm$0.4$\sigma$) \\ 5.3$\pm$0.5 (1.7$\pm$0.2$\sigma$)} &
                            \makecell{X \\ X } \\  
                                                    
    \hline 
    \multirow{4}{*}{3 yr} & 
                            \makecell{Type Ia \\ Type Ib} &
                            \makecell{\textbf{3.00$\pm$0.03 (5.0$\pm$0.2$\sigma$)} \\ 3.0$\pm$0.1 (2.4$\pm$0.5$\sigma$)} & 
                            \makecell{ \\ } &
                            \makecell{\textbf{3.00$\pm$0.03 (5.2$\pm$0.2$\sigma$)} \\ 3.0$\pm$0.1 (2.6$\pm$0.5$\sigma$)} & 
                            \makecell{ \\ } &
                            \makecell{\textbf{3.00$\pm$0.03 (5.0$\pm$0.2$\sigma$)} \\ 3.0$\pm$0.3 (2.0$\pm$0.6$\sigma$)} &
                            \makecell{ \\ } \\ 
                            \cline{2-8} 
                            & \makecell{Type IIa \\ Type IIb} &
                            \makecell{\textbf{3.00$\pm$0.03 (5.0$\pm$0.3$\sigma$)} \\ 3.0$\pm$0.3 (2.4$\pm$0.6$\sigma$)} & 
                            \makecell{ \\ } &
                            \makecell{\textbf{3.00$\pm$0.04 (5.2$\pm$0.3$\sigma$)} \\ 3.0$\pm$0.5 (2.6$\pm$0.7$\sigma$)} & 
                            \makecell{ \\ } &
                            \makecell{\textbf{3.00$\pm$0.04 (5.0$\pm$0.3$\sigma$)} \\ 3.1$\pm$0.8 (2.0$\pm$0.7$\sigma$)} &
                            \makecell{ \\ } \\  
                            \cline{2-8}
                            & \makecell{Type IIIa \\ Type IIIb} &
                            \makecell{3.0$\pm$0.3 (3.6$\pm$0.7$\sigma$) \\ 3.0$\pm$0.8 (2.0$\pm$0.5$\sigma$)} & 
                            \makecell{ \\ } &
                            \makecell{3.0$\pm$0.4 (3.9$\pm$1.1$\sigma$) \\ 3.1$\pm$0.7 (2.1$\pm$0.9$\sigma$)} & 
                            \makecell{ \\ } &
                            \makecell{3.1$\pm$1.3 (3.3$\pm$1.0$\sigma$) \\ 5.0$\pm$0.9 (1.4$\pm$0.5$\sigma$)} &
                            \makecell{ \\ } \\ 
                            \cline{2-8}
                            & \makecell{Type IVa \\ Type IVb} &
                            \makecell{3.0$\pm$0.4 (3.5$\pm$0.8$\sigma$) \\ 3.1$\pm$0.9 (2.1$\pm$0.6$\sigma$)} & 
                            \makecell{ \\ } &
                            \makecell{3.0$\pm$0.6 (4.0$\pm$1.3$\sigma$) \\ 3.1$\pm$0.9 (1.3$\pm$1.0$\sigma$)} &
                            \makecell{ \\ } &
                            \makecell{4.6$\pm$1.3 (2.1$\pm$1.0$\sigma$) \\ 5.5$\pm$0.9 (1.5$\pm$0.3$\sigma$)} &
                            \makecell{ \\ } \\   
                            \cline{2-8}
                            & \makecell{Type Va \\ Type Vb} &
                            \makecell{3.0$\pm$0.7 (3.2$\pm$0.8$\sigma$) \\ 3.2$\pm$1.0 (2.3$\pm$0.5$\sigma$)} & 
                            \makecell{ \\ } &
                            \makecell{3.1$\pm$0.8 (3.5$\pm$1.4$\sigma$) \\ 3.1$\pm$1.0 (1.1$\pm$0.9$\sigma$)} & 
                            \makecell{ \\ } &
                            \makecell{5.5$\pm$0.9 (2.0$\pm$0.7$\sigma$) \\ 5.6$\pm$0.8 (1.6$\pm$0.2$\sigma$)} &
                            \makecell{ \\ } \\   
                            \cline{2-8}
                            & \makecell{Type VIa \\ Type VIb} &
                            \makecell{3.0$\pm$0.9 (3.3$\pm$0.9$\sigma$) \\ 3.3$\pm$1.1 (2.4$\pm$0.5$\sigma$)} & 
                            \makecell{ \\ } &
                            \makecell{3.1$\pm$0.8 (3.6$\pm$1.4$\sigma$) \\ 3.2$\pm$1.1 (1.3$\pm$0.7$\sigma$)} & 
                            \makecell{ \\ } &
                            \makecell{5.6$\pm$0.9 (2.1$\pm$0.5$\sigma$) \\ 5.7$\pm$0.5 (1.7$\pm$0.2$\sigma$)} &
                            \makecell{ \\ } \\ 
                            \cline{2-8}
                            &\makecell{Type VIIa \\ Type VIIb} &
                            \makecell{3.1$\pm$1.0 (3.4$\pm$0.8$\sigma$) \\ 3.3$\pm$1.2 (2.5$\pm$0.5$\sigma$)} &
                            \makecell{ \\ } &
                            \makecell{3.1$\pm$0.9 (3.2$\pm$1.4$\sigma$) \\ 3.3$\pm$1.1 (1.5$\pm$0.8$\sigma$)} & 
                            \makecell{ \\ } &
                            \makecell{5.6$\pm$1.0 (2.0$\pm$0.5$\sigma$) \\ 5.8$\pm$0.4 (1.7$\pm$0.3$\sigma$)} &
                            \makecell{ \\ } \\   
                            \cline{2-8}
                            & \makecell{Type VIIIa \\ Type VIIIb} &
                            \makecell{3.1$\pm$1.1 (3.2$\pm$0.8$\sigma$) \\ 3.3$\pm$1.2 (2.6$\pm$0.6$\sigma$)} & 
                            \makecell{ \\ } &
                            \makecell{3.2$\pm$1.0 (1.8$\pm$1.4$\sigma$) \\ 4.8$\pm$1.1 (1.6$\pm$0.7$\sigma$)} & 
                            \makecell{ \\ } &
                            \makecell{5.6$\pm$0.9 (2.0$\pm$0.4$\sigma$) \\ 5.7$\pm$0.4 (1.8$\pm$0.2$\sigma$)} &
                            \makecell{ \\ } \\              
\hline
\hline
\end{tabular}%
}
\end{table*}
\begin{table*}
\centering
\caption{The evaluation of periodicity detection for signals with periods of 2 and 3 years, considering different types of flares in Table \ref{tab:flare_types} for the methods LSP, CWT, and PDM. This test consists of conspiring only 3 possible temporal positions of the flare in the signal (start, medium, and end), in phase with an oscillation of the sinusoidal cyclic. In bold, we highlight the tests where at least two methods report a compatible period with a significance of $\geq$5$\sigma$. A compatible period is defined as achieving a period that matches the signal period within a tolerance of $\pm$0.1. All periods are expressed in years.\label{tab:experiment_coincident}}
%\resizeboX$\pm$X{\columnwidth}{!}
{%
\begin{tabular}{l|ccccc}
\hline
\hline
Period [yr] & Position & Type of Flare & LSP & CWT & PDM \\
\hline
		\multirow{4}{*}{2 yr} & Start & 
                            \makecell{Type Ia \\ Type Ib \\ Type IIa \\ Type IIb \\ Type IIIa \\ Type IIIb \\ Type IVa \\ Type IVb
                            \\ Type Va \\ Type Vb \\ Type VIa \\ Type VIb \\ Type VIIa \\ Type VIIb \\ Type VIIIa \\ Type VIIIb} &
                            \makecell{\textbf{2.00$\pm$0.01 (5.2$\pm$0.2$\sigma$)} \\ 2.00$\pm$0.01 (3.3$\pm$0.2$\sigma$) \\ \textbf{2.00$\pm$0.02 (5.2$\pm$0.2$\sigma$)} \\ 2.00$\pm$0.02 (3.6$\pm$0.2$\sigma$) \\ 2.00$\pm$0.02 (4.5$\pm$0.3$\sigma$) \\ 2.00$\pm$0.02 (2.9$\pm$0.2$\sigma$)
                            \\ 2.00$\pm$0.02 (4.4$\pm$0.2$\sigma$) \\ 2.00$\pm$0.02 (3.0$\pm$0.2$\sigma$) \\ 2.00$\pm$0.02 (4.0$\pm$0.3$\sigma$) \\ 2.00$\pm$0.02 (2.9$\pm$0.3$\sigma$) \\ 2.00$\pm$0.02 (3.6$\pm$0.2$\sigma$) \\ 1.9$\pm$0.9 (2.7$\pm$0.5$\sigma$) \\ 1.9$\pm$0.4 (3.2$\pm$0.4$\sigma$) \\ 5.2$\pm$1.4 (1.4$\pm$0.5$\sigma$) \\ 2.2$\pm$0.8 (2.6$\pm$0.4$\sigma$) \\ 5.2$\pm$0.6 (1.5$\pm$0.4$\sigma$)} &
                            \makecell{\textbf{2.00$\pm$0.03 (5.3$\pm$0.2$\sigma$)} \\ 2.00$\pm$0.03 (3.3$\pm$0.1$\sigma$) \\ \textbf{2.00$\pm$0.03 (5.3$\pm$0.2$\sigma$)} \\ 2.00$\pm$0.03 (3.3$\pm$0.1$\sigma$) \\ 2.00$\pm$0.03 (4.5$\pm$0.2$\sigma$) \\ 1.90$\pm$0.03 (0.9$\pm$0.6$\sigma$)
                            \\ 2.0$\pm$0.03 (4.5$\pm$0.2$\sigma$) \\ 1.80$\pm$0.02 (0.8$\pm$0.2$\sigma$) \\ 2.00$\pm$0.02 (3.7$\pm$1.4$\sigma$) \\ 1.80$\pm$0.03 (1.0$\pm$0.2$\sigma$) \\ 1.90$\pm$0.03 (1.0$\pm$0.6$\sigma$) \\ 4.8$\pm$0.7 (1.2$\pm$0.2$\sigma$) \\ 1.90$\pm$0.02 (1.3$\pm$0.1$\sigma$) \\ 4.8$\pm$0.3 (1.4$\pm$0.1$\sigma$) \\ 3.8$\pm$1.2 (1.3$\pm$0.2$\sigma$) \\ 5.0$\pm$0.6 (1.5$\pm$0.2$\sigma$)} &
                            \makecell{\textbf{2.00$\pm$0.03 (5.3$\pm$0.2$\sigma$)} \\ 2.00$\pm$0.04 (4.2$\pm$0.5$\sigma$) \\ \textbf{2.00$\pm$0.04 (5.3$\pm$0.2$\sigma$)} \\ 2.0$\pm$0.7 (4.3$\pm$0.7$\sigma$) \\ 2.0$\pm$0.7 (4.5$\pm$0.9$\sigma$) \\ 3.9$\pm$0.8 (1.7$\pm$1.1$\sigma$)
                            \\ 3.9$\pm$0.8 (2.4$\pm$1.2$\sigma$) \\ 3.9$\pm$0.1 (1.8$\pm$0.1$\sigma$) \\ 3.8$\pm$0.1 (2.2$\pm$0.1$\sigma$) \\ 5.5$\pm$0.8 (1.3$\pm$0.2$\sigma$) \\ 3.8$\pm$0.7 (2.1$\pm$0.3$\sigma$) \\ 5.8$\pm$0.6 (1.4$\pm$0.3$\sigma$) \\ 5.7$\pm$0.6 (1.5$\pm$0.3$\sigma$) \\ 5.8$\pm$0.4 (1.5$\pm$0.2$\sigma$) \\ 5.9$\pm$0.5 (1.4$\pm$0.1$\sigma$) \\ 5.8$\pm$0.5 (1.5$\pm$0.2$\sigma$)} \\\cline{2-6}      
                           & Medium & 
                           \makecell{Type Ia \\ Type Ib \\ Type IIa \\ Type IIb \\ Type IIIa \\ Type IIIb \\ Type IVa \\ Type IVb
                            \\ Type Va \\ Type Vb \\ Type VIa \\ Type VIb \\ Type VIIa \\ Type VIIb \\ Type VIIIa \\ Type VIIIb} &
                            \makecell{\textbf{2.00$\pm$0.01 (5.3$\pm$0.1$\sigma$)} \\ 2.00$\pm$0.02 (3.4$\pm$0.2$\sigma$) \\ \textbf{2.00$\pm$0.01 (5.4$\pm$0.1$\sigma$)} \\ 2.00$\pm$0.02 (3.6$\pm$0.3$\sigma$) \\ \textbf{2.00$\pm$0.02 (4.6$\pm$0.3$\sigma$)} \\ 2.00$\pm$0.02 (2.9$\pm$0.2$\sigma$)
                            \\ 2.00$\pm$0.02 (4.5$\pm$0.3$\sigma$) \\ 2.00$\pm$0.02 (2.9$\pm$0.1$\sigma$) \\ 2.00$\pm$0.02 (4.0$\pm$0.2$\sigma$) \\ 2.1$\pm$0.1 (2.7$\pm$0.1$\sigma$) \\ 2.00$\pm$0.02 (3.5$\pm$0.3$\sigma$) \\ 3.4$\pm$0.7 (1.8$\pm$0.3$\sigma$) \\ 2.0$\pm$0.8 (2.6$\pm$0.3$\sigma$) \\ 3.5$\pm$0.4 (1.9$\pm$0.2$\sigma$) \\ 3.5$\pm$1.0 (1.9$\pm$0.3$\sigma$) \\ 3.7$\pm$0.9 (1.9$\pm$0.3$\sigma$)} &
                            \makecell{\textbf{2.00$\pm$0.03 (5.3$\pm$0.1$\sigma$)} \\ 2.00$\pm$0.02 (3.7$\pm$0.1$\sigma$) \\ \textbf{2.00$\pm$0.02 (5.3$\pm$0.1$\sigma$)} \\ 2.00$\pm$0.02 (4.0$\pm$0.1$\sigma$) \\ \textbf{2.00$\pm$0.02 (5.2$\pm$0.2$\sigma$)} \\ 2.00$\pm$0.02 (3.3$\pm$0.1$\sigma$)
                            \\ 2.00$\pm$0.02 (5.2$\pm$0.2$\sigma$) \\ 2.00$\pm$0.02 (3.3$\pm$0.4$\sigma$) \\ 2.00$\pm$0.02 (4.5$\pm$0.8$\sigma$) \\ 2.00$\pm$0.03 (1.5$\pm$0.1$\sigma$) \\ 2.00$\pm$0.02 (2.6$\pm$1.5$\sigma$) \\ 2.1$\pm$0.6 (1.4$\pm$0.2$\sigma$) \\ 2.1$\pm$0.7 (1.4$\pm$0.1$\sigma$) \\ 3.3$\pm$0.2 (1.6$\pm$0.2$\sigma$) \\ 3.2$\pm$0.7 (1.7$\pm$0.1$\sigma$) \\ 3.4$\pm$0.1 (1.8$\pm$0.2$\sigma$)} &
                            \makecell{\textbf{2.00$\pm$0.02 (5.3$\pm$0.2$\sigma$)} \\ 2.0$\pm$0.4 (4.1$\pm$1.2$\sigma$) \\ \textbf{2.00$\pm$0.02 (5.3$\pm$0.2$\sigma$)} \\ 3.9$\pm$1.0 (2.0$\pm$1.5$\sigma$) \\ \textbf{2.0$\pm$0.6 (5.1$\pm$0.3$\sigma$)} \\ 4.0$\pm$0.2 (1.8$\pm$0.1$\sigma$)
                            \\ 4.0$\pm$1.0 (2.7$\pm$1.5$\sigma$) \\ 4.0$\pm$0.2 (1.9$\pm$0.3$\sigma$) \\ 4.0$\pm$0.7 (2.4$\pm$1.0$\sigma$) \\ 3.9$\pm$0.2 (1.9$\pm$0.1$\sigma$) \\ 3.9$\pm$0.2 (2.1$\pm$0.2$\sigma$) \\ 3.9$\pm$0.1 (1.8$\pm$0.1$\sigma$) \\ 4.1$\pm$0.1 (1.8$\pm$0.1$\sigma$) \\ 4.1$\pm$0.6 (1.5$\pm$0.2$\sigma$) \\ 5.0$\pm$0.6 (1.3$\pm$0.2$\sigma$) \\ 5.3$\pm$0.2 (1.3$\pm$0.1$\sigma$)} \\\cline{2-6}
                            & End & 
                            \makecell{Type Ia \\ Type Ib \\ Type IIa \\ Type IIb \\ Type IIIa \\ Type IIIb \\ Type IVa \\ Type IVb
                            \\ Type Va \\ Type Vb \\ Type VIa \\ Type VIb \\ Type VIIa \\ Type VIIb \\ Type VIIIa \\ Type VIIIb} &
                            \makecell{\textbf{2.00$\pm$0.01 (5.3$\pm$0.2$\sigma$)} \\ 2.00$\pm$0.01 (3.4$\pm$0.2$\sigma$) \\ \textbf{2.00$\pm$0.01 (5.4$\pm$0.1$\sigma$)} \\ 2.00$\pm$0.01 (3.6$\pm$0.2$\sigma$) \\ 2.00$\pm$0.02 (4.9$\pm$0.3$\sigma$) \\ 2.00$\pm$0.01 (2.9$\pm$0.2$\sigma$)
                            \\ 2.00$\pm$0.02 (4.7$\pm$0.2$\sigma$) \\ 2.00$\pm$0.02 (3.0$\pm$0.2$\sigma$) \\ 2.00$\pm$0.01 (4.0$\pm$0.2$\sigma$) \\ 2.1$\pm$0.3 (2.6$\pm$0.2$\sigma$) \\ 2.00$\pm$0.02 (3.5$\pm$0.2$\sigma$) \\ 3.4$\pm$0.6 (1.8$\pm$0.3$\sigma$) \\ 2.1$\pm$1.1 (2.6$\pm$0.5$\sigma$) \\ 3.5$\pm$0.4 (1.9$\pm$0.1$\sigma$) \\ 3.5$\pm$1.3 (1.9$\pm$0.3$\sigma$) \\ 3.6$\pm$0.8 (1.9$\pm$0.2$\sigma$)} &
                            \makecell{\textbf{2.00$\pm$0.02 (5.3$\pm$0.2$\sigma$)} \\ 2.00$\pm$0.02 (3.8$\pm$0.1$\sigma$) \\ \textbf{2.00$\pm$0.02 (5.3$\pm$0.2$\sigma$)} \\ 2.00$\pm$0.02 (3.9$\pm$0.2$\sigma$) \\ 2.00$\pm$0.02 (5.4$\pm$0.1$\sigma$) \\ 2.00$\pm$0.02 (3.2$\pm$0.1$\sigma$)
                            \\ 2.00$\pm$0.02 (5.2$\pm$0.2$\sigma$) \\ 2.00$\pm$0.02 (1.3$\pm$0.9$\sigma$) \\ 2.00$\pm$0.02 (4.6$\pm$0.2$\sigma$) \\ 2.00$\pm$0.03 (1.1$\pm$0.1$\sigma$) \\ 2.00$\pm$0.02 (1.4$\pm$1.2$\sigma$) \\ 2.1$\pm$0.7 (1.3$\pm$0.1$\sigma$) \\ 2.1$\pm$0.5 (1.4$\pm$0.1$\sigma$) \\ 4.5$\pm$0.2 (1.6$\pm$0.1$\sigma$) \\ 3.9$\pm$0.7 (1.6$\pm$0.1$\sigma$) \\ 4.8$\pm$0.2 (1.8$\pm$0.1$\sigma$)} &
                            \makecell{\textbf{2.00$\pm$0.02 (5.3$\pm$0.1$\sigma$)} \\ 2.00$\pm$0.04 (4.1$\pm$0.8$\sigma$) \\ \textbf{2.00$\pm$0.02 (5.3$\pm$0.2$\sigma$)} \\ 3.9$\pm$0.9 (2.0$\pm$1.4$\sigma$) \\ 2.0$\pm$0.8 (4.2$\pm$1.2$\sigma$) \\ 4.0$\pm$0.4 (1.7$\pm$0.4$\sigma$)
                            \\ 2.0$\pm$0.9 (4.3$\pm$1.1$\sigma$) \\ 4.0$\pm$0.1 (1.9$\pm$0.1$\sigma$) \\ 4.0$\pm$0.6 (2.3$\pm$0.9$\sigma$) \\ 3.9$\pm$0.1 (1.9$\pm$0.1$\sigma$) \\ 4.0$\pm$0.1 (2.1$\pm$0.1$\sigma$) \\ 3.9$\pm$0.2 (1.8$\pm$0.1$\sigma$) \\ 4.1$\pm$0.4 (1.8$\pm$0.2$\sigma$) \\ 4.7$\pm$0.7 (1.3$\pm$0.2$\sigma$) \\ 5.2$\pm$0.6 (1.3$\pm$0.2$\sigma$) \\ 5.4$\pm$0.2 (1.3$\pm$0.1$\sigma$)} \\                 
\hline
\hline
\end{tabular}%
}
\end{table*}
\begin{table*}
\centering
\setcounter{table}{1}
\caption{\textit{(continued).\label{tab:experiment_coincident_b}}}
%\resizeboX$\pm$X{\columnwidth}{!}
{%
\begin{tabular}{l|ccccc}
\hline
\hline
Period [yr] & Position & Type of Flare & LSP & CWT & PDM \\
\hline
\multirow{4}{*}{3 yr} & Start & 
                          \makecell{Type Ia \\ Type Ib \\ Type IIa \\ Type IIb \\ Type IIIa \\ Type IIIb \\ Type IVa \\ Type IVb
                            \\ Type Va \\ Type Vb \\ Type VIa \\ Type VIb \\ Type VIIa \\ Type VIIb \\ Type VIIIa \\ Type VIIIb} &
                            \makecell{\textbf{3.00$\pm$0.01 (5.1$\pm$0.1$\sigma$)} \\ 3.00$\pm$0.01 (2.9$\pm$0.1$\sigma$) \\ \textbf{3.00$\pm$0.02 (5.1$\pm$0.1$\sigma$)} \\ 3.00$\pm$0.01 (3.0$\pm$0.1$\sigma$) \\ 3.00$\pm$0.01 (4.2$\pm$0.3$\sigma$) \\ 3.00$\pm$0.01 (2.5$\pm$0.2$\sigma$)
                            \\ 2.9$\pm$0.1 (4.0$\pm$0.2$\sigma$) \\ 2.9$\pm$0.1 (2.5$\pm$0.1$\sigma$) \\ 2.9$\pm$0.1 (3.9$\pm$0.1$\sigma$) \\ 2.8$\pm$0.1 (2.6$\pm$0.1$\sigma$) \\ 2.8$\pm$0.1 (4.1$\pm$0.2$\sigma$) \\ 2.8$\pm$0.1 (2.7$\pm$0.1$\sigma$) \\ 2.8$\pm$0.1 (3.9$\pm$0.2$\sigma$) \\ 2.7$\pm$0.1 (2.8$\pm$0.1$\sigma$) \\ 2.7$\pm$0.1 (3.8$\pm$0.2$\sigma$) \\ 2.7$\pm$0.1 (2.9$\pm$0.1$\sigma$)} &
                            \makecell{\textbf{3.00$\pm$0.01 (5.3$\pm$0.2$\sigma$)} \\ 3.00$\pm$0.01 (2.9$\pm$0.2$\sigma$) \\ \textbf{3.00$\pm$0.01 (5.2$\pm$0.3$\sigma$)} \\ 3.00$\pm$0.04 (3.0$\pm$0.2$\sigma$) \\ 3.00$\pm$0.03 (4.3$\pm$0.2$\sigma$) \\ 2.9$\pm$0.1 (2.5$\pm$0.3$\sigma$)
                            \\ 2.9$\pm$0.1 (4.4$\pm$0.2$\sigma$) \\ 2.9$\pm$0.1 (1.6$\pm$0.9$\sigma$) \\ 2.8$\pm$0.1 (4.1$\pm$0.2$\sigma$) \\ 2.8$\pm$0.1 (0.9$\pm$0.1$\sigma$) \\ 2.8$\pm$0.1 (4.2$\pm$0.1$\sigma$) \\ 2.8$\pm$0.1 (1.1$\pm$0.2$\sigma$) \\ 2.8$\pm$0.2 (4.0$\pm$0.9$\sigma$) \\ 2.7$\pm$0.1 (1.3$\pm$0.1$\sigma$) \\ 2.7$\pm$0.1 (2.7$\pm$1.1$\sigma$) \\ 2.7$\pm$0.2 (1.4$\pm$0.2$\sigma$)} &
                            \makecell{\textbf{3.00$\pm$0.03 (5.0$\pm$0.2$\sigma$)} \\ 3.00$\pm$0.02 (2.6$\pm$0.2$\sigma$) \\ \textbf{3.00$\pm$0.02 (5.0$\pm$0.3$\sigma$)} \\ 3.00$\pm$0.02 (3.0$\pm$0.3$\sigma$) \\ 3.00$\pm$0.03 (4.0$\pm$0.2$\sigma$) \\ 5.1$\pm$0.9 (1.3$\pm$0.6$\sigma$)
                            \\ 2.9$\pm$0.9 (4.1$\pm$0.7$\sigma$) \\ 5.2$\pm$0.7 (1.4$\pm$0.2$\sigma$) \\ 5.2$\pm$0.8 (2.1$\pm$0.8$\sigma$) \\ 5.8$\pm$0.2 (1.5$\pm$0.1$\sigma$) \\ 5.7$\pm$0.2 (2.1$\pm$0.7$\sigma$) \\ 5.8$\pm$0.1 (1.6$\pm$0.3$\sigma$) \\ 5.6$\pm$0.2 (2.0$\pm$0.2$\sigma$) \\ 5.6$\pm$0.2 (1.7$\pm$0.1$\sigma$) \\ 5.5$\pm$0.2 (2.0$\pm$0.1$\sigma$) \\ 5.5$\pm$0.2 (1.7$\pm$0.1$\sigma$)} \\\cline{2-6}     
                          & Medium & 
                          \makecell{Type Ia \\ Type Ib \\ Type IIa \\ Type IIb \\ Type IIIa \\ Type IIIb \\ Type IVa \\ Type IVb
                            \\ Type Va \\ Type Vb \\ Type VIa \\ Type VIb \\ Type VIIa \\ Type VIIb \\ Type VIIIa \\ Type VIIIb} &
                            \makecell{\textbf{3.00$\pm$0.02 (5.1$\pm$0.2$\sigma$)} \\ 3.00$\pm$0.02 (3.0$\pm$0.1$\sigma$) \\ \textbf{3.00$\pm$0.03 (5.1$\pm$0.2$\sigma$)} \\ 3.00$\pm$0.02 (2.9$\pm$0.2$\sigma$) \\ 3.00$\pm$0.02 (4.0$\pm$0.1$\sigma$) \\ 2.9$\pm$0.1 (2.3$\pm$0.1$\sigma$)
                            \\ 2.9$\pm$0.1 (4.1$\pm$0.1$\sigma$) \\ 2.9$\pm$0.1 (2.6$\pm$0.2$\sigma$) \\ 2.9$\pm$0.1 (4.0$\pm$0.2$\sigma$) \\ 2.8$\pm$0.1 (2.5$\pm$0.1$\sigma$) \\ 2.8$\pm$0.1 (4.0$\pm$0.2$\sigma$) \\ 2.8$\pm$0.1 (2.7$\pm$0.1$\sigma$) \\ 2.8$\pm$0.1 (3.8$\pm$0.2$\sigma$) \\ 2.7$\pm$0.1 (2.8$\pm$0.1$\sigma$) \\ 2.7$\pm$0.1 (3.9$\pm$0.2$\sigma$) \\ 2.7$\pm$0.1 (2.8$\pm$0.1$\sigma$)} &
                            \makecell{\textbf{3.00$\pm$0.02 (5.3$\pm$0.1$\sigma$)} \\ 3.00$\pm$0.02 (3.0$\pm$0.2$\sigma$) \\ \textbf{3.00$\pm$0.02 (5.2$\pm$0.2$\sigma$)} \\ 3.00$\pm$0.04 (3.0$\pm$0.1$\sigma$) \\ 2.9$\pm$0.1 (4.3$\pm$0.2$\sigma$) \\ 2.8$\pm$0.1 (2.3$\pm$0.7$\sigma$)
                            \\ 2.8$\pm$0.1 (4.2$\pm$0.3$\sigma$) \\ 2.8$\pm$0.1 (0.9$\pm$0.7$\sigma$) \\ 2.8$\pm$0.1 (3.9$\pm$0.2$\sigma$) \\ 2.8$\pm$0.1 (0.9$\pm$0.2$\sigma$) \\ 2.8$\pm$0.1 (4.0$\pm$0.7$\sigma$) \\ 2.8$\pm$0.1 (1.1$\pm$0.3$\sigma$) \\ 2.8$\pm$0.1 (3.9$\pm$1.0$\sigma$) \\ 2.7$\pm$0.1 (1.2$\pm$0.2$\sigma$) \\ 2.7$\pm$0.1 (2.7$\pm$1.3$\sigma$) \\ 2.7$\pm$0.1 (1.4$\pm$0.1$\sigma$)} &
                            \makecell{\textbf{3.00$\pm$0.02 (5.1$\pm$0.2$\sigma$)} \\ 3.00$\pm$0.02 (2.6$\pm$0.2$\sigma$) \\ \textbf{3.00$\pm$0.02 (5.0$\pm$0.2$\sigma$)} \\ 3.00$\pm$0.02 (2.8$\pm$0.1$\sigma$) \\ 3.0$\pm$0.6 (3.9$\pm$0.5$\sigma$) \\ 5.5$\pm$0.4 (1.2$\pm$0.6$\sigma$)
                            \\ 2.9$\pm$0.1 (4.1$\pm$0.3$\sigma$) \\ 5.8$\pm$0.1 (1.4$\pm$0.1$\sigma$) \\ 5.7$\pm$0.3 (2.1$\pm$0.9$\sigma$) \\ 5.9$\pm$0.1 (1.5$\pm$0.1$\sigma$) \\ 5.7$\pm$0.1 (2.1$\pm$0.1$\sigma$) \\ 5.7$\pm$0.1 (1.6$\pm$0.1$\sigma$) \\ 5.6$\pm$0.1 (2.0$\pm$0.1$\sigma$) \\ 5.6$\pm$0.1 (1.7$\pm$0.1$\sigma$) \\ 5.6$\pm$0.1 (2.0$\pm$0.1$\sigma$) \\ 5.5$\pm$0.1 (1.7$\pm$0.1$\sigma$)} \\\cline{2-6}     
                          & End & 
                          \makecell{Type Ia \\ Type Ib \\ Type IIa \\ Type IIb \\ Type IIIa \\ Type IIIb \\ Type IVa \\ Type IVb
                            \\ Type Va \\ Type Vb \\ Type VIa \\ Type VIb \\ Type VIIa \\ Type VIIb \\ Type VIIIa \\ Type VIIIb} &
                            \makecell{\textbf{3.00$\pm$0.02 (5.1$\pm$0.2$\sigma$)} \\ 3.0$\pm$0.02 (2.8$\pm$0.1$\sigma$) \\ \textbf{3.00$\pm$0.03 (5.0$\pm$0.3$\sigma$)} \\ 3.00$\pm$0.02 (2.9$\pm$0.2$\sigma$) \\ 3.00$\pm$0.02 (4.1$\pm$0.3$\sigma$) \\ 2.9$\pm$0.1 (2.3$\pm$0.1$\sigma$)
                            \\ 2.9$\pm$0.1 (4.1$\pm$0.2$\sigma$) \\ 2.9$\pm$0.1 (2.5$\pm$0.1$\sigma$) \\ 2.9$\pm$0.1 (4.0$\pm$0.2$\sigma$) \\ 2.8$\pm$0.1 (2.6$\pm$0.1$\sigma$) \\ 2.8$\pm$0.1 (4.0$\pm$0.2$\sigma$) \\ 2.8$\pm$0.1 (2.7$\pm$0.1$\sigma$) \\ 2.8$\pm$0.1 (3.9$\pm$0.2$\sigma$) \\ 2.7$\pm$0.1 (2.8$\pm$0.1$\sigma$) \\ 2.7$\pm$0.1 (3.8$\pm$0.2$\sigma$) \\ 2.7$\pm$0.1 (2.9$\pm$0.1$\sigma$)} &
                            \makecell{\textbf{3.00$\pm$0.02 (5.4$\pm$0.1$\sigma$)} \\ 3.00$\pm$0.02 (3.0$\pm$0.1$\sigma$) \\ \textbf{3.00$\pm$0.02 (5.3$\pm$0.1$\sigma$)} \\ 3.00$\pm$0.04 (3.0$\pm$0.2$\sigma$) \\ 2.9$\pm$0.1 (4.3$\pm$0.2$\sigma$) \\ 2.8$\pm$0.1 (2.2$\pm$0.7$\sigma$)
                            \\ 2.9$\pm$0.1 (4.4$\pm$0.2$\sigma$) \\ 2.8$\pm$0.1 (0.9$\pm$0.8$\sigma$) \\ 2.8$\pm$0.1 (4.1$\pm$0.2$\sigma$) \\ 2.8$\pm$0.1 (0.9$\pm$0.2$\sigma$) \\ 2.8$\pm$0.1 (4.1$\pm$0.2$\sigma$) \\ 2.8$\pm$0.1 (1.1$\pm$0.1$\sigma$) \\ 2.8$\pm$0.1 (3.8$\pm$1.1$\sigma$) \\ 2.7$\pm$0.1 (1.2$\pm$0.1$\sigma$) \\ 2.7$\pm$0.1 (2.7$\pm$1.4$\sigma$) \\ 2.7$\pm$0.1 (1.4$\pm$0.2$\sigma$)} &
                            \makecell{\textbf{3.00$\pm$0.02 (5.1$\pm$0.2$\sigma$)} \\ 3.00$\pm$0.03 (2.5$\pm$0.1$\sigma$) \\ \textbf{3.00$\pm$0.02 (5.0$\pm$0.2$\sigma$)} \\ 3.00$\pm$0.03 (2.9$\pm$0.1$\sigma$) \\ 3.00$\pm$0.05 (4.0$\pm$0.3$\sigma$) \\ 3.1$\pm$1.1 (2.2$\pm$0.6$\sigma$)
                            \\ 2.9$\pm$0.6 (4.1$\pm$0.5$\sigma$) \\ 5.8$\pm$0.1 (1.4$\pm$0.2$\sigma$) \\ 5.4$\pm$0.6 (2.1$\pm$0.7$\sigma$) \\ 5.9$\pm$0.1 (1.5$\pm$0.1$\sigma$) \\ 5.1$\pm$0.9 (2.1$\pm$0.8$\sigma$) \\ 5.8$\pm$0.1 (1.6$\pm$0.1$\sigma$) \\ 5.6$\pm$0.1 (2.0$\pm$0.1$\sigma$) \\ 5.6$\pm$0.1 (1.6$\pm$0.2$\sigma$) \\ 5.5$\pm$0.1 (2.0$\pm$0.1$\sigma$) \\ 5.5$\pm$0.1 (1.7$\pm$0.1$\sigma$)} \\
\hline
\hline
\end{tabular}%
}
\end{table*}
\begin{table*}
\centering
\caption{The evaluation of periodicity detection for signals with periods of 2 and 3 years, considering different types of flares shown in Table \ref{tab:flare_types} for SSA. In bold, we highlight the tests where the analysis reports a compatible period with a significance of $\geq$5$\sigma$. A compatible period is defined as achieving a period that matches the signal period within a tolerance of $\pm$0.1. We also estimate the number of cycles required to achieve a period detection significance of 5$\sigma$ (only performed for the periodic signal with a period of 2 years). The symbol \textit{--} denotes that a 5$\sigma$ detection was achieved in the initial analysis, while \textit{X} values indicate that it was not possible to obtain the estimation within a limit of 10 cycles. All periods are expressed in years.\label{tab:experiment_ssa}}
%\resizebox{\columnwidth}{!}
{%
\begin{tabular}{cccccc}
\hline
\hline
Period [yr] & Flux of Flare & SSA & Cycles\\
\hline
		\multirow{4}{*}{2 yr} & 
                             \makecell{Type Ia \\ Type Ib} &
                            \makecell{\textbf{2.00$\pm$0.03 (5.2$\pm$0.3$\sigma$)} \\ \textbf{2.0$\pm$0.2 (5.0$\pm$0.4$\sigma$)}} & 
                            \makecell{-- \\ -- } \\
                            \cline{2-4} 
                            & \makecell{Type IIa \\ Type IIb} &
                            \makecell{\textbf{2.00$\pm$0.03 (5.1$\pm$0.4$\sigma$)} \\ \textbf{2.0$\pm$0.6 (5.1$\pm$0.4$\sigma$)}} & 
                            \makecell{-- \\ -- } \\  
                            \cline{2-4}
                            & \makecell{Type IIIa \\ Type IIIb} &
                            \makecell{\textbf{2.0$\pm$0.6 (5.2$\pm$0.3$\sigma$)} \\ \textbf{2.0$\pm$0.7 (5.2$\pm$0.3$\sigma$)}} & 
                            \makecell{-- \\ -- } \\ 
                            \cline{2-4}
                            & \makecell{Type IVa \\ Type IVb} &
                            \makecell{\textbf{2.0$\pm$0.8 (5.0$\pm$0.5$\sigma$)} \\ \textbf{2.0$\pm$1.0 (5.0$\pm$0.4$\sigma$)}} & 
                            \makecell{-- \\ -- } \\   
                            \cline{2-4}
                            & \makecell{Type Va \\ Type Vb} &
                            \makecell{\textbf{2.0$\pm$0.9 (5.0$\pm$0.5$\sigma$)} \\ 2.1$\pm$1.2 (2.8$\pm$1.1$\sigma$)} & 
                            \makecell{-- \\ 6 } \\   
                            \cline{2-4}
                            & \makecell{Type VIa \\ Type VIb} &
                            \makecell{2.1$\pm$0.9 (3.4$\pm$1.2$\sigma$) \\ 2.2$\pm$0.9 (2.7$\pm$0.9$\sigma$)} & 
                            \makecell{2 \\ 6 } \\ 
                            \cline{2-4}
                            &\makecell{Type VIIa \\ Type VIIb} &
                            \makecell{2.1$\pm$0.8 (3.3$\pm$1.1$\sigma$) \\ 2.4$\pm$0.8 (2.5$\pm$0.6$\sigma$)} & 
                            \makecell{3 \\ 9 } \\   
                            \cline{2-4}
                            & \makecell{Type VIIIa \\ Type VIIIb} &
                            \makecell{2.2$\pm$0.7 (3.0$\pm$1.1$\sigma$) \\ 2.5$\pm$0.8 (2.2$\pm$0.5$\sigma$)} & 
                            \makecell{3 \\ X } \\                                                  
    \hline 
    \multirow{4}{*}{3 yr} & \makecell{Type Ia \\ Type Ib} &
                            \makecell{\textbf{3.00$\pm$0.04 (5.1$\pm$0.5$\sigma$)} \\ \textbf{3.0$\pm$0.2 (5.0$\pm$0.5$\sigma$)}} & 
                            \makecell{  \\  } \\ 
                            \cline{2-3}
                            & \makecell{Type IIa \\ Type IIb} &
                            \makecell{\textbf{3.00$\pm$0.07 (5.2$\pm$0.3$\sigma$)} \\ \textbf{3.0$\pm$0.7 (5.1$\pm$0.5$\sigma$)}} & 
                            \makecell{  \\  } \\
                            \cline{2-3}
                            & \makecell{Type IIIa \\ Type IIIb} &
                            \makecell{\textbf{3.0$\pm$0.5 (5.2$\pm$0.5$\sigma$)} \\ 2.9$\pm$1.1 (4.4$\pm$1.1$\sigma$)} & 
                            \makecell{  \\  } \\ 
                            \cline{2-3}
                            & \makecell{Type IVa \\ Type IVb} &
                            \makecell{\textbf{3.0$\pm$0.6 (5.1$\pm$0.4$\sigma$)} \\ 2.9$\pm$1.2 (3.5$\pm$1.1$\sigma$)} & 
                            \makecell{  \\  } \\ 
                            \cline{2-3}
                            & \makecell{Type Va \\ Type Vb} &
                            \makecell{3.1$\pm$0.9 (4.3$\pm$1.2$\sigma$) \\ 3.1$\pm$1.2 (3.2$\pm$1.0$\sigma$)} & 
                            \makecell{  \\  } \\ 
                            \cline{2-3}
                            & \makecell{Type VIa \\ Type VIb} &
                            \makecell{3.1$\pm$0.9 (4.2$\pm$1.2$\sigma$) \\ 3.3$\pm$1.1 (2.8$\pm$0.9$\sigma$)} & 
                            \makecell{  \\  } \\ 
                            \cline{2-3}
                            &\makecell{Type VIIa \\ Type VIIb} &
                            \makecell{3.3$\pm$1.2 (3.7$\pm$1.2$\sigma$) \\ 3.6$\pm$1.2 (2.5$\pm$0.8$\sigma$)} & 
                            \makecell{  \\  } \\ 
                            \cline{2-3}
                            & \makecell{Type VIIIa \\ Type VIIIb} &
                            \makecell{3.0$\pm$0.9 (3.2$\pm$1.2$\sigma$) \\ 3.6$\pm$1.1 (2.3$\pm$0.6$\sigma$)} & 
                            \makecell{  \\  } \\               
\hline
\hline
\end{tabular}%
}
\end{table*}
\begin{table*}
\centering
\caption{The evaluation of periodicity detection for signals with periods of 2 and 3 years, considering different types of flares in Table \ref{tab:flare_types} for the SSA. This test consists of conspiring only 3 possible temporal positions of the flare in the signal (start, medium, and end), coincident with an oscillation of the sinusoidal cyclic. In bold, we highlight the tests where the analysis reports a compatible period with a significance of $\geq$5$\sigma$. A compatible period is defined as achieving a period that matches the signal period within a tolerance of $\pm$0.1. All periods are expressed in years.\label{tab:experiment_coincident_SSA}}
%\resizeboX$\pm$X{\columnwidth}{!}
{%
\begin{tabular}{l|cccc}
\hline
\hline
Period [yr] & Position & Type of Flare & SSA \\
\hline
		\multirow{3}{*}{2 yr} & Start & 
                            \makecell{Type Ia \\ Type Ib \\ Type IIa \\ Type IIb \\ Type IIIa \\ Type IIIb \\ Type IVa \\ Type IVb
                            \\ Type Va \\ Type Vb \\ Type VIa \\ Type VIb \\ Type VIIa \\ Type VIIb \\ Type VIIIa \\ Type VIIIb} &
                            \makecell{\textbf{2.00$\pm$0.02 (5.1$\pm$0.3$\sigma$)} \\ \textbf{2.00$\pm$0.03 (5.1$\pm$0.3$\sigma$)} \\ \textbf{2.00$\pm$0.02 (5.1$\pm$0.4$\sigma$)} \\ \textbf{2.00$\pm$0.03 (5.1$\pm$0.3$\sigma$)} \\ 2.00$\pm$0.02 (4.8$\pm$0.8$\sigma$) \\ 2.00$\pm$0.03 (4.2$\pm$0.8$\sigma$)
                            \\ 2.00$\pm$0.02 (4.4$\pm$0.2$\sigma$) \\ 2.0$\pm$0.1 (4.0$\pm$0.9$\sigma$) \\ 2.00$\pm$0.02 (4.0$\pm$0.3$\sigma$) \\ 2.0$\pm$0.1 (3.1$\pm$0.9$\sigma$) \\ 2.00$\pm$0.02 (3.6$\pm$0.2$\sigma$) \\ 2.0$\pm$0.2 (2.5$\pm$0.7$\sigma$) \\ 1.9$\pm$0.4 (3.2$\pm$0.4$\sigma$) \\ 2.0$\pm$0.5 (2.0$\pm$0.5$\sigma$) \\ 2.1$\pm$0.5 (2.6$\pm$0.4$\sigma$) \\ 2.2$\pm$0.6 (1.5$\pm$0.6$\sigma$)} 
                            \\\cline{2-4}      
                           & Medium & 
                           \makecell{Type Ia \\ Type Ib \\ Type IIa \\ Type IIb \\ Type IIIa \\ Type IIIb \\ Type IVa \\ Type IVb
                            \\ Type Va \\ Type Vb \\ Type VIa \\ Type VIb \\ Type VIIa \\ Type VIIb \\ Type VIIIa \\ Type VIIIb} &
                            \makecell{\textbf{2.00$\pm$0.01 (5.2$\pm$0.2$\sigma$)} \\ \textbf{2.00$\pm$0.03 (5.1$\pm$0.2$\sigma$)} \\ \textbf{2.00$\pm$0.01 (5.1$\pm$0.1$\sigma$)} \\ \textbf{2.00$\pm$0.02 (5.2$\pm$0.1$\sigma$)} \\ \textbf{2.00$\pm$0.02 (5.0$\pm$0.3$\sigma$)} \\ \textbf{2.00$\pm$0.02 (5.1$\pm$0.1$\sigma$)}
                            \\ 2.00$\pm$0.03 (4.7$\pm$0.3$\sigma$) \\ 2.00$\pm$0.02 (4.7$\pm$0.5$\sigma$) \\ 2.00$\pm$0.02 (4.2$\pm$0.6$\sigma$) \\ 2.1$\pm$0.1 (4.0$\pm$0.5$\sigma$) \\ 2.00$\pm$0.03 (4.0$\pm$0.6$\sigma$) \\ 2.1$\pm$0.1 (3.8$\pm$0.6$\sigma$) \\ 2.0$\pm$0.1 (3.8$\pm$0.5$\sigma$) \\ 2.0$\pm$0.3 (3.2$\pm$0.4$\sigma$) \\ 2.1$\pm$0.2 (3.0$\pm$0.9$\sigma$) \\ 2.1$\pm$0.3 (2.7$\pm$0.9$\sigma$)} 
                             \\\cline{2-4}
                            & End & 
                            \makecell{Type Ia \\ Type Ib \\ Type IIa \\ Type IIb \\ Type IIIa \\ Type IIIb \\ Type IVa \\ Type IVb
                            \\ Type Va \\ Type Vb \\ Type VIa \\ Type VIb \\ Type VIIa \\ Type VIIb \\ Type VIIIa \\ Type VIIIb} &
                            \makecell{\textbf{2.00$\pm$0.02 (5.2$\pm$0.1$\sigma$)} \\ \textbf{2.00$\pm$0.02 (5.2$\pm$0.1$\sigma$)} \\ \textbf{2.00$\pm$0.02 (5.1$\pm$0.2$\sigma$)} \\ \textbf{2.00$\pm$0.02 (5.1$\pm$0.1$\sigma$)} \\ \textbf{2.00$\pm$0.03 (5.0$\pm$0.5$\sigma$)} \\ 2.00$\pm$0.03 (3.9$\pm$0.6$\sigma$)
                            \\ 2.00$\pm$0.02 (4.9$\pm$0.3$\sigma$) \\ 2.0$\pm$0.2 (3.3$\pm$0.6$\sigma$) \\ 2.00$\pm$0.02 (4.7$\pm$0.5$\sigma$) \\ 2.0$\pm$0.1 (2.8$\pm$0.4$\sigma$) \\ 2.00$\pm$0.03 (4.8$\pm$0.6$\sigma$) \\ 2.0$\pm$0.1 (2.7$\pm$0.4$\sigma$) \\ 2.00$\pm$0.02 (4.2$\pm$0.6$\sigma$) \\ 2.0$\pm$0.3 (2.6$\pm$0.4$\sigma$) \\ 2.00$\pm$0.03 (3.9$\pm$0.8$\sigma$) \\ 2.0$\pm$0.5 (2.5$\pm$0.5$\sigma$)} \\ 
\hline
\hline
\end{tabular}%
}
\end{table*}
\begin{table*}
\centering
\setcounter{table}{3}
\caption{\textit{(continued).\label{tab:experiment_coincident_SSA_b}}}
%\resizeboX$\pm$X{\columnwidth}{!}
{%
\begin{tabular}{l|cccc}
\hline
\hline
Period [yr] & Position & Type of Flare & SSA \\
\hline
		\multirow{3}{*}{3 yr} & Start & 
                            \makecell{Type Ia \\ Type Ib \\ Type IIa \\ Type IIb \\ Type IIIa \\ Type IIIb \\ Type IVa \\ Type IVb
                            \\ Type Va \\ Type Vb \\ Type VIa \\ Type VIb \\ Type VIIa \\ Type VIIb \\ Type VIIIa \\ Type VIIIb} &
                            \makecell{\textbf{3.00$\pm$0.03 (5.2$\pm$0.2$\sigma$)} \\ \textbf{3.00$\pm$0.04 (5.2$\pm$0.2$\sigma$)} \\ \textbf{3.00$\pm$0.03 (5.2$\pm$0.2$\sigma$)} \\ 3.00$\pm$0.04 (4.8$\pm$0.3$\sigma$) \\ \textbf{3.00$\pm$0.04 (5.2$\pm$0.4$\sigma$)} \\ 3.0$\pm$0.1 (2.8$\pm$0.5$\sigma$) \\ \textbf{3.00$\pm$0.04 (5.2$\pm$0.4$\sigma$)} \\ 2.9$\pm$0.1 (2.6$\pm$0.4$\sigma$) \\ 3.00$\pm$0.04 (4.7$\pm$0.5$\sigma$) \\ 2.9$\pm$0.3 (2.4$\pm$0.4$\sigma$) \\ 2.9$\pm$0.1 (4.5$\pm$0.6$\sigma$) \\ 2.8$\pm$0.4 (2.5$\pm$0.4$\sigma$) \\ 2.8$\pm$0.1 (4.2$\pm$0.6$\sigma$) \\ 2.7$\pm$0.6 (2.6$\pm$0.7$\sigma$) \\ 2.7$\pm$0.3 (4.2$\pm$0.7$\sigma$) \\ 2.7$\pm$0.8 (2.5$\pm$0.9$\sigma$)} 
                            \\\cline{2-4}      
                           & Medium & 
                           \makecell{Type Ia \\ Type Ib \\ Type IIa \\ Type IIb \\ Type IIIa \\ Type IIIb \\ Type IVa \\ Type IVb
                            \\ Type Va \\ Type Vb \\ Type VIa \\ Type VIb \\ Type VIIa \\ Type VIIb \\ Type VIIIa \\ Type VIIIb} &
                            \makecell{\textbf{3.00$\pm$0.02 (5.2$\pm$0.1$\sigma$)} \\ 3.00$\pm$0.05 (4.0$\pm$0.5$\sigma$) \\ \textbf{3.00$\pm$0.03 (5.2$\pm$0.2$\sigma$)} \\ 3.00$\pm$0.05 (4.0$\pm$0.5$\sigma$) \\ \textbf{3.00$\pm$0.04 (5.2$\pm$0.4$\sigma$)} \\ 3.0$\pm$0.6 (3.1$\pm$0.6$\sigma$) \\ \textbf{3.00$\pm$0.04 (5.1$\pm$0.5$\sigma$)} \\ 3.1$\pm$0.4 (2.7$\pm$0.5$\sigma$) \\ 3.00$\pm$0.04 (4.7$\pm$0.5$\sigma$) \\ 3.1$\pm$0.3 (2.7$\pm$0.4$\sigma$) \\ 3.0$\pm$0.2 (4.5$\pm$0.7$\sigma$) \\ 3.2$\pm$0.2 (2.7$\pm$0.5$\sigma$) \\ 2.9$\pm$0.4 (4.0$\pm$0.7$\sigma$) \\ 2.8$\pm$0.8 (2.4$\pm$0.3$\sigma$) \\ 2.8$\pm$0.6 (3.8$\pm$0.8$\sigma$) \\ 2.9$\pm$0.5 (2.4$\pm$0.4$\sigma$)} 
                             \\\cline{2-4}
                            & End & 
                            \makecell{Type Ia \\ Type Ib \\ Type IIa \\ Type IIb \\ Type IIIa \\ Type IIIb \\ Type IVa \\ Type IVb
                            \\ Type Va \\ Type Vb \\ Type VIa \\ Type VIb \\ Type VIIa \\ Type VIIb \\ Type VIIIa \\ Type VIIIb} &
                            \makecell{\textbf{3.00$\pm$0.02 (5.2$\pm$0.1$\sigma$)} \\ \textbf{3.00$\pm$0.05 (5.2$\pm$0.2$\sigma$)} \\ \textbf{3.00$\pm$0.03 (5.2$\pm$0.2$\sigma$)} \\ \textbf{3.0$\pm$0.1 (5.0$\pm$0.4$\sigma$)} \\ \textbf{3.00$\pm$0.04 (5.0$\pm$0.4$\sigma$)} \\ 3.0$\pm$0.1 (2.9$\pm$0.5$\sigma$) \\ \textbf{3.0$\pm$0.1 (5.0$\pm$0.4$\sigma$)} \\ 3.0$\pm$0.1 (2.7$\pm$0.4$\sigma$) \\ 3.0$\pm$0.1 (4.6$\pm$0.6$\sigma$) \\ 2.9$\pm$0.3 (2.5$\pm$0.4$\sigma$) \\ 2.9$\pm$0.2 (4.5$\pm$0.6$\sigma$) \\ 2.8$\pm$0.7 (2.2$\pm$0.6$\sigma$) \\ 2.7$\pm$0.4 (4.3$\pm$0.7$\sigma$) \\ 3.1$\pm$0.4 (2.5$\pm$0.8$\sigma$) \\ 2.7$\pm$0.5 (3.9$\pm$0.8$\sigma$) \\ 3.2$\pm$0.5 (2.4$\pm$0.9$\sigma$)} \\ 
\hline
\hline
\end{tabular}%
}
\end{table*}

\begin{table*}
\centering
\caption{The evaluation of periodicity detection for signals with periods of 2 and 3 years applying the sigma clipping technique with 3$\sigma$ of threshold, considering different types of flares shown in Table \ref{tab:flare_types}. In bold, we highlight the tests where at least two methods report a compatible period with a significance of $\geq$5$\sigma$. A compatible period is defined as achieving a period that matches the signal period within a tolerance of $\pm$0.1. All periods are expressed in years. We also estimate the number of cycles required to achieve a period detection significance of 5$\sigma$ (only performed for the periodic signal with a period of 2 years). The symbol \textit{--} denotes that a 5$\sigma$ detection was achieved in the initial analysis, while \textit{X} values indicate that it was not possible to obtain the estimation within a limit of 10 cycles.
\label{tab:experiment_general_sigma_cipping}}
%\resizebox{\columnwidth}{!}
{%
\begin{tabular}{ccccccccc}
\hline
\hline
Period [yr] & Flux of Flare & LSP & Cycles & CWT & Cycles & PDM & Cycles \\
\hline
		\multirow{4}{*}{2 yr} & 
                            \makecell{Type Ia \\ Type Ib} &
                            \makecell{\textbf{2.00$\pm$0.02 (5.4$\pm$0.3$\sigma$)} \\ \textbf{2.00$\pm$0.02 (5.4$\pm$0.3$\sigma$)}} & 
                            \makecell{-- \\ -- } &
                            \makecell{\textbf{2.00$\pm$0.02 (5.3$\pm$0.2$\sigma$)} \\ \textbf{2.00$\pm$0.02 (5.3$\pm$0.2$\sigma$)}} &  
                            \makecell{-- \\ -- } &
                            \makecell{\textbf{2.00$\pm$0.03 (5.3$\pm$0.2$\sigma$)} \\ \textbf{2.00$\pm$0.03 (5.3$\pm$0.2$\sigma$)}} & 
                            \makecell{-- \\ -- } \\ \cline{2-8} 
                            & \makecell{Type IIa \\ Type IIb} &
                            \makecell{\textbf{2.00$\pm$0.02 (5.4$\pm$0.3$\sigma$)} \\ \textbf{2.00$\pm$0.02 (5.3$\pm$0.3$\sigma$)}} &  
                            \makecell{-- \\ -- } &
                            \makecell{\textbf{2.00$\pm$0.02 (5.3$\pm$0.2$\sigma$)} \\ \textbf{2.00$\pm$0.02 (5.3$\pm$0.2$\sigma$)}} &  
                            \makecell{-- \\ -- } &
                            \makecell{\textbf{2.00$\pm$0.02 (5.3$\pm$0.3$\sigma$)} \\ \textbf{2.00$\pm$0.02 (5.3$\pm$0.3$\sigma$)}} & 
                            \makecell{-- \\ -- } \\ 
                            \cline{2-8}
                            & \makecell{Type IIIa \\ Type IIIb} &
                            \makecell{\textbf{2.00$\pm$0.02 (5.3$\pm$0.3$\sigma$)} \\ \textbf{2.00$\pm$0.02 (5.3$\pm$0.3$\sigma$)}} & 
                            \makecell{ -- \\ -- } &
                            \makecell{\textbf{2.00$\pm$0.03 (5.3$\pm$0.2$\sigma$)} \\ \textbf{1.9$\pm$0.1 (5.2$\pm$0.3$\sigma$)}}&  
                            \makecell{ -- \\ -- } &
                            \makecell{\textbf{2.0$\pm$0.1 (5.2$\pm$0.3$\sigma$)} \\ \textbf{2.0$\pm$0.1 (5.2$\pm$0.3$\sigma$)}} & 
                            \makecell{ -- \\ -- } \\ 
                            \cline{2-8}
                            & \makecell{Type IVa \\ Type IVb} &
                            \makecell{\textbf{2.00$\pm$0.02 (5.2$\pm$0.4$\sigma$)} \\ \textbf{2.00$\pm$0.02 (5.2$\pm$0.3$\sigma$)}} &  
                            \makecell{ -- \\ -- } &
                            \makecell{\textbf{1.9$\pm$0.1 (5.2$\pm$0.3$\sigma$)} \\ \textbf{1.9$\pm$0.1 (5.2$\pm$0.3$\sigma$)}} &  
                            \makecell{ -- \\ -- } &
                            \makecell{\textbf{2.0$\pm$0.2 (5.1$\pm$0.4$\sigma$)} \\ \textbf{2.0$\pm$0.1 (5.2$\pm$0.3$\sigma$)}} & 
                            \makecell{ -- \\ -- } \\ 
                            \cline{2-8}
                            & \makecell{Type Va \\ Type Vb} &
                            \makecell{\textbf{2.00$\pm$0.02 (5.0$\pm$0.4$\sigma$)} \\ \textbf{2.00$\pm$0.02 (5.0$\pm$0.3$\sigma$)}} &  
                            \makecell{ -- \\ -- } &
                            \makecell{\textbf{1.9$\pm$0.1 (5.2$\pm$0.2$\sigma$)} \\ \textbf{1.9$\pm$0.1 (5.0$\pm$0.5$\sigma$)}} & 
                            \makecell{ -- \\ -- } &
                            \makecell{\textbf{2.0$\pm$0.3 (5.0$\pm$0.6$\sigma$)} \\ \textbf{2.0$\pm$0.3 (5.1$\pm$0.5$\sigma$)}} & 
                            \makecell{ -- \\ -- } \\ 
                            \cline{2-8}
                            & \makecell{Type VIa \\ Type VIb} &
                            \makecell{2.00$\pm$0.03 (4.9$\pm$0.4$\sigma$) \\ 2.00$\pm$0.02 (4.8$\pm$0.3$\sigma$)} & 
                            \makecell{ 1 \\ 1 } &
                            \makecell{\textbf{1.9$\pm$0.1 (5.0$\pm$0.5$\sigma$)} \\ 2.0$\pm$0.1 (4.5$\pm$0.6$\sigma$)} &  
                            \makecell{ -- \\ 2 } &
                            \makecell{\textbf{2.0$\pm$0.5 (5.0$\pm$0.7$\sigma$)} \\ 2.0$\pm$0.4 (5.1$\pm$0.6$\sigma$)} & 
                            \makecell{ -- \\ -- } \\ 
                            \cline{2-8}
                            &\makecell{Type VIIa \\ Type VIIb} &
                            \makecell{2.00$\pm$0.03 (4.7$\pm$0.3$\sigma$) \\ 2.00$\pm$0.03 (4.6$\pm$0.3$\sigma$)} &  
                            \makecell{ 1 \\ 1 } &
                            \makecell{1.9$\pm$0.2 (4.8$\pm$0.6$\sigma$) \\ 2.1$\pm$0.2 (4.4$\pm$0.6$\sigma$)} & 
                            \makecell{1 \\ 3 } &
                            \makecell{2.0$\pm$0.9 (4.7$\pm$0.9$\sigma$) \\ 2.0$\pm$0.4 (4.8$\pm$0.7$\sigma$)} & 
                            \makecell{1 \\ 1 } \\ 
                            \cline{2-8}
                            & \makecell{Type VIIIa \\ Type VIIIb} &
                            \makecell{2.00$\pm$0.03 (4.4$\pm$0.4$\sigma$) \\ 2.00$\pm$0.03 (4.5$\pm$0.4$\sigma$)} &  
                            \makecell{1 \\ 1 } &
                            \makecell{2.0$\pm$0.2 (4.5$\pm$0.6$\sigma$) \\ 2.1$\pm$0.2 (4.6$\pm$0.6$\sigma$)} &  
                            \makecell{2 \\ 2 } &
                            \makecell{2.0$\pm$1.2 (4.2$\pm$1.5$\sigma$) \\ 2.0$\pm$0.6 (4.5$\pm$1.0$\sigma$)} & 
                            \makecell{1 \\ 1 } \\  
                                                    
    \hline 
    \multirow{4}{*}{3 yr} & 
                            \makecell{Type Ia \\ Type Ib} &
                            \makecell{\textbf{3.00$\pm$0.03 (5.0$\pm$0.2$\sigma$)} \\ \textbf{3.00$\pm$0.03 (5.0$\pm$0.2$\sigma$)}} &  
                            \makecell{ \\ } &
                            \makecell{\textbf{3.00$\pm$0.04 (5.3$\pm$0.2$\sigma$)} \\ \textbf{3.00$\pm$0.04 (5.3$\pm$0.2$\sigma$)}} &  
                            \makecell{ \\ } &
                            \makecell{\textbf{3.00$\pm$0.03 (5.3$\pm$0.2$\sigma$)} \\ \textbf{3.00$\pm$0.03 (5.3$\pm$0.2$\sigma$)}} & 
                            \makecell{ \\ } \\ 
                            \cline{2-8} 
                            & \makecell{Type IIa \\ Type IIb} &
                            \makecell{\textbf{3.00$\pm$0.03 (5.0$\pm$0.2$\sigma$)} \\ \textbf{3.00$\pm$0.03 (5.0$\pm$0.2$\sigma$)}} &  
                            \makecell{ \\ } &
                            \makecell{\textbf{3.00$\pm$0.05 (5.2$\pm$0.2$\sigma$)} \\ \textbf{3.00$\pm$0.06 (5.3$\pm$0.3$\sigma$)}} &  
                            \makecell{ \\ } &
                            \makecell{\textbf{3.00$\pm$0.03 (5.2$\pm$0.3$\sigma$)} \\ \textbf{3.00$\pm$0.03 (5.2$\pm$0.2$\sigma$)}} & 
                            \makecell{ \\ } \\  
                            \cline{2-8}
                            & \makecell{Type IIIa \\ Type IIIb} &
                            \makecell{\textbf{3.00$\pm$0.03 (5.0$\pm$0.2$\sigma$)} \\ \textbf{3.00$\pm$0.03 (5.0$\pm$0.2$\sigma$)}} &  
                            \makecell{ \\ } &
                            \makecell{\textbf{2.9$\pm$0.1 (5.2$\pm$0.2$\sigma$)} \\ \textbf{2.9$\pm$0.1 (5.2$\pm$0.3$\sigma$)}} &  
                            \makecell{ \\ } &
                            \makecell{\textbf{3.00$\pm$0.03 (5.1$\pm$0.3$\sigma$)} \\ \textbf{3.00$\pm$0.03 (5.0$\pm$0.3$\sigma$)}} & 
                            \makecell{ \\ } \\ 
                            \cline{2-8}
                            & \makecell{Type IVa \\ Type IVb} &
                            \makecell{3.00$\pm$0.04 (4.8$\pm$0.3$\sigma$) \\ 3.00$\pm$0.03 (4.8$\pm$0.3$\sigma$)} &  
                            \makecell{ \\ } &
                            \makecell{2.8$\pm$0.1 (5.1$\pm$0.3$\sigma$) \\ 2.8$\pm$0.1 (5.1$\pm$0.4$\sigma$)} & 
                            \makecell{ \\ } &
                            \makecell{3.0$\pm$0.1 (5.0$\pm$0.4$\sigma$) \\ 3.00$\pm$0.03 (5.0$\pm$0.3$\sigma$)} & 
                            \makecell{ \\ } \\   
                            \cline{2-8}
                            & \makecell{Type Va \\ Type Vb} &
                            \makecell{3.00$\pm$0.04 (4.6$\pm$0.3$\sigma$) \\ 3.00$\pm$0.04 (4.7$\pm$0.3$\sigma$)} &  
                            \makecell{ \\ } &
                            \makecell{2.8$\pm$0.1 (5.0$\pm$0.4$\sigma$) \\ 2.8$\pm$0.1 (4.9$\pm$0.3$\sigma$)} &  
                            \makecell{ \\ } &
                            \makecell{3.0$\pm$0.2 (4.8$\pm$0.4$\sigma$) \\ 3.00$\pm$0.03 (4.9$\pm$0.3$\sigma$)} & 
                            \makecell{ \\ } \\   
                            \cline{2-8}
                            & \makecell{Type VIa \\ Type VIb} &
                            \makecell{3.00$\pm$0.05 (4.5$\pm$0.5$\sigma$) \\ 3.00$\pm$0.04 (4.4$\pm$0.3$\sigma$)} &  
                            \makecell{ \\ } &
                            \makecell{2.8$\pm$0.1 (5.0$\pm$0.5$\sigma$) \\ 2.8$\pm$0.1 (4.8$\pm$0.4$\sigma$)} &  
                            \makecell{ \\ } &
                            \makecell{3.0$\pm$0.3 (4.6$\pm$0.5$\sigma$) \\ 3.0$\pm$0.1 (4.6$\pm$0.3$\sigma$)} &  
                            \makecell{ \\ } \\ 
                            \cline{2-8}
                            &\makecell{Type VIIa \\ Type VIIb} &
                            \makecell{3.00$\pm$0.04 (4.3$\pm$0.4$\sigma$) \\ 3.00$\pm$0.04 (4.2$\pm$0.3$\sigma$)} & 
                            \makecell{ \\ } &
                            \makecell{2.8$\pm$0.2 (4.8$\pm$0.6$\sigma$) \\ 2.7$\pm$0.3 (4.4$\pm$0.5$\sigma$)} &  
                            \makecell{ \\ } &
                            \makecell{3.0$\pm$0.5 (4.4$\pm$0.5$\sigma$) \\ 3.0$\pm$0.3 (4.4$\pm$0.4$\sigma$)} & 
                            \makecell{ \\ } \\   
                            \cline{2-8}
                            & \makecell{Type VIIIa \\ Type VIIIb} &
                            \makecell{3.00$\pm$0.06 (4.1$\pm$0.4$\sigma$) \\ 3.00$\pm$0.06 (4.0$\pm$0.3$\sigma$)} &  
                            \makecell{ \\ } &
                            \makecell{2.7$\pm$0.4 (4.6$\pm$0.7$\sigma$) \\ 2.8$\pm$0.4 (4.2$\pm$0.6$\sigma$)} &  
                            \makecell{ \\ } &
                            \makecell{3.0$\pm$0.8 (4.2$\pm$0.7$\sigma$) \\ 3.0$\pm$0.3 (4.2$\pm$0.4$\sigma$)} & 
                            \makecell{ \\ } \\              
\hline
\hline
\end{tabular}%
}
\end{table*}
\begin{table*}
\centering
\caption{The evaluation of periodicity detection for signals with periods of 2 and 3 years applying the sigma clipping technique with 3$\sigma$ of threshold, considering different types of flares in Table \ref{tab:flare_types} for the methods LSP, CWT, and PDM. This test consists of conspiring only 3 possible temporal positions of the flare in the signal (start, medium, and end), in phase with an oscillation of the sinusoidal cyclic. In bold, we highlight the tests where at least two methods report a compatible period with a significance of $\geq$5$\sigma$. A compatible period is defined as achieving a period that matches the signal period within a tolerance of $\pm$0.1. All periods are expressed in years.\label{tab:experiment_coincident_sigma_clipping}}
%\resizeboX$\pm$X{\columnwidth}{!}
{%
\begin{tabular}{l|ccccc}
\hline
\hline
Period [yr] & Position & Type of Flare & LSP & CWT & PDM \\
\hline
		\multirow{4}{*}{2 yr} & Start & 
                            \makecell{Type Ia \\ Type Ib \\ Type IIa \\ Type IIb \\ Type IIIa \\ Type IIIb \\ Type IVa \\ Type IVb
                            \\ Type Va \\ Type Vb \\ Type VIa \\ Type VIb \\ Type VIIa \\ Type VIIb \\ Type VIIIa \\ Type VIIIb} &
                            \makecell{\textbf{2.00$\pm$0.02 (5.3$\pm$0.3$\sigma$)} \\ \textbf{2.00$\pm$0.02 (5.4$\pm$0.2$\sigma$)} \\ \textbf{2.00$\pm$0.02 (5.3$\pm$0.2$\sigma$)} \\ \textbf{2.00$\pm$0.02 (5.3$\pm$0.3$\sigma$)} \\ \textbf{2.00$\pm$0.02 (5.3$\pm$0.3$\sigma$)} \\ \textbf{2.00$\pm$0.02 (5.3$\pm$0.3$\sigma$)}
                            \\ \textbf{2.00$\pm$0.02 (5.3$\pm$0.3$\sigma$)} \\ \textbf{2.00$\pm$0.02 (5.3$\pm$0.3$\sigma$)} \\ \textbf{2.00$\pm$0.02 (5.1$\pm$0.2$\sigma$)} \\ \textbf{2.00$\pm$0.02 (5.2$\pm$0.2$\sigma$)} \\ 2.00$\pm$0.03 (4.9$\pm$0.4$\sigma$) \\ \textbf{2.00$\pm$0.02 (5.0$\pm$0.3$\sigma$)} \\ 2.00$\pm$0.03 (4.7$\pm$0.4$\sigma$) \\ \textbf{2.00$\pm$0.03 (5.0$\pm$0.3$\sigma$)} \\ 2.00$\pm$0.03 (4.6$\pm$0.3$\sigma$) \\ 2.00$\pm$0.02 (4.9$\pm$0.3$\sigma$)} &
                            \makecell{\textbf{2.00$\pm$0.03 (5.3$\pm$0.3$\sigma$)} \\ \textbf{2.00$\pm$0.03 (5.3$\pm$0.3$\sigma$)} \\ \textbf{2.00$\pm$0.03 (5.3$\pm$0.3$\sigma$)} \\ \textbf{2.00$\pm$0.03 (5.3$\pm$0.3$\sigma$)} \\ \textbf{2.00$\pm$0.03 (5.3$\pm$0.3$\sigma$)} \\ \textbf{2.00$\pm$0.03 (5.3$\pm$0.3$\sigma$)}
                            \\ \textbf{2.00$\pm$0.02 (5.3$\pm$0.3$\sigma$)} \\ \textbf{2.00$\pm$0.02 (5.3$\pm$0.3$\sigma$)} \\ \textbf{2.00$\pm$0.02 (5.3$\pm$0.3$\sigma$)} \\ \textbf{2.00$\pm$0.02 (5.3$\pm$0.3$\sigma$)} \\ \textbf{2.00$\pm$0.02 (5.3$\pm$0.3$\sigma$)}\\ \textbf{2.00$\pm$0.02 (5.3$\pm$0.3$\sigma$)} \\ \textbf{2.00$\pm$0.02 (5.1$\pm$0.2$\sigma$)} \\ \textbf{2.00$\pm$0.02 (5.3$\pm$0.3$\sigma$)} \\ \textbf{2.00$\pm$0.02 (5.3$\pm$0.3$\sigma$)} \\ \textbf{2.00$\pm$0.02 (5.3$\pm$0.3$\sigma$)}} &
                            \makecell{\textbf{2.00$\pm$0.03 (5.3$\pm$0.2$\sigma$)} \\ \textbf{2.00$\pm$0.02 (5.2$\pm$0.3$\sigma$)} \\ \textbf{2.00$\pm$0.03 (5.3$\pm$0.2$\sigma$)} \\ \textbf{2.00$\pm$0.03 (5.3$\pm$0.2$\sigma$)} \\ \textbf{2.00$\pm$0.04 (5.3$\pm$0.2$\sigma$)} \\ \textbf{2.00$\pm$0.03 (5.3$\pm$0.2$\sigma$)}
                            \\ \textbf{2.0$\pm$0.2 (5.3$\pm$0.2$\sigma$)} \\ \textbf{2.0$\pm$0.2 (5.3$\pm$0.3$\sigma$)} \\ \textbf{1.9$\pm$0.3 (5.1$\pm$0.2$\sigma$)} \\ \textbf{2.0$\pm$0.2 (5.3$\pm$0.2$\sigma$)} \\ \textbf{2.0$\pm$0.8 (5.1$\pm$0.8$\sigma$)} \\ \textbf{2.0$\pm$0.4 (5.0$\pm$0.8$\sigma$)} \\ \textbf{2.0$\pm$0.8 (5.0$\pm$1.0$\sigma$)} \\ \textbf{2.0$\pm$0.5 (5.0$\pm$0.8$\sigma$)} \\ \textbf{2.0$\pm$1.0 (5.0$\pm$1.2$\sigma$)} \\ \textbf{2.0$\pm$0.6 (5.0$\pm$0.5$\sigma$)}} \\\cline{2-6}      
                           & Medium & 
                           \makecell{Type Ia \\ Type Ib \\ Type IIa \\ Type IIb \\ Type IIIa \\ Type IIIb \\ Type IVa \\ Type IVb
                            \\ Type Va \\ Type Vb \\ Type VIa \\ Type VIb \\ Type VIIa \\ Type VIIb \\ Type VIIIa \\ Type VIIIb} &
                            \makecell{\textbf{2.00$\pm$0.02 (5.4$\pm$0.2$\sigma$)} \\ \textbf{2.00$\pm$0.02 (5.4$\pm$0.2$\sigma$)} \\ \textbf{2.00$\pm$0.02 (5.4$\pm$0.2$\sigma$)} \\ \textbf{2.00$\pm$0.02 (5.4$\pm$0.2$\sigma$)} \\ \textbf{2.00$\pm$0.02 (5.4$\pm$0.2$\sigma$)} \\ \textbf{2.00$\pm$0.02 (5.4$\pm$0.2$\sigma$)}
                            \\ \textbf{2.00$\pm$0.02 (5.4$\pm$0.2$\sigma$)} \\ \textbf{2.00$\pm$0.02 (5.2$\pm$0.2$\sigma$)} \\ \textbf{2.00$\pm$0.02 (5.2$\pm$0.3$\sigma$)} \\ \textbf{2.00$\pm$0.02 (5.1$\pm$0.3$\sigma$)} \\ 2.00$\pm$0.02 (4.8$\pm$0.3$\sigma$) \\ 2.00$\pm$0.02 (4.9$\pm$0.3$\sigma$) \\ 2.00$\pm$0.02 (4.5$\pm$0.4$\sigma$) \\ 2.00$\pm$0.02 (4.6$\pm$0.3$\sigma$) \\ 2.00$\pm$0.02 (4.3$\pm$0.4$\sigma$) \\ 2.00$\pm$0.02 (4.6$\pm$0.3$\sigma$)} &
                            \makecell{\textbf{2.00$\pm$0.01 (5.3$\pm$0.3$\sigma$)} \\ \textbf{2.00$\pm$0.01 (5.3$\pm$0.2$\sigma$)} \\ \textbf{2.00$\pm$0.01 (5.4$\pm$0.3$\sigma$)} \\ \textbf{2.00$\pm$0.02 (5.4$\pm$0.3$\sigma$)} \\ \textbf{2.00$\pm$0.01 (5.4$\pm$0.3$\sigma$)} \\ \textbf{2.00$\pm$0.01 (5.4$\pm$0.3$\sigma$)}
                            \\ \textbf{1.9$\pm$0.1 (5.3$\pm$0.3$\sigma$)} \\ 1.8$\pm$0.1 (5.2$\pm$0.2$\sigma$) \\ 1.8$\pm$0.1 (5.2$\pm$0.2$\sigma$) \\ 1.8$\pm$0.1 (5.0$\pm$0.4$\sigma$)\\ 1.8$\pm$0.1 (5.2$\pm$0.3$\sigma$) \\ 1.8$\pm$0.2 (4.5$\pm$0.4$\sigma$) \\ 1.8$\pm$0.2 (4.4$\pm$0.4$\sigma$) \\ 2.3$\pm$0.2 (4.0$\pm$0.3$\sigma$) \\ 2.3$\pm$0.2 (4.0$\pm$0.4$\sigma$) \\ 2.3$\pm$0.3 (4.5$\pm$0.3$\sigma$)} &
                            \makecell{\textbf{2.00$\pm$0.03 (5.3$\pm$0.3$\sigma$)} \\ \textbf{2.00$\pm$0.03 (5.3$\pm$0.3$\sigma$)} \\ \textbf{2.00$\pm$0.03 (5.3$\pm$0.2$\sigma$)} \\ \textbf{2.00$\pm$0.03 (5.2$\pm$0.2$\sigma$)} \\ \textbf{2.00$\pm$0.03 (5.3$\pm$0.2$\sigma$)} \\ \textbf{2.00$\pm$0.03 (5.3$\pm$0.2$\sigma$)}
                            \\ \textbf{2.0$\pm$0.2 (5.3$\pm$0.2$\sigma$)} \\ \textbf{2.0$\pm$0.1 (5.3$\pm$0.3$\sigma$)} \\ \textbf{1.9$\pm$0.2 (5.1$\pm$0.6$\sigma$)} \\ \textbf{2.0$\pm$0.2 (5.2$\pm$0.4$\sigma$)} \\ 2.0$\pm$0.5 (5.1$\pm$0.8$\sigma$) \\ 2.0$\pm$0.2 (5.1$\pm$0.5$\sigma$) \\ 2.0$\pm$0.8 (5.0$\pm$1.1$\sigma$) \\ 2.0$\pm$0.5 (5.1$\pm$0.7$\sigma$) \\ 2.0$\pm$0.8 (5.0$\pm$1.1$\sigma$) \\ 2.0$\pm$0.4 (5.0$\pm$0.6$\sigma$)} \\\cline{2-6}
                            & End & 
                            \makecell{Type Ia \\ Type Ib \\ Type IIa \\ Type IIb \\ Type IIIa \\ Type IIIb \\ Type IVa \\ Type IVb
                            \\ Type Va \\ Type Vb \\ Type VIa \\ Type VIb \\ Type VIIa \\ Type VIIb \\ Type VIIIa \\ Type VIIIb} &
                            \makecell{\textbf{2.00$\pm$0.02 (5.4$\pm$0.3$\sigma$)} \\ \textbf{2.00$\pm$0.02 (5.4$\pm$0.3$\sigma$)} \\ \textbf{2.00$\pm$0.02 (5.4$\pm$0.3$\sigma$)} \\ \textbf{2.00$\pm$0.02 (5.4$\pm$0.3$\sigma$)} \\ \textbf{2.00$\pm$0.02 (5.4$\pm$0.3$\sigma$)} \\ \textbf{2.00$\pm$0.02 (5.4$\pm$0.3$\sigma$)}
                            \\ \textbf{2.00$\pm$0.02 (5.4$\pm$0.3$\sigma$)} \\ \textbf{2.00$\pm$0.02 (5.4$\pm$0.3$\sigma$)} \\ \textbf{2.00$\pm$0.02 (5.1$\pm$0.3$\sigma$)} \\ \textbf{2.00$\pm$0.02 (5.1$\pm$0.3$\sigma$)} \\ 2.00$\pm$0.02 (4.8$\pm$0.3$\sigma$) \\ 2.00$\pm$0.02 (4.9$\pm$0.3$\sigma$) \\ 2.00$\pm$0.02 (4.5$\pm$0.4$\sigma$) \\ 2.00$\pm$0.02 (4.6$\pm$0.3$\sigma$) \\ 2.00$\pm$0.02 (4.3$\pm$0.4$\sigma$) \\ 2.00$\pm$0.02 (4.6$\pm$0.3$\sigma$)} &
                            \makecell{\textbf{2.00$\pm$0.01 (5.3$\pm$0.2$\sigma$)} \\ \textbf{2.00$\pm$0.01 (5.3$\pm$0.2$\sigma$)} \\ \textbf{2.00$\pm$0.02 (5.4$\pm$0.2$\sigma$)} \\ \textbf{2.00$\pm$0.03 (5.4$\pm$0.2$\sigma$)} \\ \textbf{2.00$\pm$0.02 (5.3$\pm$0.3$\sigma$)} \\ \textbf{2.00$\pm$0.02 (5.3$\pm$0.3$\sigma$)}
                            \\ \textbf{1.9$\pm$0.1 (5.4$\pm$0.3$\sigma$)} \\ 1.8$\pm$0.1 (5.4$\pm$0.3$\sigma$) \\ 1.8$\pm$0.1 (5.3$\pm$0.3$\sigma$) \\ 1.8$\pm$0.1 (5.1$\pm$0.3$\sigma$) \\ 1.8$\pm$0.1 (5.1$\pm$0.3$\sigma$) \\ 1.8$\pm$0.2 (4.5$\pm$0.4$\sigma$) \\ 1.8$\pm$0.2 (4.4$\pm$0.4$\sigma$) \\ 2.3$\pm$0.1 (4.0$\pm$0.3$\sigma$) \\ 2.3$\pm$0.2 (4.0$\pm$0.3$\sigma$) \\ 2.2$\pm$0.1 (4.5$\pm$0.3$\sigma$)} &
                            \makecell{\textbf{2.00$\pm$0.03 (5.3$\pm$0.3$\sigma$)} \\ \textbf{2.00$\pm$0.03 (5.4$\pm$0.2$\sigma$)} \\ \textbf{2.00$\pm$0.03 (5.3$\pm$0.2$\sigma$)} \\ \textbf{2.00$\pm$0.03 (5.3$\pm$0.2$\sigma$)} \\ \textbf{2.0$\pm$0.2 (5.3$\pm$0.3$\sigma$)} \\ \textbf{2.0$\pm$0.2 (5.3$\pm$0.3$\sigma$)}
                            \\ \textbf{2.0$\pm$0.3 (5.2$\pm$0.6$\sigma$)} \\ \textbf{2.0$\pm$0.3 (5.3$\pm$0.4$\sigma$)} \\ \textbf{1.9$\pm$0.2 (5.1$\pm$0.5$\sigma$)} \\ \textbf{2.0$\pm$0.4 (5.1$\pm$0.5$\sigma$)} \\ 2.0$\pm$0.8 (5.1$\pm$1.0$\sigma$) \\ 2.0$\pm$0.4 (5.2$\pm$0.6$\sigma$) \\ 2.0$\pm$0.6 (5.1$\pm$0.9$\sigma$) \\ 2.0$\pm$0.5 (5.1$\pm$0.8$\sigma$) \\ 2.0$\pm$0.7 (5.2$\pm$1.1$\sigma$) \\ 2.0$\pm$0.3 (5.2$\pm$0.5$\sigma$)} \\                 
\hline
\hline
\end{tabular}%
}
\end{table*}
\begin{table*}
\centering
\setcounter{table}{5}
\caption{\textit{(continued).\label{tab:experiment_coincident_sigma_clipping_b}}}
%\resizeboX$\pm$X{\columnwidth}{!}
{%
\begin{tabular}{l|ccccc}
\hline
\hline
Period [yr] & Position & Type of Flare & LSP & CWT & PDM \\
\hline
		\multirow{4}{*}{3 yr} & Start & 
                            \makecell{Type Ia \\ Type Ib \\ Type IIa \\ Type IIb \\ Type IIIa \\ Type IIIb \\ Type IVa \\ Type IVb
                            \\ Type Va \\ Type Vb \\ Type VIa \\ Type VIb \\ Type VIIa \\ Type VIIb \\ Type VIIIa \\ Type VIIIb} &
                            \makecell{\textbf{3.00$\pm$0.03 (5.0$\pm$0.2$\sigma$)} \\ \textbf{3.00$\pm$0.03 (5.0$\pm$0.2$\sigma$)} \\ \textbf{3.00$\pm$0.03 (5.0$\pm$0.2$\sigma$)} \\ \textbf{3.00$\pm$0.03 (5.0$\pm$0.2$\sigma$)} \\ \textbf{3.00$\pm$0.03 (5.0$\pm$0.2$\sigma$)} \\ \textbf{3.00$\pm$0.03 (5.0$\pm$0.2$\sigma$)}
                            \\ \textbf{3.00$\pm$0.03 (5.0$\pm$0.2$\sigma$)} \\ \textbf{3.00$\pm$0.03 (5.0$\pm$0.2$\sigma$)} \\ 3.00$\pm$0.03 (4.9$\pm$0.2$\sigma$) \\ 3.00$\pm$0.03 (4.8$\pm$0.2$\sigma$) \\ 3.00$\pm$0.03 (4.8$\pm$0.2$\sigma$) \\ 3.00$\pm$0.03 (4.7$\pm$0.2$\sigma$) \\ 3.00$\pm$0.03 (4.6$\pm$0.2$\sigma$) \\ 3.00$\pm$0.03 (4.4$\pm$0.2$\sigma$) \\ 3.00$\pm$0.03 (4.3$\pm$0.2$\sigma$) \\ 3.00$\pm$0.03 (4.1$\pm$0.2$\sigma$)} &
                            \makecell{\textbf{2.9$\pm$0.1 (5.4$\pm$0.2$\sigma$)} \\ \textbf{2.9$\pm$0.1 (5.4$\pm$0.2$\sigma$)} \\ \textbf{2.9$\pm$0.1 (5.4$\pm$0.2$\sigma$)} \\ 2.8$\pm$0.1 (5.4$\pm$0.2$\sigma$) \\ 2.8$\pm$0.1 (5.4$\pm$0.2$\sigma$) \\ 2.8$\pm$0.1 (5.3$\pm$0.2$\sigma$)
                            \\ 2.8$\pm$0.1 (5.3$\pm$0.2$\sigma$) \\ 2.8$\pm$0.1 (5.3$\pm$0.2$\sigma$) \\ 2.8$\pm$0.1 (5.4$\pm$0.2$\sigma$) \\ 2.7$\pm$0.1 (5.1$\pm$0.2$\sigma$) \\ 2.7$\pm$0.1 (5.2$\pm$0.2$\sigma$) \\ 2.6$\pm$0.1 (5.0$\pm$0.2$\sigma$) \\ 2.7$\pm$0.1 (5.0$\pm$0.2$\sigma$) \\ 2.6$\pm$0.1 (4.7$\pm$0.3$\sigma$) \\ 2.7$\pm$0.1 (4.8$\pm$0.2$\sigma$) \\ 2.5$\pm$0.1 (4.3$\pm$0.3$\sigma$)} &
                            \makecell{\textbf{3.00$\pm$0.03 (5.2$\pm$0.2$\sigma$)} \\ \textbf{3.00$\pm$0.03 (5.3$\pm$0.3$\sigma$)} \\ \textbf{3.00$\pm$0.03 (5.2$\pm$0.3$\sigma$)} \\ \textbf{3.00$\pm$0.02 (5.3$\pm$0.3$\sigma$)} \\ \textbf{3.00$\pm$0.03 (5.1$\pm$0.2$\sigma$)} \\ \textbf{3.00$\pm$0.03 (5.1$\pm$0.2$\sigma$)}
                            \\ \textbf{3.00$\pm$0.02 (5.1$\pm$0.2$\sigma$)} \\ \textbf{3.00$\pm$0.03 (5.0$\pm$0.2$\sigma$)} \\ 3.00$\pm$0.03 (5.0$\pm$0.2$\sigma$) \\ 3.00$\pm$0.03 (5.0$\pm$0.2$\sigma$) \\ 3.00$\pm$0.03 (4.9$\pm$0.3$\sigma$) \\ 3.00$\pm$0.03 (4.8$\pm$0.2$\sigma$) \\ 2.9$\pm$0.1 (4.7$\pm$0.2$\sigma$) \\ 3.00$\pm$0.03 (4.5$\pm$0.3$\sigma$) \\ 2.9$\pm$0.3 (4.5$\pm$0.3$\sigma$) \\ 3.00$\pm$0.03 (4.3$\pm$0.3$\sigma$)} \\\cline{2-6}      
                           & Medium & 
                           \makecell{Type Ia \\ Type Ib \\ Type IIa \\ Type IIb \\ Type IIIa \\ Type IIIb \\ Type IVa \\ Type IVb
                            \\ Type Va \\ Type Vb \\ Type VIa \\ Type VIb \\ Type VIIa \\ Type VIIb \\ Type VIIIa \\ Type VIIIb} &
                            \makecell{\textbf{3.00$\pm$0.03 (5.0$\pm$0.2$\sigma$)} \\ \textbf{3.00$\pm$0.03 (5.0$\pm$0.2$\sigma$)} \\ \textbf{3.00$\pm$0.03 (5.0$\pm$0.1$\sigma$)} \\ \textbf{3.00$\pm$0.03 (5.0$\pm$0.1$\sigma$)} \\ \textbf{3.00$\pm$0.03 (5.0$\pm$0.2$\sigma$)} \\ \textbf{3.00$\pm$0.03 (5.0$\pm$0.2$\sigma$)}
                            \\ \textbf{3.00$\pm$0.03 (5.0$\pm$0.2$\sigma$)} \\ \textbf{3.00$\pm$0.03 (5.0$\pm$0.2$\sigma$)} \\ 3.00$\pm$0.03 (4.9$\pm$0.2$\sigma$) \\ 3.00$\pm$0.03 (4.8$\pm$0.2$\sigma$) \\ 3.00$\pm$0.03 (4.7$\pm$0.2$\sigma$) \\ 3.00$\pm$0.03 (4.6$\pm$0.2$\sigma$) \\ 3.00$\pm$0.03 (4.4$\pm$0.2$\sigma$) \\ 3.00$\pm$0.03 (4.3$\pm$0.2$\sigma$) \\ 3.00$\pm$0.03 (4.1$\pm$0.3$\sigma$) \\ 3.00$\pm$0.03 (4.1$\pm$0.3$\sigma$)} &
                            \makecell{\textbf{2.9$\pm$0.1 (5.3$\pm$0.3$\sigma$)} \\ \textbf{2.9$\pm$0.1 (5.3$\pm$0.3$\sigma$)} \\ \textbf{2.9$\pm$0.1 (5.3$\pm$0.3$\sigma$)} \\ 2.8$\pm$0.1 (5.3$\pm$0.3$\sigma$) \\ 2.8$\pm$0.1 (5.3$\pm$0.2$\sigma$) \\ 2.8$\pm$0.1 (5.3$\pm$0.3$\sigma$)
                            \\ 2.8$\pm$0.1 (5.3$\pm$0.3$\sigma$) \\ 2.8$\pm$0.1 (5.3$\pm$0.3$\sigma$) \\ 2.8$\pm$0.1 (5.2$\pm$0.2$\sigma$) \\ 2.7$\pm$0.1 (5.1$\pm$0.3$\sigma$) \\ 2.7$\pm$0.1 (5.2$\pm$0.2$\sigma$) \\ 2.6$\pm$0.1 (5.0$\pm$0.3$\sigma$) \\ 2.7$\pm$0.1 (5.1$\pm$0.2$\sigma$) \\ 2.6$\pm$0.1 (4.7$\pm$0.3$\sigma$) \\ 2.7$\pm$0.1 (4.8$\pm$0.3$\sigma$) \\ 2.5$\pm$0.1 (4.3$\pm$0.3$\sigma$)} &
                            \makecell{\textbf{3.00$\pm$0.03 (5.3$\pm$0.2$\sigma$)} \\ \textbf{3.00$\pm$0.03 (5.3$\pm$0.2$\sigma$)} \\ \textbf{3.00$\pm$0.03 (5.3$\pm$0.2$\sigma$)} \\ \textbf{3.00$\pm$0.03 (5.3$\pm$0.2$\sigma$)} \\ \textbf{3.00$\pm$0.03 (5.2$\pm$0.2$\sigma$)} \\ \textbf{3.00$\pm$0.03 (5.1$\pm$0.2$\sigma$)}
                            \\ \textbf{3.00$\pm$0.03 (5.1$\pm$0.2$\sigma$)} \\ \textbf{3.00$\pm$0.03 (5.0$\pm$0.2$\sigma$)} \\ 3.00$\pm$0.03 (5.0$\pm$0.2$\sigma$) \\ 3.00$\pm$0.03 (5.0$\pm$0.2$\sigma$) \\ 3.00$\pm$0.02 (4.9$\pm$0.2$\sigma$) \\ 3.00$\pm$0.02 (4.8$\pm$0.2$\sigma$) \\ 2.9$\pm$0.1 (4.7$\pm$0.3$\sigma$) \\ 3.00$\pm$0.02 (4.5$\pm$0.3$\sigma$) \\ 2.9$\pm$0.4 (4.5$\pm$0.4$\sigma$) \\ 3.00$\pm$0.03 (4.3$\pm$0.3$\sigma$)} \\\cline{2-6}
                            & End & 
                            \makecell{Type Ia \\ Type Ib \\ Type IIa \\ Type IIb \\ Type IIIa \\ Type IIIb \\ Type IVa \\ Type IVb
                            \\ Type Va \\ Type Vb \\ Type VIa \\ Type VIb \\ Type VIIa \\ Type VIIb \\ Type VIIIa \\ Type VIIIb} &
                            \makecell{\textbf{3.00$\pm$0.03 (5.0$\pm$0.1$\sigma$)} \\ \textbf{3.00$\pm$0.03 (5.0$\pm$0.2$\sigma$)} \\ \textbf{3.00$\pm$0.03 (5.0$\pm$0.2$\sigma$)} \\ \textbf{3.00$\pm$0.03 (5.0$\pm$0.2$\sigma$)} \\ \textbf{3.00$\pm$0.03 (5.0$\pm$0.2$\sigma$)} \\ \textbf{3.00$\pm$0.03 (5.0$\pm$0.2$\sigma$)}
                            \\ \textbf{3.00$\pm$0.03 (5.0$\pm$0.2$\sigma$)} \\ 3.00$\pm$0.03 (4.9$\pm$0.2$\sigma$) \\ 3.00$\pm$0.03 (4.9$\pm$0.2$\sigma$) \\ 3.00$\pm$0.03 (4.8$\pm$0.2$\sigma$) \\ 3.00$\pm$0.03 (4.8$\pm$0.2$\sigma$) \\ 3.00$\pm$0.03 (4.6$\pm$0.2$\sigma$) \\ 3.00$\pm$0.03 (4.6$\pm$0.2$\sigma$) \\ 3.00$\pm$0.03 (4.4$\pm$0.2$\sigma$) \\ 3.00$\pm$0.03 (4.3$\pm$0.2$\sigma$) \\ 3.00$\pm$0.03 (4.1$\pm$0.2$\sigma$)} &
                            \makecell{\textbf{3.00$\pm$0.03 (5.4$\pm$0.2$\sigma$)} \\ \textbf{3.00$\pm$0.03 (5.4$\pm$0.2$\sigma$)} \\ \textbf{3.00$\pm$0.03 (5.4$\pm$0.2$\sigma$)} \\ \textbf{3.00$\pm$0.04 (5.4$\pm$0.2$\sigma$)} \\ \textbf{3.00$\pm$0.03 (5.4$\pm$0.2$\sigma$)} \\ \textbf{3.00$\pm$0.03 (5.4$\pm$0.2$\sigma$)}
                            \\ \textbf{3.00$\pm$0.05 (5.3$\pm$0.2$\sigma$)} \\ \textbf{3.00$\pm$0.04 (5.4$\pm$0.2$\sigma$)} \\ \textbf{3.00$\pm$0.05 (5.2$\pm$0.2$\sigma$)} \\ \textbf{3.00$\pm$0.04 (5.3$\pm$0.2$\sigma$)} \\ 2.9$\pm$0.1 (5.1$\pm$0.2$\sigma$) \\ 2.9$\pm$0.1 (5.1$\pm$0.2$\sigma$) \\ 2.9$\pm$0.1 (5.0$\pm$0.2$\sigma$) \\ 2.9$\pm$0.1 (4.9$\pm$0.3$\sigma$) \\ 2.8$\pm$0.1 (4.6$\pm$0.3$\sigma$) \\ 2.9$\pm$0.1 (4.6$\pm$0.3$\sigma$)} &
                            \makecell{\textbf{3.00$\pm$0.03 (5.3$\pm$0.3$\sigma$)} \\ \textbf{3.00$\pm$0.03 (5.3$\pm$0.3$\sigma$)} \\ \textbf{3.00$\pm$0.03 (5.2$\pm$0.2$\sigma$)} \\ \textbf{3.00$\pm$0.03 (5.2$\pm$0.3$\sigma$)} \\ \textbf{3.00$\pm$0.03 (5.2$\pm$0.2$\sigma$)} \\ \textbf{3.00$\pm$0.03 (5.1$\pm$0.2$\sigma$)}
                            \\ \textbf{3.00$\pm$0.03 (5.0$\pm$0.2$\sigma$)} \\ \textbf{3.00$\pm$0.03 (5.0$\pm$0.2$\sigma$)} \\ \textbf{3.00$\pm$0.03 (5.0$\pm$0.3$\sigma$)} \\ \textbf{3.00$\pm$0.03 (5.0$\pm$0.2$\sigma$)} \\ 3.00$\pm$0.03 (4.9$\pm$0.3$\sigma$) \\ 3.00$\pm$0.03 (4.8$\pm$0.2$\sigma$) \\ 3.00$\pm$0.03 (4.7$\pm$0.2$\sigma$) \\ 3.00$\pm$0.03 (4.5$\pm$0.3$\sigma$) \\ 3.00$\pm$0.03 (4.5$\pm$0.3$\sigma$) \\ 3.00$\pm$0.03 (4.3$\pm$0.3$\sigma$)} \\                 
\hline
\hline
\end{tabular}%
}
\end{table*}

\clearpage

% Don't change these lines
\bsp	% typesetting comment
\label{lastpage}
\end{document}